\documentclass[aps,prx,10pt,twocolumn,superscriptaddress,nobibnotes]{revtex4-2}

\usepackage{amsthm, color, mathtools, amsmath, amssymb, mathtools, graphicx, float, verbatim, xcolor, placeins, needspace}
\usepackage[caption=false]{subfig}
\usepackage{enumerate}
\usepackage{braket}
\usepackage{url}
\usepackage{algorithm}
\usepackage{algpseudocode}
\usepackage{tikz}
\usetikzlibrary{fit,calc}
\usepackage{microtype} 
\usepackage[british]{babel} 	
\usepackage[unicode=true,bookmarks=true,bookmarksnumbered=false,bookmarksopen=false,breaklinks=false,pdfborder={0 0 1}, backref=false,colorlinks=true]{hyperref}
\addto\captionsbritish{}
\addto\captionsbritish{}
\usepackage{orcidlink}
\usepackage{xr-hyper}

\newcommand{\todai}{Department of Physics, Graduate School of Science, The University of Tokyo, Hongo 7-3-1, Bunkyo-ku, Tokyo, 113-0033, Japan}
\newcommand{\perimeter}{Perimeter Institute for Theoretical Physics, 31 Caroline Street North, Waterloo, Ontario, N2L 2Y5, Canada}
\newcommand{\iqc}{Institute for Quantum Computing, University of Waterloo, 200 University Avenue West, Waterloo, Ontario, N2L 3G1, Canada}
\newcommand{\umontreal}{Department of Computer Science and Operations Research, Universit{\' e} de Montr{\' e}al, Montr{\' e}al, Qu{\' e}bec, H3T 1J4, Canada}
\newcommand{\mila}{Mila -- Qu{\' e}bec AI Institute, Montr{\' e}al, Qu{\' e}bec, H2S 3H1, Canada}
\newcommand{\manchester}{Department of Physics \& Astronomy, University of Manchester, Manchester M13 9PL, United Kingdom}
\newcommand{\transcale}{Trans-Scale Quantum Science Institute, The University of Tokyo, Hongo 7-3-1, Bunkyo-ku, Tokyo 113-0033, Japan}

\newtheorem{de}{Definition}
\newtheorem{lem}{Lemma}

\newtheorem{Theorem}{Theorem}

\colorlet{pink}{red!40}

\newcounter{algsubstate}

\newenvironment{algsubstates}
  {\setcounter{algsubstate}{0}%
   \renewcommand{\State}{%
     \refstepcounter{algsubstate}%
     \Statex {\footnotesize \arabic{ALG@line}.\arabic{algsubstate}:}\space}}
  {}

\makeatletter
\newcommand*{\addFileDependency}[1]{
\typeout{(#1)}
\@addtofilelist{#1}
\IfFileExists{#1}{}{\typeout{No file #1.}}
}\makeatother

\newcommand*{\myexternaldocument}[1]{%
\externaldocument{#1}%
\addFileDependency{#1.tex}%
\addFileDependency{#1.aux}%
}

\myexternaldocument{prr_supp}

\renewcommand{\thealgorithm}{\arabic{algorithm}}
\let\oldcomment\algorithmiccomment
\renewcommand{\algorithmiccomment}[1]{\textsf{ \oldcomment{#1}}}

\definecolor{darkgreen}{RGB}{0,200,107}

\definecolor{darkorange}{RGB}{255,100,0}

\begin{document}

\title{Universal algorithm for transforming Hamiltonian eigenvalues}

\author{Tatsuki Odake\,\orcidlink{0009-0009-6844-9959}}
\affiliation{\todai}

\author{Hl\'er Kristj\'ansson\,\orcidlink{0000-0003-4465-2863}}
\affiliation{\perimeter}
\affiliation{\iqc}
\affiliation{\todai}
\affiliation{\umontreal}
\affiliation{\mila}

\author{Philip Taranto\,\orcidlink{0000-0002-4247-3901}}
\affiliation{\manchester}
\affiliation{\todai}

\author{Mio Murao\,\orcidlink{0000-0001-7861-1774}}
\email{murao@phys.s.u-tokyo.ac.jp}
\affiliation{\todai}
\affiliation{\transcale}

\begin{abstract}
Manipulating Hamiltonians governing physical systems has found a broad range of applications, from quantum chemistry to semiconductor design. In this work, we provide a new way of manipulating Hamiltonians, by transforming their eigenvalues while keeping their eigenstates unchanged. We develop a universal algorithm that deterministically implements any desired (suitably differentiable) function on the eigenvalues of any unknown Hamiltonian, whose positive-time and negative-time dynamics are given as a black box. Our algorithm uses correlated randomness to efficiently combine two subroutines---namely controlization and Fourier series simulation---exemplifying a general compilation procedure that we develop. The time complexity of our algorithm is significantly reduced via said compilation technique compared to a na{\"i}ve concatenation of the subroutines and outperforms similar methods based on the quantum singular value transformation.
\end{abstract}

\maketitle

\section{Introduction}\label{sec::introduction}

The physical properties of a system are determined by its Hamiltonian. Thus, realizing a Hamiltonian that exhibits desireable properties in a given context is an important task in fields ranging from condensed matter physics and materials science to quantum chemistry and simulation; e.g., an important problem in materials science is the discovery of new materials for specific tasks \cite{jain2013commentary,tang2019efficient,Alexeev_2023}. 

In many cases, such properties depend on the eigenvalues or eigenstates of the Hamiltonian. The eigenstates of a Hamiltonian can be straightforwardly transformed by applying a unitary operation on the Hamiltonian. However, tuning the eigenvalues of a Hamiltonian while keeping the eigenstates fixed---as addressed by the \textbf{quantum singular value transformation (QSVT)} \cite{low2019hamiltonian, gilyen2019quantum,martyn2021grand}---is a much more involved problem. The ability to transform Hamiltonian eigenvalues in an efficient manner would open up a new way of manipulating the physical properties of a system, leading to increased flexibility in the simulation, control, and design of quantum systems.

In this work, we propose a quantum algorithm to transform the eigenvalues of a Hamiltonian by any given (suitably differentiable) function, while keeping the eigenstates unchanged, given access to the positive-time and negative-time Hamiltonian dynamics. In the standard setting of Hamiltonian simulation, a classical description or access to a block-encoded version of the Hamiltonian to be transformed is assumed~\cite{suzuki1990fractal,suzuki1991general,campbell2019random,berry2015simulating,low2017optimal,low2019hamiltonian,childs2018toward}. However, in many cases of interest---such as Hamiltonian learning~\cite{bakshi2024structure}---the description of the Hamiltonian of the system is not given but rather only its dynamics is accessible, especially for many-body or complex systems. 

Our algorithm provides an alternative \textit{universal} Hamiltonian simulation method that does not require knowledge of the Hamiltonian, whose \textit{dynamics} is given as a black box \cite{odake2023higher}. Although several works, including algorithms based on QSVT, are compatible with this black-box setting, they typically assume access to controlled versions of the Hamiltonian dynamics \cite{poulin2009sampling,zhang2012randomized,zhang2015time,Wang_2022,kosugi2022imaginary} or to an oracle encoding information of the Hamiltonian, such as a block encoding \cite{low2019hamiltonian} or a quantum walk oracle \cite{childs2012linear,berry2015hamiltonian,childs2017quantum}. 

Crucially, our approach returns to the more natural setting where only the \textit{uncontrolled} Hamiltonian dynamics is accessible. In practice, many physical systems, such as quantum simulators or analog quantum computers, naturally exhibit evolution under Hamiltonians that are difficult or impossible to control precisely. This is especially relevant in quantum simulation tasks, where one may have limited access to the system's Hamiltonian or lack the ability to directly perform gate-based transformations. In such cases, leveraging the system's inherent Hamiltonian dynamics without requiring full control can reduce resource overhead and offer a more efficient path for implementing quantum algorithms~\cite{Hangleiter_2023}.
Possible applications of our method include constructing oracles for Grover’s algorithm \cite{grover1997quantum}, and we envisage that further developments in this direction could find applications more broadly in condensed matter physics, chemistry, and material science, for example in tuning energy gaps to enable precise transitions between energy levels~\cite{zare1998laser} and in the design of semiconductors \cite{greiner2012universal}.

\FloatBarrier

\section{Summary of Main Results }\label{sec::summary}

\begin{figure}[t]
\includegraphics[width=9cm]{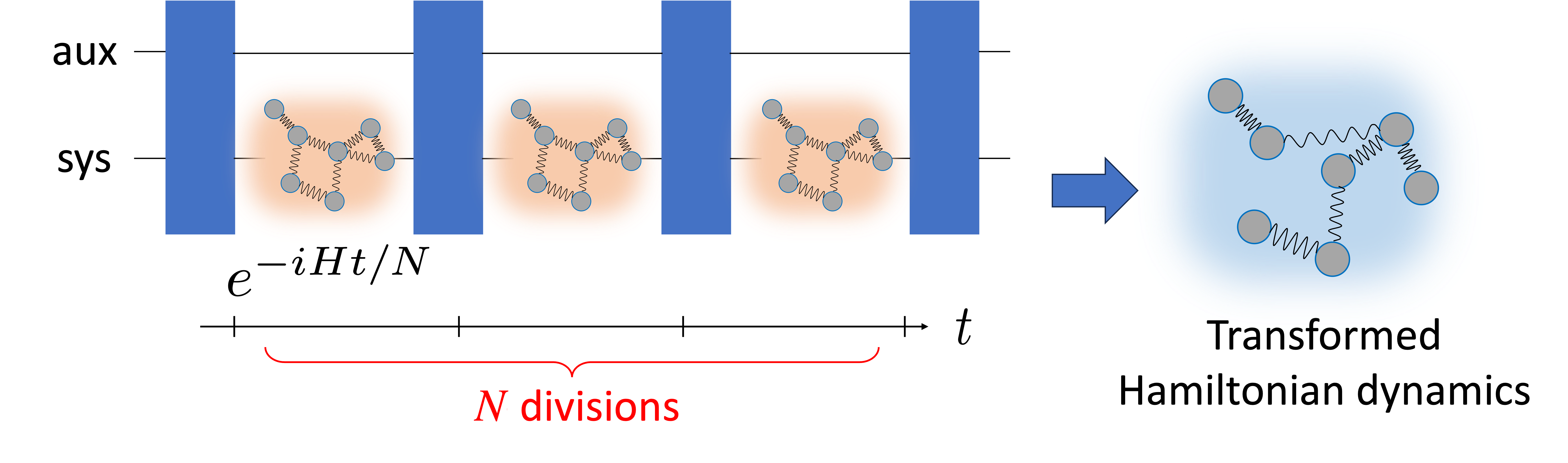} 
\caption{\emph{Higher-order quantum transformation.---}By appropriately 
interspersing fixed quantum operations involving an auxiliary system (blue) throughout a Hamiltonian dynamics $e^{-iHt}$ that has been divided into $N$ smaller portions $e^{-iHt/N}$, the physical properties of a system can be transformed as desired.\vspace{0.3cm}}\label{fig::higher_order}
\end{figure}

To achieve our universal Hamiltonian eigenvalue transformation, we employ the framework of higher-order quantum transformations, a research area that has attracted significant attention in recent years \cite{chiribella2008quantum,miyazaki2019complex,dong2019controlled, quintino2019probabilistic,quintino2019reversing,yoshida2022reversing,chiribella2019quantum,chiribella2013quantum,pollock2018non,oreshkov2012quantum,bai2020efficient}. The adjective ``higher-order'' here refers to the fact that such processes take \emph{quantum maps} as input, returning a transformed quantum map as output, e.g., by appropriately applying quantum operations before and after the dynamics. Here, we will consider the situation where the input and output quantum maps correspond to Hamiltonian dynamics parameterized by a time $t$ (see Fig.~\ref{fig::higher_order}). In this case, the ability to divide the map in time and intersperse it with gate operations provides extra power for implementing desired transformations~\cite{odake2023higher}. Leveraging this framework, we construct a method to implement functions of unknown Hamiltonian dynamics by first controlizing the black-box dynamics \cite{dong2019controlled} and then applying a novel algorithm for Fourier series simulation on the controlled black-box dynamics, which implements the desired function on the eigenvalues. Although previous work demonstrates the ability to apply \emph{linear} functions to unknown Hamiltonian dynamics~\cite{odake2023higher}, our present method permits the implementation of arbitrary (sufficiently smooth) functions; this includes non-linear functions that may be pertinent in simulating many-body or chaotic dynamics or even designing processes with desired non-linear features.

Furthermore, our broader methodology shows how component subroutines can be compiled to build complex and efficient algorithms from simpler ones. Concatenating subroutines---as we do here with controlization and Fourier series simulation---provides an instance of quantum functional programming~\cite{selinger2004towards,valiron2005typed,chiribella2013quantum}, which endows quantum algorithms with a modular flexibility similar to their classical counterparts. Yet, whenever two subroutines that rely on random sampling are concatenated directly, the overall time complexity scales poorly due to the independence of the two sampling distributions. To overcome this issue, we develop a general \textit{compilation} procedure for the concatenation of randomized algorithms, which makes use of \emph{correlated} randomness as a key resource to optimize the algorithm at a global level. Crucially, this means that our overall compiled algorithm is, loosely speaking, greater than the sum of its parts in the sense that it arises from correlated (rather than independent) subroutines. Moving forward, we expect this tool to be leveraged to optimize various algorithms built upon concatenated randomized subroutines.

We highlight the power of our method by demonstrating a \textit{compiled} algorithm that outperforms the na{\" i}ve concatenation of controlization and Fourier series simulation; moreover, it outperforms other methods based upon combining controlization and QSVT-based algorithms. As such, our compilation method goes beyond the existing techniques of quantum functional programming, leading to more efficient compiled algorithms. By introducing the notion of compilation to functional programming in the context of quantum information processing, we align it with classical software design principles that allow code to be compiled and executed at various levels of abstraction, opening the possibilities for computations to be deployed across diverse applications. 

\FloatBarrier

\section{Outline of Paper}\label{sec::outline}

\begin{figure}[t]
\centering
\includegraphics[width=\linewidth]{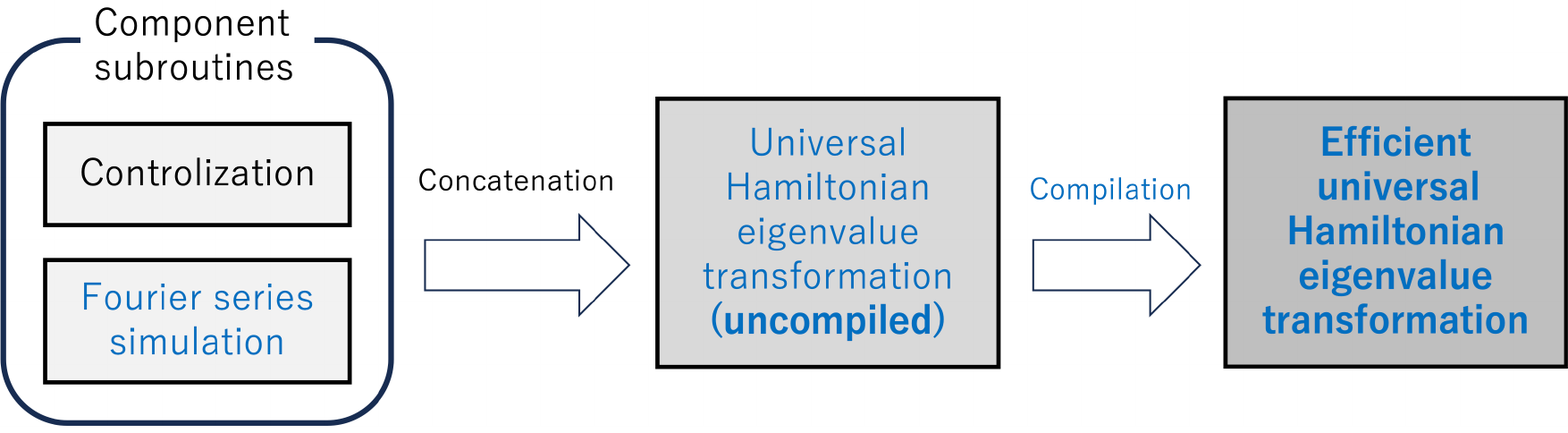}
\caption{\emph{Description of our work.---}Tasks written in blue denote our contributions. We first develop an \textit{uncompiled} algorithm for a universal Hamiltonian eigenvalue transformation (UHET) by concatenating two subroutines, controlization and Fourier series simulation. We then present a \textit{compiled} version of the UHET algorithm, which improves upon the efficiency. \label{fig::logic_flow}}
\end{figure}

The main results of our work are schematically depicted in Fig.~\ref{fig::logic_flow} and summarized as follows.

\begin{enumerate}
    \item In Sec.~\ref{sec::uncompiled}, we develop an algorithm that performs a universal Hamiltonian eigenvalue transformation by concatenating two subroutines, namely controlization~\cite{dong2019controlled} and Fourier series simulation. The controlization subroutine adds control to an unknown Hamiltonian dynamics $e^{-iH\tau}$. By approximating the Fourier series of a desired function $f$, one can then use this controlled dynamics to simulate the dynamics associated to the transformed Hamiltonian $f(H)$. Both subroutines are implemented efficiently using a randomized Hamiltonian simulation technique.
    \item In Sec.~\ref{sec::compiled}, we present a method, hereby dubbed \textit{compilation}, which provides a framework for constructing a more efficient circuit for any quantum algorithm constructed by concatenating subroutines that each make use of randomized Hamiltonian simulation. 
    We achieve this by employing the resource of temporally correlated randomness in order to optimize the overall task at hand rather than the individual modules as per the \textit{uncompiled} algorithm. We show that this leads to a more efficient algorithm for transforming unknown Hamiltonian dynamics.
    \item In Sec.~\ref{sec::qsvt}, we compare the performance of our two new algorithms for transforming eigenvalues of an unknown Hamiltonian with a method based upon the Quantum Singular Value Transformation (QSVT) \cite{gilyen2019quantum, low2019hamiltonian,martyn2021grand}. We show that our compiled algorithm has a better time complexity than the QSVT-based algorithm, which in turn outperforms our uncompiled algorithm in this regard. 
\end{enumerate}

We begin by formalizing the task and setting.

\section{Task: Universal Hamiltonian Eigenvalue Transformation (UHET)}\label{sec::uhet}

The goal of \textbf{universal Hamiltonian eigenvalue transformation (UHET)} is to simulate the dynamics of a desired function of an unknown input Hamiltonian $H$. Here, the term ``simulate'' refers to implementing the corresponding Hamiltonian dynamics for an arbitrary evolution time up to an acceptable approximation error. More precisely, given a function $f:[-1,1]\to \mathbb{R}$, time $t>0$, and precision $\epsilon >0$, a universal Hamiltonian eigenvalue transformation uses an \emph{unknown} input dynamics $e^{\pm iH\tau}\ (\tau>0)$ to approximate a desired transformed dynamics $e^{-if(H_0)t}$ up to precision $\epsilon$ and for all $t$. 

Here, $H$ is associated to an $n$-qubit Hilbert space $\mathcal{H}$ and $H_0:=H-({\rm tr}(H)/2^n)I$ is its traceless part, which we assume to be upper bounded $\|H_0\|_{\rm op}\leq1$. Note that taking only the traceless part of the Hamiltonian leads to no loss of generality, since for $H_1=H_2+\alpha I$ where $H_1,\ H_2$ are Hamiltonians and $\alpha\in \mathbb{R}$, we have that $e^{-iH_1t}=e^{-i\alpha t}e^{-iH_2t}$, and thus the dynamics corresponding to $H_1$ and $H_2$ are equivalent (up to a global phase). Furthermore, as long as an upper bound $\Delta_H$ of the difference between the maximum and the minimum energy eigenvalues of the Hamiltonian is known, one can always rescale the Hamiltonian via $H\to H/\Delta_H$ and redefine the function $f$ as $f\to f_H$ such that $f_H(x):=f(\Delta_H x)$, justifying the assumption $\|H_0\|_{\rm op}\leq1$. We note that $f(H_0)$ is defined in the usual way, i.e., by applying $f$ to the spectrum of $H_0$. Lastly, we assume access to both the positive and negative-time dynamics.

As mentioned previously, transforming Hamiltonians has applications ranging from condensed matter physics to quantum information processing. In many situations considered, a classical description of the initial seed Hamiltonian is known or at least access to a block encoding of it is given, e.g., in algorithms based upon QSVT~\cite{gilyen2019quantum,low2019hamiltonian,martyn2021grand}. However, from a physics perspective, this scenario is not always justified; for instance, one may have access to a time-evolving many-body system without knowing any description of the dynamics \emph{a priori}. Our setting of \emph{universal} Hamiltonian transformations allows such scenarios: the classical description of the input Hamiltonian $H$ can be completely unknown, as long as its (positive- and negative-time) dynamics $e^{\pm iH\tau}$ is accessible. See the \textbf{Supplemental Material (SM)}, App.~\ref{app::prelim} for details and preliminary definitions.

Throughout the remainder of this article, we will present various methods to achieve UHET. At their core, all such methods make use of classical randomness to simulate the desired dynamics by sampling many rounds of evolution; being approximate, one must therefore take the iteration number $N$ sufficiently large to ensure a desired accuracy $\epsilon >0$. In order to assess the performance of these algorithms, we will employ two related figures of merit. The first we denote \emph{time complexity}, which is defined in terms of the gate depth of the circuit that implements the algorithm when decomposed into single qubit and $\texttt{CNOT}$ gates, \emph{not} counting the calls to unknown dynamics $e^{\pm iH\tau}$ used as oracles. The second quantifier is the \emph{total evolution time}, which refers to the sum of the absolute values $|\tau|$ of the evolution times $\tau$ of the Hamiltonian dynamics $e^{-iH\tau}$ used throughout the algorithm.

To quantify said error in any such scheme, we make use of the following distance measures. First, note that simulating any Hamiltonian dynamics for a specified amount of time (ideally) leads to a unitary operation. Thus, we consider the situation where one attempts to simulate such a unitary operation $\mathcal{U}:\mathcal{L}(\mathcal{H})\to \mathcal{L}(\mathcal{H})$ by a random protocol $\sum_jp_j\mathcal{F}_j$, where the index $j$ is chosen with a probability $p_j$ and $\mathcal{F}_j:\mathcal{L}(\mathcal{H})\to \mathcal{L}(\mathcal{H})$ is the corresponding quantum operation. If the input state to said dynamics is not fixed (i.e., remains arbitrary), then we use the following error measure for the quantum operation:
\begin{gather}
    \underset{\substack{ 
    \mathrm{dim}(\mathcal{H}^{\prime}) \\
    \ket{\psi}\in \mathcal{H}\otimes \mathcal{H}^{\prime}\\
    \|\ket{\psi}\|=1 }}{\mathrm{sup}}
        \| \mathcal{U}\otimes \mathcal{I}_{\mathcal{H}'}(\ket{\psi}\bra{\psi})-\sum_j p_j \mathcal{F}_j\otimes\mathcal{I}_{\mathcal{H}'}(\ket{\psi}\bra{\psi}) \|_1.
    \label{measure_ope}
    \end{gather}
Here, $\mathcal{I}$ represents the identity channel, $\mathcal{H}^\prime$ is an auxiliary Hilbert space of arbitrary dimension and $\| \cdot \|_1$ denotes the 1-norm. If, on the other hand, the input state is specified to be some known $\ket{\psi}\in \mathcal{H}$, then the accuracy of any protocol can be determined by comparing the post-transformation state with the ideal case, and hence we use the following error measure for quantum states:
    \begin{align}
    \|\mathcal{U}(\ket{\psi}\bra{\psi})-\sum_j p_j\mathcal{F}_j(\ket{\psi}\bra{\psi})\|_1 .
    \label{measure_sta}
\end{align}
When the average state $\sum_jp_j\mathcal{F}_j(\ket{\psi}\bra{\psi})$ approximates the target state $\mathcal{U}(\ket{\psi}\bra{\psi})$ [in terms of Eq.~(\ref{measure_sta})] with an error that is upper bounded by a fixed constant $\epsilon$ for \emph{any} input state $\ket{\psi}$, then the mean square of the approximation is upper bounded by $2\epsilon$ for any input, i.e.,
\begin{align}
    \label{varivari}
    \sum_j p_j
    \|\mathcal{U}(\ket{\psi}\bra{\psi})-\mathcal{F}_j(\ket{\psi}\bra{\psi})\|_1^2
    \leq 2\epsilon 
\end{align}
(see the SM, App.~\ref{app::uhetvarivari}).

\FloatBarrier

\section{Uncompiled UHET Algorithm}\label{sec::uncompiled}

We now move to present our first algorithm that performs the task of UHET. This algorithm is based upon a random protocol for simulating Hamiltonian dynamics. 
Our method can be seen as an extension of a Hamiltonian simulation technique called \textbf{qDRIFT}~\cite{campbell2019random} to the case where the Hamiltonian is unknown. 
Using this technique as a basis, we concatenate two subroutines, namely controlization and Fourier series simulation, to develop our overall UHET algorithm. Finally, we present a resource analysis of the time complexity and total evolution time of the protocol. Additional technical details are provided in the SM, App.~\ref{app::uncompiled}.

\subsection{Efficient Hamiltonian Simulation (qDRIFT)}\label{subsec::qdrift}

\begin{figure}[t]
        \includegraphics[width=0.5\linewidth]{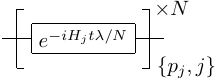}
        \caption{\emph{qDRIFT primitive.---}Circuit representation of the qDRIFT protocol for approximating $e^{-i(\sum_jh_jH_j)t}$ by randomly implementing $e^{-i H_j t \lambda /N}$ with probability $p_j = h_j / \lambda$, for each $j$ and for $N$ iterations, where $\lambda = \sum_k h_k$.
        \label{fig::basic_random_circuit}}
\end{figure}

Suppose one has access to a set of Hamiltonian dynamics, i.e., the ability to perform $e^{-iH_j\tau}\ (\tau >0)$ for a set of Hamiltonians $\{H_j\}$. Assuming w.l.o.g. that the Hamiltonians are normalized $\|H_j\|_{\rm op}=1$, any dynamics of the form $e^{-i(\sum_j h_j H_j) t}\ (t>0)$ for a set of positive coefficients $h_j>0$ can be approximated by the following protocol (see Fig.~\ref{fig::basic_random_circuit}):
\needspace{10em}
\begin{enumerate}
    \item Define $\lambda:=\sum_k h_k$ and the probability distribution $p_j:=h_j/\lambda$, from which an index $j$ is randomly sampled.
    \item Apply the dynamics $e^{-i H_j t \lambda /N}$.
    \item Repeat steps (1)--(2) $N$ times.
\end{enumerate}
This method is based on the Hamiltonian simulation technique qDRIFT~\cite{campbell2019random}, which makes use of the Trotter-Suzuki decomposition~\cite{suzuki1991general} of $e^{-i(\sum_j h_jH_j)t}$ to approximate $e^{-i(\sum_j p_jH_j)\delta t}$ for a small time interval $\delta t:=t\lambda /N$ up to the first order of $\delta t$ via a probabilistic mixture of dynamics $e^{-iH_j \delta t}$ with probability $p_j$. In order to suppress the approximation error below $\epsilon$, it is sufficient to take the iteration number
\begin{align}\label{eq::iterationnumber}
    N(\lambda,t,\epsilon):={\rm ceil}[\max (10\lambda^2t^2/\epsilon, 5\lambda t/2)]
\end{align}
(see the SM, App.~\ref{app::qdriftiterations}). In this work, we employ such a probabilistic Trotter-Suzuki technique in all algorithms and subroutines (except the QSVT-based subroutine described in Sec.~\ref{sec::qsvt}) since such methods typically exhibit reduced time complexity when compared to deterministic counterparts due to their independence of the number of terms in $\sum_j h_jH_j$.

We now describe the two subroutines that leverage this qDRIFT primitive for \textit{unknown} Hamiltonians, namely controlization~\cite{dong2019controlled} and Fourier series simulation, to perform UHET upon their concatenation. We refer to such a straightforward concatenation as \emph{uncompiled}, in contrast to a later algorithm which we dub \emph{compiled} that optimizes UHET globally (see Sec.~\ref{sec::compiled}). 

\FloatBarrier

\subsection{Controlization}\label{subsec::controlization}

\begin{figure}[tb]
        \includegraphics[width=0.8\linewidth]{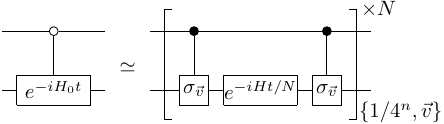}
        \caption{\emph{Controlization circuit.---}By randomly applying controlled Pauli gates $\texttt{ctrl}(\sigma_{\vec{v}})$ before and after portions of unknown Hamiltonian dynamics $e^{-iHt/N}$ a sufficient number of times, said dynamics can be controlized, i.e., simulate $\texttt{ctrl}_0(e^{-iH_0t})$, where $H_0$ is the traceless part of $H$. The black dot represents control such that the corresponding gate is applied whenever the control qubit is in the state $\ket{1}$; the white dot represents the case where the controlled gate is enacted whenever the control qubit is in the state $\ket{0}$. \label{fig::contr_N}}
\end{figure}

\textbf{Controlization} is a procedure that adds control to an unknown Hamiltonian dynamics~\cite{dong2019controlled}. Such a primitive is especially crucial for analog or hybrid quantum computers and/or quantum simulation, where one may have access to some Hamiltonian dynamics, but not necessarily be able to determine a description of it. Controlization takes a finite number of queries to the dynamics $e^{-iH\tau}$ as an input and outputs a random unitary operator approximating ${\tt ctrl}(e^{-iH_0t})\in \mathcal{L}(\mathcal{H}_c\otimes \mathcal{H})$, where $H_0$ represents the traceless part of $H$, $\mathcal{H}_c$ is the Hilbert space associated to the control qubit, and ${\tt ctrl}(e^{-iH_0t}):=\ket{0}\bra{0}\otimes I+\ket{1}\bra{1}\otimes e^{-iH_0t}$ represents controlled-$e^{-iH_0t}$. Our following \emph{uncompiled algorithm} for UHET first makes use of a variant of controlization that simulates ${\tt ctrl}_0(e^{-iH_0t}):=\ket{0}\bra{0}\otimes e^{-iH_0t}+\ket{1}\bra{1}\otimes I$ by reversing the roles of $\ket{0}$ and $\ket{1}$ of the control qubit, as presented in Subroutine \ref{subr::contr} and the circuit in Fig.~\ref{fig::contr_N}.

Intuitively, controlization makes use of a randomized algorithm to approximately implement a unitary dynamics to simulate the Hamiltonian
\begin{align}
    \label{cont_desc}
    & \sum_{\vec{v}\in \{0,1,2,3\}^n}
    \frac{1}{4^n}
    \left(
        \begin{array}{cc}
            I&0\\
            0&\sigma_{\vec{v}}\\
        \end{array}
    \right)
    \left(
        \begin{array}{cc}
            H&0\\
            0&H\\
        \end{array}
    \right)
    \left(
        \begin{array}{cc}
            I&0\\
            0&\sigma_{\vec{v}}\\
        \end{array}
    \right)
    \nonumber\\
    &= 
    \left(
        \begin{array}{cc}
            H_0&0\\
            0&0\\
        \end{array}
    \right)
    +\frac{{\rm tr}(H)}{2^n}I,
\end{align}
where $\vec{v}:= (v_1,\ldots ,v_n)\in \{0,1,2,3\}^n$ and $\sigma_{\vec{v}}:=\sigma_{v_1}\otimes \cdots \otimes \sigma_{v_n}$. 
Equation~(\ref{cont_desc}) follows from the identity $\frac{1}{4^n}\sum_{\vec{v}}\sigma_{\vec{v}}H\sigma_{\vec{v}}
=\frac{{\rm tr}(H)}{2^n}I$. Each term ${\tt ctrl}(\sigma_{\vec{v}})(I\otimes H){\tt ctrl}(\sigma_{\vec{v}})$ in the sum on the l.h.s.\ of Eq.~(\ref{cont_desc}) can be implemented by applying the dynamics $e^{-i H \tau}$ in between the gate $\texttt{ctrl}(\sigma_{\vec{v}})$, which follows from the identity $Ue^{-iHt}U^{\dagger}=e^{-i(UHU^{\dagger})t}$ for a general Hamiltonian $H$ and unitary $U$. When exponentiated, the r.h.s.\ of Eq.\ (\ref{cont_desc}) yields the desired controlled Hamiltonian dynamics (up to a global phase contribution from the second term, which is irrelevant).

The time complexity $\Theta(t^2n/\epsilon)$ follows from the fact that $N:=N(1, t, \epsilon)$ is $\Theta(t^2/\epsilon)$ and that implementing ${\tt ctrl}(\sigma_{\vec{v}})$ takes $\Theta(n)$ time, since each such operation can be decomposed in terms of up to $n$ two-qubit gates (more specifically, controlled single-Pauli gates). The total evolution time is simply $(t/N)\times N=t$. 

\begin{algorithm}[H]
    \floatname{algorithm}{Subroutine}
    \caption{Controlization}
    \label{subr::contr}
    \begin{algorithmic}[1]
        \Statex{\textbf{Input:}}        
        \begin{itemize}
            \item A finite number of queries to a black-box Hamiltonian dynamics $e^{-iH\tau}$ of a seed Hamiltonian $H$ normalized as $\|H_0\|_{\mathrm{op}}\leq 1$, where $H_0$ is the traceless part of $H$, i.e., $H_0:=H-(1/2^n)\mathrm{tr}(H)I$, with $\tau>0$
            \item Allowed error $\epsilon >0$
            \item Time $t>0$
        \end{itemize}
        \Statex{\textbf{Output:}}
        A random unitary operator approximating
        \begin{align}
            \left(
                \begin{array}{cc}
                    e^{-iH_0t}&0\\
                    0&I\\
                \end{array}
            \right)
            ={\tt ctrl}_0(e^{-iH_0t})\in \mathcal{L}(\mathcal{H}_c\otimes \mathcal{H})
        \end{align}
        with an error according to Eq.~(\ref{measure_ope}) upper bounded by $\epsilon$
    \Statex \hrulefill
    	\Statex{\textbf{Time complexity:}}
    	$\Theta(t^2n/\epsilon)$
            \Statex{\textbf{Total evolution time:}}
            $t$
    	\Statex{\textbf{Used Resources:}}
    	\Statex \hskip1.0em System: $n$-qubit system $\mathcal{H}$ and one auxiliary qubit $\mathcal{H}_c$
    	\Statex \hskip1.0em Gates: $e^{-iH\tau}$ and Clifford gates on $\mathcal{H}_c\otimes\mathcal{H}$
    	\Statex \hrulefill
        \Statex{\textbf{Procedure:}}
        \Statex \hspace{-1.5em}\textit{Pre-processing:} 
        \State Compute $N:=N(1, t, \epsilon)$ using $N(\lambda, t, \epsilon)$ from Eq.~\eqref{eq::iterationnumber} 
        \Statex \hspace{-1.5em}\textit{Main Process:} 
        \State{Initialize}
        $U_{\rm current}\gets I$
        \For{$m=1,\ldots ,N$}
        \State Uniformly randomly choose $\vec{v}\in \{0,1,2,3\}^n$
        \State $U_{\rm current}\gets  {\tt ctrl}(\sigma_{\vec{v}})(I\otimes e^{-iHt/N}){\tt ctrl}(\sigma_{\vec{v}})U_{\rm current}$
        \EndFor
        \State {\textbf{Return}} $U_{\rm current}$
    \end{algorithmic}
\end{algorithm}

Note finally that the time complexity for the case where a deterministic Trotter-Suzuki technique is used instead of the randomized one is lower bounded by the number of terms $4^n$ in the sum, which grows exponentially with the number of qubits $n$. Lastly, if the Hamiltonian is restricted to being a $k$-local Hamiltonian, then the size of the set of operations that must be sampled from in order to controlize the dynamics can be significantly reduced~\cite{Chowdhury_2024}.

\begin{figure}[t]
        \includegraphics[width=9cm]{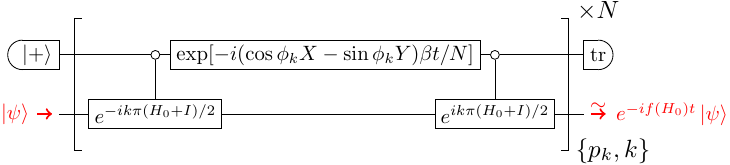}
        \caption{\emph{Fourier series simulation circuit.---}By randomly applying ${\tt ctrl}_0(e^{\pm ik\pi (H_0+I)/2})$ before and after the dynamics $e^{-i(\cos \phi_k X - \sin \phi_k Y) \beta t/N}$, the circuit simulates the Fourier series of the desired transformation function $\tilde{f}[(H_0+I)/2]=f(H_0)$. \label{fig::fouri_subr}}
\end{figure}

\subsection{Fourier Series Simulation}\label{subsec::fss}

We now move to describe the second subroutine that comprises our uncompiled UHET algorithm, namely \textbf{Fourier series simulation}. This algorithm transforms eigenvalues of an unknown Hamiltonian, using the theory of Fourier series at its core. A Fourier expansion of a function $f:[-1,1]\to \mathbb{R}$ is given by $\sum_{k=-\infty}^{\infty}c_ke^{i\pi kx}$ for $c_k:=(1/2)\int_{-1}^1{\rm d}xf(x)e^{-i\pi k x}$, which converges to $f(x)$ for a wide range of functions (see, e.g., \cite{walker2003fourier}). Note that Fourier series approximations have been applied in various contexts of quantum information processing, e.g., Ref.~\cite{childs2017quantum} for solving systems of linear equations; however, most such applications are concerned with approximating a state $\ket{x}$, whereas we use the series to represent a Hamiltonian $H$. The procedure of Fourier series simulation of Hamiltonian dynamics is described in Subroutine \ref{subr::fouri}, with a circuit representation in Fig. \ref{fig::fouri_subr}. 

\begin{figure}[t]%
\centering
\subfloat[$f(x)=x$\label{fig::ftildea}]{\includegraphics[keepaspectratio, scale=0.16]{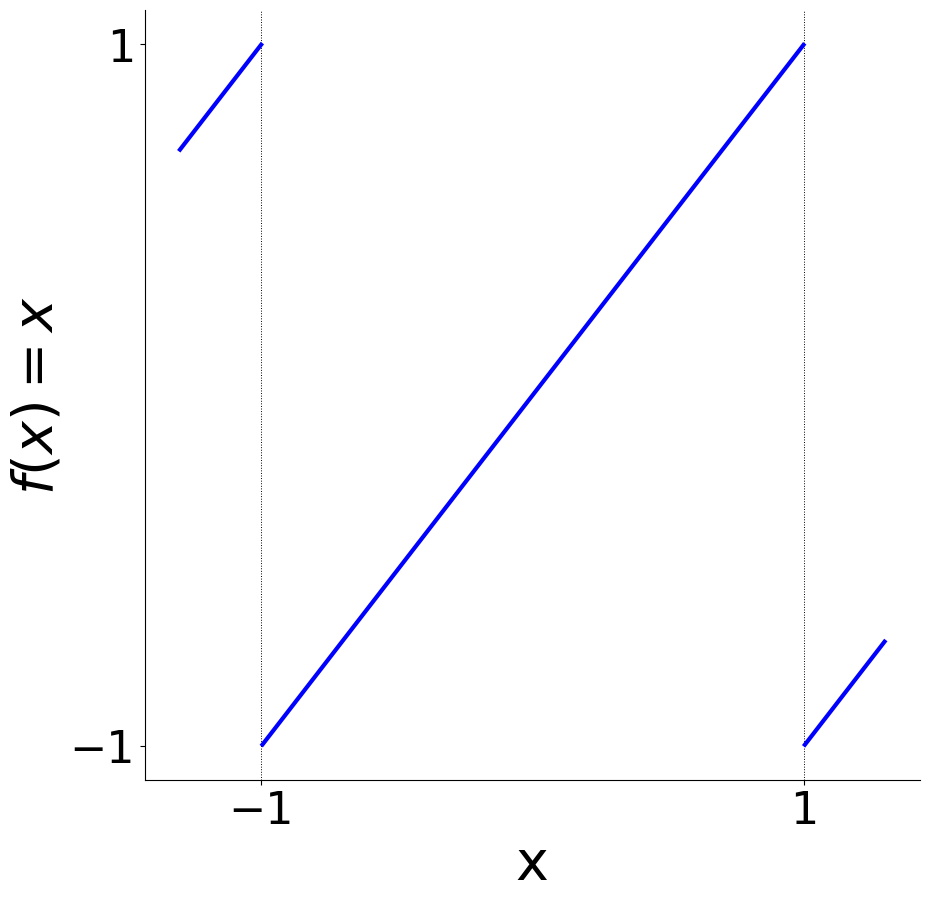}}\qquad
\subfloat[$\tilde{f}$ corresponding to $f$\label{fig::ftildeb}]{\includegraphics[keepaspectratio, scale=0.16]{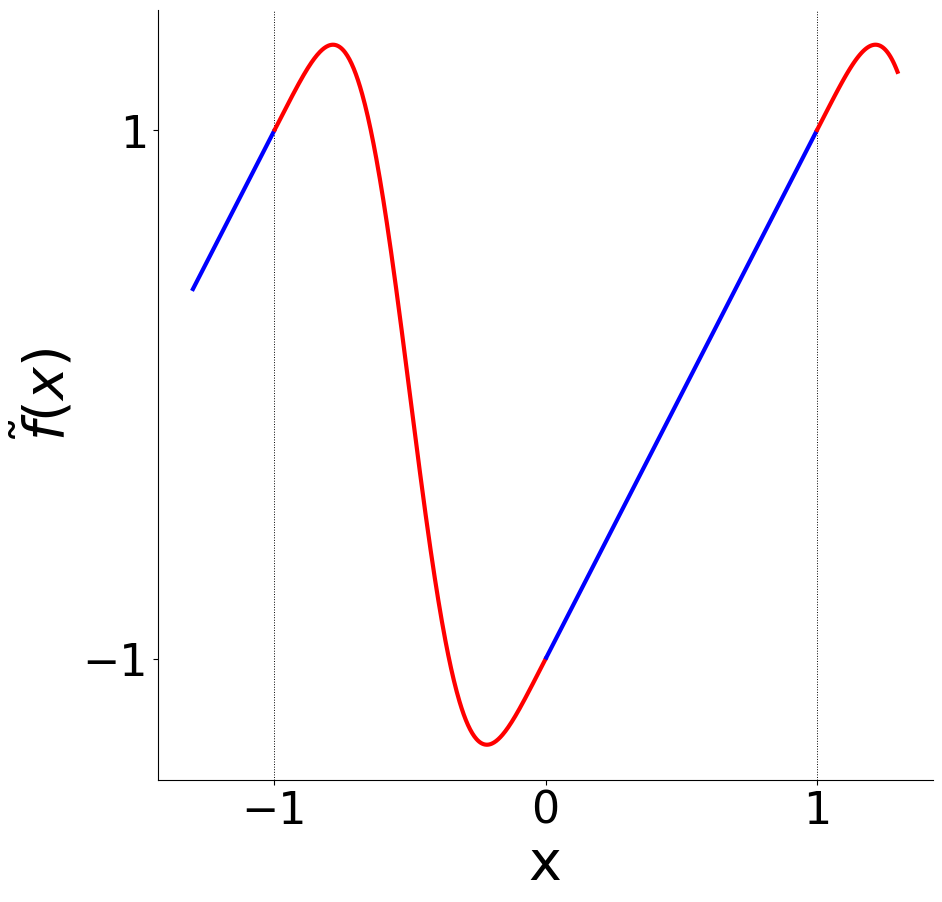}}
\caption{\emph{Comparison between the function $f$ and corresponding $\tilde{f}$.---}The function $\tilde{f}$ is defined by merging $f$ (blue) with an additional function $g_f$ (red, see the SM, App.~\ref{app:funcs}). The absolute values of the Fourier coefficients $|\tilde{c}_k|$ of $\tilde{f}$ converge faster than those of $f$. \label{fig::ftilde}}
\end{figure}

Fourier series simulation makes use of a controlled dynamics ${\tt ctrl}_0(e^{\pm iH_0\tau})\,(\tau>0)$ (which can be constructed via Subroutine \ref{subr::contr}) to implement the target Hamiltonian dynamics $e^{-if(H_0)t}\,(t>0)$. First, the desired function $f$ is deformed to a periodically smooth function $\tilde{f}$ as
\begin{align}
    &\tilde{f}(x):=
    \begin{cases}
        g_f(x) \quad &x\in [-1,0]\\
        f(2x-1) \quad &x\in [0,1]\\
    \end{cases} 
    \label{def_tilde_f} ,
\end{align}
with $g_f$ defined in the SM, App.~\ref{app:funcs} (see Fig.~\ref{fig::ftilde}). Due to its periodic smoothness, the absolute values of the Fourier coefficients of $\tilde{f}$ quickly shrink to 0, which avoids rapid growth of the cutoff number $K$ in terms of the error $\epsilon$. 

We then compute the Fourier coefficients $\tilde{c}_k:=(1/2)\int_{-1}^1{\rm d}xe^{-ik\pi x}\tilde{f}(x)$ for $k\in \{-K, -K+1,\ldots , K\}$, where the cutoff number $K$ satisfies
        \begin{align}
            \label{stop2}
            \left|
            \tilde{f}(x)-\sum_{k=-K}^K \tilde{c}_k e^{ik\pi x}
            \right|
            <\frac{\epsilon}{4t}
            \quad \forall \, x\in [-1,1] \, ,
        \end{align}
and the iteration number $N:=N(\beta, t, \epsilon/2)$ [using Eq.~\eqref{eq::iterationnumber}] for $\beta:=\sum_{k=-K}^K|\tilde{c}_k|$. 

\begin{algorithm}[H]
    \floatname{algorithm}{Subroutine}
    \caption{Fourier Series Simulation}
    \label{subr::fouri}
    \begin{algorithmic}[1]
        \Statex{\textbf{Input:}}
        \begin{itemize}
            \item A finite number of queries to ${\tt ctrl}_0(e^{\pm iH_0\tau})\in \mathcal{L}(\mathcal{H}_c\otimes \mathcal{H})\ (\tau >0)$ where $H_0\in \mathcal{L}(\mathcal{H})$ is a traceless Hamiltonian normalized as $\|H_0\|_{\mathrm{op}}\leq 1$
            \item A class $C^3$ (three times continuously differentiable) function $f:[-1,1]\to \mathbb{R}$, such that $f^{(4)}$  is piecewise $C^2$ 
            (see the SM, App.~\ref{app::jsmooth})
            \item Input state $\ket{\psi}\in \mathcal{H}$
            \item Allowed error $\epsilon >0$
            \item Time $t>0$
        \end{itemize}
        \Statex{\textbf{Output:}}
        A state approximating $e^{-if(H_0)t}\ket{\psi}\ (t>0)$ with an error according to Eq.~(\ref{measure_sta}) upper bounded by $\epsilon$, and mean squared error according to Eq.~(\ref{varivari}) upper bounded by $2\epsilon$
    \Statex \hrulefill
    	\Statex{\textbf{Time complexity}} $\Theta (\beta^2t^2/\epsilon)$
            \Statex{\textbf{Total evolution time:}}
            $O (C_{2,f}t^2/\epsilon)$ for an $f$-dependent constant $C_{2,f}$ which is independent of $n,\ t,$ and $\epsilon$
    	\Statex{\textbf{Used Resources:}}
    	\Statex \hskip1.0em System: $n$-qubit system $\mathcal{H}$ and one auxiliary qubit $\mathcal{H}_c$
    	\Statex \hskip1.0em Gates: ${\tt ctrl}(e^{\pm iH_0\tau})$ and single-qubit gates on $\mathcal{H}_c$
    	\Statex \hrulefill
        \Statex{\textbf{Procedure:}}
        \Statex \hspace{-1.5em}\textit{Pre-processing:} 
        \State Define modified version $\tilde{f}$ of $f$ as in Eq.~(\ref{def_tilde_f})
        \State Compute Fourier coefficients $\tilde{c}_k:=(1/2)\int_{-1}^1{\rm d}xe^{-ik\pi x}\tilde{f}(x)$ for $k\in \{-K, -K+1,\ldots , K\}$ where the cutoff number $K$ satisfies
        \begin{align}
            \left|
            \tilde{f}(x)-\sum_{k=-K}^K \tilde{c}_k e^{ik\pi x}
            \right|
            <\frac{\epsilon}{4t}
            \quad \forall \ x\in [-1,1]
            \nonumber
        \end{align}
        \State Compute $N:=N(\beta, t, \epsilon/2)$ using $N(\lambda, t, \epsilon)$ from Eq.~\eqref{eq::iterationnumber}, for $\beta:=\sum_{k=-K}^K|\tilde{c}_k|$
        \Statex \hspace{-1.5em}\textit{Main Process:} 
        \State{Initialize}
        $\ket{{\rm current}}\gets \ket{+}\otimes \ket{\psi}$
        \For{$m=1,\ldots ,N$}
        \State Randomly choose $k\in \{-K,\ldots ,K\}$ with probability $p_k:=|\tilde{c}_k|/\beta$
        \State Define $Q:={\tt ctrl}_0 (e^{-ik\pi H_0/2})$ and $R:={\tt ctrl}_0 (e^{ik\pi H_0/2})$
        \State $\ket{{\rm current}}\gets (e^{ik\pi Z/4}\otimes I)R(e^{-i [\cos\phi_kX-\sin\phi_kY]\beta t/N}\otimes I) Q(e^{-ik\pi Z/4}\otimes I)\ket{{\rm current}}$ for $\phi_k$ defined by $\tilde{c}_k=|\tilde{c}_k|e^{i\phi_k}$
        \EndFor
        \State Trace out $\mathcal{H}_c$ of $\ket{\rm current}$
        \State {\textbf{Return}} $\ket{\rm current}$
    \end{algorithmic}
\end{algorithm}

The remainder of Subroutine \ref{subr::fouri}  simulates the Fourier-transformed Hamiltonian of $H_0$, which can be understood from the following analysis. The unitary applied in step 8, for a random $k$ chosen with probability $p_k=|\tilde{c}_k|/\beta$, can be written as a Hamiltonian evolution given by
\begin{align}
    &(e^{ik\pi Z/4}\otimes I)
    \left(
        \begin{array}{cc}
            e^{ik\pi H_0/2}&0\\
            0&I\\
        \end{array}
    \right) (e^{{-i(\cos \phi_k X -\sin \phi_k Y)\beta t/N}}\otimes I) 
    \nonumber\\
    & \cdot \left(
        \begin{array}{cc}
            e^{-ik\pi H_0/2}&0\\
            0&I\\
        \end{array}
    \right)
    (e^{-ik\pi Z/4}\otimes I)\\
    &=\exp
    \left[ -i
        \left(
            \begin{array}{cc}
                0&e^{i\phi_k}e^{ik\pi (H_0+I)/2}\\
                e^{-i\phi_k}e^{-ik\pi (H_0+I)/{2}}&0\\
            \end{array}
        \right) \beta t/N \, 
    \right]. \nonumber
\end{align}

As such, steps 5 to 9 of Subroutine \ref{subr::fouri} correspond to simulating $e^{-iH't}$ for the following Hamiltonian $H'$
\begin{align}
\label{qdrift_out}
H'&:=\sum_{k=-K}^{K}|\tilde{c}_k|
\left(
    \begin{array}{cc}
        0&e^{i\phi_k}e^{ik\pi (H_0+I)/2}\\
        e^{-i\phi_k}e^{-ik\pi (H_0+I)/2}&0\\
    \end{array}
\right)
\nonumber\\
\simeq &
\sum_{k=-\infty}^{\infty}|\tilde{c}_k|
\left(
    \begin{array}{cc}
        0&e^{i\phi_k}e^{ik\pi (H_0+I)/2}\\
        e^{-i\phi_k}e^{-ik\pi (H_0+I)/2}&0\\
    \end{array}
\right)
\nonumber\\
=&
\left(
    \begin{array}{cc}
        0&f(H_0)\\
        f(H_0)&0\\
    \end{array}
\right) 
=
X\otimes f(H_0) \, ,
\end{align} 
which follows from the fact that
\begin{align}
    &\sum_{k=-\infty}^{\infty}|\tilde{c}_k|e^{i\phi_k}{e^{ik\pi (H_0+I)/2}}
    \nonumber\\
    =&
    \sum_{k=-\infty}^{\infty}
    \sum_{m}\tilde{c}_k{e^{ik\pi (E_m+1)/2}}
    \ket{E_m}\bra{E_m}
    \nonumber\\
    =&
    \sum_m\tilde{f}((E_m+1)/2)\ket{E_m}\bra{E_m}=
    f(H_0) \, ,
    \label{core_fouri}
\end{align}
where $H_0$ is diagonalized as $H_0:=\sum_mE_m\ket{E_m}\bra{E_m}$.
By taking the initial state as $\ket{+}\otimes \ket{\psi}$ in step 4 and tracing out the $\mathcal{H}_c$ subsystem in step 10, the dynamics $e^{-if(H_0)t}$ is applied to the input state $\ket{\psi}$. 

We provide a full error analysis of Subroutine \ref{subr::fouri} in the SM, App.~\ref{app:error_nokori}.
The time complexity, which is equal to the iteration number $N$, is evaluated as $\Theta (\beta^2 t^2/\epsilon)$. The total evolution time [of the controlled Hamiltonian dynamics ${\rm ctrl}(e^{\pm iH\tau})$] is evaluated as (Number of iterations $N$)$\times$(Average evolution time of each iteration). Given that $N$ is $\Theta(\beta^2 t^2/\epsilon)$ and that each iteration has evolution time $\sum_{k}p_k \Theta(|k|)$ on average, the total evolution time is
\begin{align}
    &\Theta\left(\frac{\beta^2 t^2}{\epsilon}\right)\times 
    \sum_{k}p_k \Theta(|k|) =
    \nonumber\\
    &O\left(
    \left(\sum_{k=-\infty}^{\infty} |\tilde{c}_k| \right)\left(
    \sum_{k=-\infty}^{\infty}|\tilde{c}_k||k|
    \right)\frac{t^2}{\epsilon}\right)
    =:O\left(C_{2,f}\frac{t^2}{\epsilon}\right) \, ,
\end{align}
where $C_{2,f}$ is a parameter that depends on $f$ but is independent of $n,t,\epsilon$.
Note that the sum $\sum_{k=-\infty}^{\infty}|\tilde{c}_k||k|$ is guaranteed to converge due to the periodic smoothness of $\tilde{f}$ (see the SM, App.~\ref{app::fourierconvergence}).

\FloatBarrier

\subsection{Uncompiled UHET Algorithm}\label{subsec::uncompiled}

\begin{figure}[t]%
\centering
\subfloat[Outer layer: Fourier series simulation (Subroutine~\ref{subr::fouri})]{\includegraphics[width=9cm]{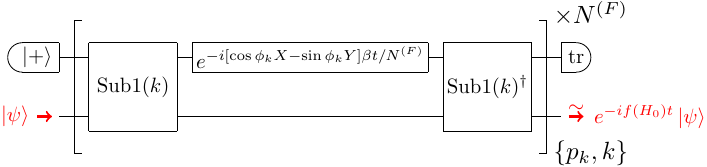}\label{fig::uncompiled_circuit}}\\
\subfloat[Inner layer: The procedure inside square brackets corresponds to controlization (Subroutine~\ref{subr::contr}).]{\includegraphics[width=\linewidth]{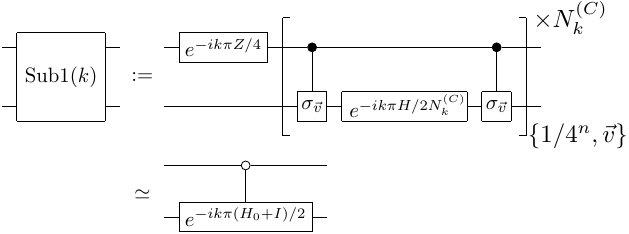}\label{fig::sub1_circuit}}
\caption{\emph{Circuit representation of Algorithm \ref{alg::umcompi}.---}The uncompiled UHET algorithm comprises an ``outer'' layer (a) that implements the Fourier series simulation Subroutine~\ref{subr::fouri} upon the output of the ``inner'' layer (b), which itself controlizes the seed Hamiltonian dynamics via Subroutine~\ref{subr::contr}.}
\label{fig::uncompiledalgorithm}
\end{figure}

We now present the uncompiled UHET Algorithm \ref{alg::umcompi}, which results from concatenating the previous two subroutines and is depicted in Fig.~\ref{fig::uncompiledalgorithm}. We first construct ${\tt ctrl}_0(e^{\pm ik\pi (H_0+I)/2})$ from the input dynamics $e^{\pm iH\tau}$ via controlization and then perform the Fourier series simulation to simulate the desired dynamics $e^{-if(H_0)t}$. Thus, Algorithm \ref{alg::umcompi} is a direct concatenation of Subroutines \ref{subr::contr} and \ref{subr::fouri}, and we therefore deem it \textbf{uncompiled}.

As a consequence of its concatenated structure, two layers of iterations are used in Algorithm \ref{alg::umcompi}: $N^{(C)}_k$ for the controlization part and $N^{(F)}$ for the Fourier series simulation. We choose these numbers such that the total error of each subroutine is bounded from above by $\epsilon/2$ (so that overall error is upper bounded by $\epsilon$). We begin by fixing the allowed error of the Fourier series simulation to be $\epsilon/2$. Then, fixing that for controlization to be $\epsilon /(4 N^{(F)})$, it follows that iterating the outer layer of the procedure (see Fig.~\ref{fig::uncompiled_circuit}) $N^{(F)}$ times implies that the total error due to controlization is upper bounded by $[\epsilon/(4N^{(F)})] \cdot 2N^{(F)}=\epsilon/2$. 

We now analyze the time complexity of Algorithm \ref{alg::umcompi}. We split this cost into two parts. First, there is the \emph{pre-processing step}, i.e., the processes that only need to be run once for a given set of inputs. This corresponds to the time complexity for computing the Fourier coefficients $\tilde{c}_k$ and cutoff number $K$ until Eq.~(\ref{stop2}) is satisfied, plus the time complexity for computing $N^{(F)}$. We define the sum of these two time complexities as $T_{3}$. The \emph{main process} takes $N^{(F)}\times$ (average time complexity of each iteration), the latter of which in turn depends upon $N^{(C)}_k$.

\setcounter{algorithm}{0}

\begin{algorithm}[H]
    \floatname{algorithm}{Algorithm}
    \caption{Universal Hamiltonian eigenvalue transformation (uncompiled)}
    \label{alg::umcompi}
    \begin{algorithmic}[1]
        \Statex{\textbf{Input:}}
        \begin{itemize}
            \item A finite number of queries to a black-box Hamiltonian dynamics $e^{\pm iH\tau}$ of a seed Hamiltonian $H$ normalized as $\|H_0\|_{\rm op}\leq 1$ where $H_0$ is the traceless part of $H$, i.e., $H_0:=H-(1/2^n)\mathrm{tr}(H)I$, with $\tau>0$
            \item A class $C^3$ (three times continuously differentiable) function $f:[-1,1]\to \mathbb{R}$, such that $f^{(4)}$  is piecewise $C^2$ 
            (see the SM, App.\ \ref{app::jsmooth})
            \item Input state $\ket{\psi}\in \mathcal{H}$
            \item Allowed error $\epsilon >0$
            \item Time $t>0$
        \end{itemize}
        \Statex{\textbf{Output:}}
         A state approximating $e^{-if(H_0)t}\ket{\psi}\ (t>0)$ with an error according to Eq.~(\ref{measure_sta}) upper bounded by $\epsilon$; additionally, the mean squared error according to  Eq.~(\ref{varivari}) is upper bounded by $2\epsilon$)
    \Statex \hrulefill
    	\Statex{\textbf{Time complexity}}
     \Statex \hskip1.0em Pre-processing (only once): $T_{3}$
     \Statex \hskip1.0em Main Process:
    $O(C_{3,f}t^4n/\epsilon^3)$ for an $f$-dependent constant $C_{3,f}$ which is independent of $n,\ t,$ and $\epsilon$
    \Statex{\textbf{Total evolution time (main process):}}
    $O (C_{2,f}t^2/\epsilon)$
    	\Statex{\textbf{Used Resources:}}
    	\Statex \hskip1.0em System: $n$-qubit system $\mathcal{H}$ and one auxiliary qubit $\mathcal{H}_c$
    	\Statex \hskip1.0em Gates: $e^{\pm iH\tau}$, single-qubit gate on $\mathcal{H}_c$, and Clifford gates on $\mathcal{H}_c\otimes\mathcal{H}$
     \Statex \hrulefill
        \Statex{\textbf{Procedure:}}
        \Statex \hspace{-1.5em}\textit{Pre-processing:} 
        \State Run Pre-processing of Subroutine \ref{subr::fouri} for allowed error $\epsilon /2$ to obtain iteration number $N^{(F)}:=N(\beta, t, \epsilon /4)$, cutoff number $K$ and Fourier coefficients $\{\tilde{c}_k$\}
        \Statex \hspace{-1.5em}\textit{Main Process:}
  \State Run Main Process of Subroutine \ref{subr::fouri} with $N^{(F)}$ iterations using $K$ and $\{\tilde{c}_k\}$ obtained before, with step 7 modified to:
  \setcounter{ALG@line}{7}
  \begin{algsubstates}
\State Run Pre-processing of Subroutine \ref{subr::contr} for allowed error $\epsilon /4N^{(F)}$ to obtain iteration number $N^{(C)}_k:=N(1, k\pi/2, \epsilon /4N^{(F)})$
  \State Run Main Process of Subroutine \ref{subr::contr} with $N^{(C)}_k$ iterations and time $k\pi/2$ for $H$ to obtain unitary $Q'$
  \State  Run Main Process of Subroutine \ref{subr::contr} with $N^{(C)}_k$ iterations and time $k\pi/2$ for $-H$ to obtain unitary $R'$ 
  \State  Define $Q:=Q'$ and $R:= R'$  
  \end{algsubstates}
    \end{algorithmic}
\end{algorithm}

The iteration number $N^{(F)}=N(\beta , t, \epsilon/{4})$ scales as $O[(\sum_{k=-\infty}^{\infty}|\tilde{c}_k|)^2t^2/\epsilon]$ (note that $\sum_{k=-\infty}^{\infty}|\tilde{c}_k|$ is an upper bound of $\beta$ which is independent of $\epsilon$). Furthermore, the average time complexity  of each iteration is $\sum_{k=-K}^{K}p_k\times$(time complexity of circuit inside Fig. \ref{fig::uncompiled_circuit} for each $k$), which scales as
\begin{align}
    \sum_{k=-K}^{K}&p_k\Theta(n N^{(C)}_k)=
    \sum_{k=-K}^{K}p_k\Theta\left(\frac{\beta^2 t^2 k^2n}{\epsilon^2}\right)
    \nonumber\\
    &
   \leq \Theta
   \left(\left(\sum_{k=-\infty}^{\infty}|\tilde{c}_k|k^2\right)
    \left(\sum_{k=-\infty}^{\infty}|\tilde{c}_k|\right)\frac{t^2n}{\epsilon^2}\right).
\end{align}

The inequality is obtained by replacing $K$ with $\infty$ in order to remove the $\epsilon$ dependence. Taking the product of the scaling of the two iteration layers, the overall time complexity scaling of the main process of Algorithm \ref{alg::umcompi} is upper bounded as
\begin{align}
    O \left( \left(\sum_{k=-\infty}^{\infty}|\tilde{c}_k|\right)^3\left(\sum_{k=-\infty}^{\infty}|\tilde{c}_k|k^2\right)\frac{t^4n}{\epsilon^3}\right)
    =:O \left(C_{3,f}\frac{t^4n}{\epsilon^3}\right)
    .
\end{align}
Again, the sum $\sum_{k=-\infty}^{\infty}|\tilde{c}_k||k|^2$ here is guaranteed to converge due to the periodic smoothness of $\tilde{f}$ (see the SM, App.~\ref{app::fourierconvergence}). Lastly, the total evolution time is evaluated in the same manner as in Subroutine \ref{subr::fouri} to be $O (C_{2,f}t^2/\epsilon)$.

\FloatBarrier

\section{Compiled UHET Algorithm}\label{sec::compiled}

We now move to introduce a more efficient algorithm that implements UHET. At its core, this algorithm is inspired by the components that make up Algorithm~\ref{alg::umcompi}, but rather than optimizing the subroutines independently, here we \emph{compile} the algorithm at the level of the overall task by making use of \emph{correlated classical randomness} to provide a more efficient implementation. We begin by describing a general framework of this novel notion of compilation, which can be used in many situations in which random Hamiltonian simulation subroutines are concatenated to perform a particular task. We subsequently apply this technique specifically to the task of UHET, compiling Algorithm \ref{alg::umcompi} in order to construct the better Algorithm \ref{alg::main}. Finally, we provide a resource analysis. Details are provided in the SM, App.~\ref{app::compiled}.

\FloatBarrier

\subsection{General Method of Compilation}\label{subsec::generalcompilation}

\begin{figure}%
\centering
\subfloat[Outer layer: In each round of the outer layer (within the square brackets), the inner layer is called.]{\includegraphics[width=0.7\linewidth]{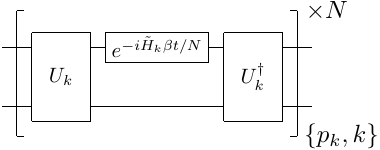}\label{fig::before1}}\\
\subfloat[Inner layer: This circuit simulates $U_k$ defined in Eq.~\eqref{eq::unitarysimulationmain} by random Hamiltonian simulation.]{\includegraphics[width=\linewidth]{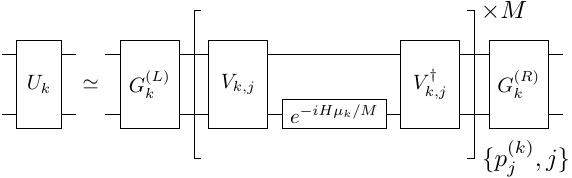}\label{fig::before2}}
\caption{\emph{Circuit representation of a general algorithm involving two layers of Hamiltonian simulation.---}Our method of compilation can be applied to any general two-layer protocol in which each subroutine uses the method of random Hamiltonian simulation.}
\label{fig::compiledalgorithm}
\end{figure}

We begin by abstracting the key structural components of the uncompiled algorithm above. Notably, the algorithm consists of two independent layers, namely those depicted in Figs.~\ref{fig::uncompiled_circuit} and~\ref{fig::sub1_circuit}. These circuits are a special case of the more general form depicted in Figs.~\ref{fig::before1} and~\ref{fig::before2} respectively. The \emph{outer layer} comprises portions of the input dynamics sandwiched between a unitary $U_k(H)$ that depends upon the input Hamiltonian $H$ and is iterated some $N\in \mathbb{Z}_{>0}$ times; for the sake of conciseness, we will simply write $U_k$ instead of explicitly writing $U_k(H)$. Since
\begin{align}
    U_k^{\dagger}e^{-i\tfrac{\beta t}{N} \tilde{H}_k}U_k=e^{-i \tfrac{\beta t}{N} U_k^{\dagger}\tilde{H}_kU_k},
\end{align}
this process corresponds to the simulation of the randomized dynamics $e^{-i (\sum_k p_k\beta U_k^{\dagger}\tilde{H}_kU_k) t}$. In this simulation, $U_k$ is itself approximated using the \emph{inner layer} procedure depicted in Fig.~\ref{fig::before2}. Here, $j$ is a random index (depending on the choice of $k$) sampled from the probability distribution $\{p_j^{(k)}\}_j$, $\mu_k\in \mathbb{R}$, iteration number $M\in \mathbb{Z}_{>0}$ (which also depends on $k$, but we omit the subscript $k$ for ease of notation), and $V_{k,j},\ G^{(L)}_k$, and $G^{(R)}_k$ are unitaries such that 
\begin{align}\label{eq::unitarysimulationmain}
    U_k=G^{(R)}_ke^{-i(\sum_j p^{(k)}_j\mu_kV_{k,j}^{\dagger}HV_{k,j})}G^{(L)}_k. 
\end{align}
Evidently, Algorithm \ref{alg::umcompi} corresponds to the special case where $\tilde{H}_k=\cos\phi_kX-\sin\phi_kY$, $\beta=\sum_{k=-K}^K|\tilde{c}_k|$, $p_k=|\tilde{c}_k|/\beta$, $U_k={\tt ctrl}_0(e^{-ik\pi (H_0+I)/2})$,  $M=N_k^{(C)}$, $j=\vec{v}$, $\mu_k=k\pi/2$, $p^{(k)}_{\vec{v}}=1/4^n$, $V_{k,j}={\tt ctrl}(\sigma_{\vec{v}})$, $G^{(L)}_k=e^{-ik\pi Z/4}\otimes I$, and $G^{(R)}_k=I\otimes I$. 

Having abstracted the key features of our previous algorithm, we are now in a position to introduce the general notion of \emph{compilation}. The structure described above consists of two \emph{independent} layers of random Hamiltonian simulation; the method of compilation harnesses \emph{correlated randomness} to correlate said layers in such a way that the error accumulation is reduced. In general, in order to reduce the approximation error of this overall procedure below $\epsilon$, we must ensure that both errors introduced due to the simulations depicted in Figs.~\ref{fig::before1} and~\ref{fig::before2} are at most of $O(\epsilon)$. Subsequently, both iteration numbers $N$ and $M$ must have a $1/\epsilon$ dependence, which leads to an accumulation of $1/\epsilon$ dependence for each round of concatenated subroutines. 

\begin{figure}%
\centering
\subfloat[Instead of taking the two layers independently, here we correlate the outer (Fig. \ref{fig::before1}) and inner layer (Fig.~\ref{fig::before2}) and run the overall scheme $N$ times.]{\includegraphics[width=0.9\linewidth]{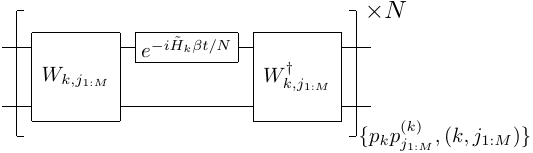}\label{fig::after1}}\\
\subfloat[The subroutine constructs the unitary $W_{k,j_{1:M}}$ from $M$ uses of the seed Hamiltonian dynamics.]{\includegraphics[width=\linewidth]{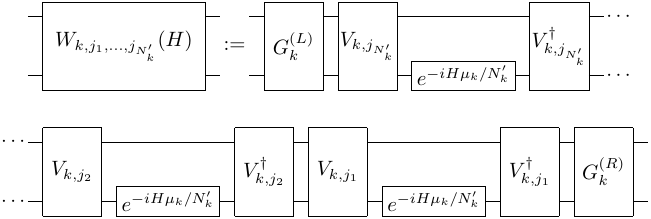}\label{fig::after2}}
\caption{\emph{Intermediate stage of compilation.---}By correlating the layers, an additional error is introduced due to the finite iteration of the previously inner layer. However, we show how this error can be compensated for by choosing $W_{k,j_{1:M}}$ appropriately, therefore constructing an efficient compiled algorithm.}
\label{fig::compiledalgorithm2}
\end{figure}

We do this by first modifying the circuit in Fig.~\ref{fig::before1} to an \emph{intermediate circuit} shown in Fig.~\ref{fig::after1}, where $W_{k,j_{1:M}}(H) =: W_{k,j_{1:M}}$ is written with its dependence on $H$ implicit and is itself defined as in Fig.~\ref{fig::after2}; similarly, we compress the labels $j_{1:M}:=\{ j_1,\hdots,j_M\}$ and $p_{j_{1:M}}^{(k)}:= p_{j_1}^{(k)}\hdots p_{j_M}^{(k)}$ for ease of notation. Finally, an additional modification---which depends upon the task at hand---can then be applied to said intermediate circuit in order to compensate for error accumulation. This leads to the \emph{compiled algorithm}, which we will analyze explicitly for the case of UHET in the coming section.

In the intermediate circuit, the random indices $j_{1:M}$ are correlated in such a way that the components before and after the dynamics $e^{-i\tilde{H}_k\beta t/N}$ are inversions of each other. Therefore, the circuit in Fig. \ref{fig::compiledalgorithm2} simulates $e^{-iH_{\rm new}t}$, where
\begin{align}
    H_{\rm new}:= \beta \sum_{k,j_{1:M}} p_k\,p_{j_{1:M}}^{(k)} W_{k,j_{1:M}}^{\dagger}\tilde{H}_k W_{k,j_{1:M}}.
    \label{H_new_def}
\end{align}
It can be shown using analysis of qDRIFT \cite{campbell2019random} that 
\begin{align}
    \sum_{j_{1:M}} p_{j_{1:M}}^{(k)} W_{k,j_{1:M}}^{\dagger}\tilde{H}_k W_{k,j_{1:M}}
 = U_k^{\dagger}\tilde{H}_kU_k+O\left(\frac{\mu_k^2}{M}\right) ,    \label{eq::intermid_err}
\end{align}
since the error analysis for the qDRIFT procedure remains valid when replacing the simulation of a density operator with that of a Hamiltonian. 

In order to suppress the inner layer error $O(\mu_k^2 / M)$---due to the random sampling of $j$ in each of the $N$ iterations of the outer layer---below $O(\epsilon /N)$, the iteration number $M$ must be chosen as $\Omega(\mu_k^2 N/\epsilon)$. Consequently, since $N$ itself must have $1/\epsilon$ dependence (in order to suppress the overall error below $\epsilon$, as explained previously) the overall time complexity of the intermediate circuit $N \cdot M$ scales according to $1/\epsilon^3$. Next, by compiling the subroutines of the algorithm, one can reduce this resource scaling to $1/\epsilon$. This compilation procedure can be achieved by setting $M$ as $O(\mu_k^2)$---removing the dependence on $\epsilon$ here comes at a cost of increasing the error. However, said $O(\mu_k^2/M)=O(1)$ inner layer error can be compensated for by introducing an additional modification to the procedure shown in Fig. \ref{fig::compiledalgorithm2}, as we will discuss in the coming section for the case of UHET. The total time complexity will be proportional to $N \cdot M$ which scales in terms of $\epsilon$ in the same way as $N$. Therefore, the total time complexity scales as $O(1/\epsilon)$ in terms of $\epsilon$.

\subsection{Compiled UHET Algorithm}\label{subsec::compiledalgorithm}

We now apply this general method of compilation to the task of UHET, thereby providing a more efficient procedure than Algorithm~\ref{alg::umcompi}. The \textbf{compiled algorithm} is presented in full in Algorithm~\ref{alg::main}, with details regarding the error and time complexity provided throughout the SM, App.~\ref{app::compiled}.

With respect to the circuit depicted in Fig.~\ref{fig::after1}, here we choose each iteration of the circuit to correspond to a short time-evolution by the following Hamiltonian: 
\begin{widetext}
\begin{align}
    &\left( \frac{1}{4}
    \right)^{n M}
    \sum_{\vec{v}_{1:M}} {W}_{k,\vec{v}_{1:M}}^{\dagger}([\cos \phi_k X -\sin \phi_k Y]\otimes I) {W}_{k,\vec{v}_{1:M}}
    \nonumber\\
    &= A_{k,M}(e^{i\theta_{k,M} Z/2}\otimes I)
    \left(
    \begin{array}{cc}
        0&e^{i\phi_k}e^{ik\pi (H_0+I)/2}\\
        e^{-i\phi_k}e^{-ik\pi (H_0+I)/2}&0\\
    \end{array}
    \right)
    (e^{-i\theta_{k, M}Z/2}\otimes I)
    \label{Atheta}
\end{align}  
\end{widetext}

\noindent where the iteration number $M$ can depend upon $k$, $A_{k,M}=1-O(k^2/M)>0$, $\theta_{k,M}=O(k^3/M^2) \in \mathbb{R}$ and

\begin{align}
    \label{our_two}
    &{W}_{k,\vec{v}_{1:M}}^{(M)}:=
    \left[\prod_{l=1}^{M}{\tt ctrl}(\sigma_{\vec{v}_l})(I\otimes e^{-i (k \pi H)/(2M)}){\tt ctrl}(\sigma_{\vec{v}_l})\right]
    \nonumber\\
    &\quad \quad \quad \quad \quad \quad  \times (e^{-ik\pi Z/4}\otimes I) .
\end{align}
For simplicity, we sometimes suppress the superscript $(M)$ and write ${W}_{k,\vec{v}_{1:M}}$. We derive Eq.~(\ref{Atheta}) in the SM, App.~\ref{app::pratheta}.

As mentioned earlier, to improve the scaling behavior with respect to the uncompiled algorithm, we must modify the circuit to account for the error introduced by the finite iteration number $M$ of the inner layer. The factor $A_{k,M}$ can be compensated for by modifying the probability distribution $p_k$ and the iteration number of the outer layer $N$; the rotation errors of the form $e^{\pm i\theta_{k,M}Z/2}\otimes I$ can be corrected via an inverse rotation. Thus, the general compilation procedure can be applied to reduce the time complexity. In particular, we take $M$ as $10k^2$ such that $A_{k,M}=1-O(k^2/M) > 1/2$ independently of $k$, which follows from a lower bound of $A_{k,M}$ [see Eq.~(\ref{eq::Atheta_bound}) in the SM, App.~\ref{app::pratheta}].

If a classical description (or block encoding) of the dynamics were given, then the compensation parameters $A_{k,M}$ and $\theta_{k,M}$ could be explicitly calculated via Eq.~\eqref{Atheta}. However, since we only have access to the black-box dynamics $e^{\pm iH\tau}$, we must construct a circuit that efficiently estimates them without relying on explicit knowledge of $H$. We present such a method that makes use of robust phase estimation~\cite{kimmel2015robust} to obtain estimates $(\hat{A}_k, \hat{\theta}_k)$ of the error parameters $(A_{k, 10k^2}, \theta_{k, 10k^2})$ for $k\in \{-K,\ldots ,K\}$ (where $K$ is a cutoff number) in the SM, App.~\ref{app:subrs}. This completes the pre-processing step of Algorithm~\ref{alg::main}.

\begin{algorithm}[H]
    \floatname{algorithm}{Algorithm}
    \caption{Efficient universal Hamiltonian eigenvalue transformation (Compiled)}
    \label{alg::main}
    \begin{algorithmic}[1]
        \Statex{\textbf{Input:}}
        \begin{itemize}
            \item A finite number of queries to a black-box Hamiltonian dynamics $e^{\pm iH\tau}$ of a seed Hamiltonian $H$ normalized as $\|H_0\|_{\mathrm{op}}\leq 1$, where $H_0$ is the traceless part of $H$, i.e., $H_0:=H-(1/2^n)\mathrm{tr}(H)I$, with $\tau>0$
            \item A class $C^3$ (three times continuously differentiable) function $f:[-1,1]\to \mathbb{R}$, such that $f^{(4)}$
            is piecewise $C^2$ (see the SM, App.~\ref{app::jsmooth})
            \item Input state $\ket{\psi}\in \mathcal{H}$
            \item Allowed error $\epsilon >0$
            \item Time $t>0$
        \end{itemize}
        \Statex{\textbf{Output:}}
         A state approximating $e^{-if(H_0)t}\ket{\psi}\ (t>0)$ with an error in terms Eq.~(\ref{measure_sta}) upper-bounded by $\epsilon$ (also, the mean square of error in terms of Eq.~(\ref{varivari}) upper-bounded by $2\epsilon$)
    \Statex \hrulefill
    	\Statex{\textbf{Time complexity:}}
            \Statex  Pre-processing (only once): $O(\grave{K}^3t^3n/\epsilon^3)+T_{\ref{alg::main}},\ \grave{K}=O[(t/\epsilon)^{1/3}]$
            \Statex  Main Process: $O(C_{4,f} t^2n/\epsilon)$ for an $f$-dependent constant $C_{4,f}$ which is independent of $n,\ t,$ and $\epsilon$
            \Statex{\textbf{Total evolution time (main process):}}
            $O (C_{2,f}t^2/\epsilon)$
    	\Statex{\textbf{Used Resources:}}
    	\Statex \hskip1.0em System: $n$-qubit system $\mathcal{H}$ and one auxiliary qubit $\mathcal{H}_c$
    	\Statex \hskip1.0em Gates: $e^{\pm iH\tau}$, single-qubit gate on $\mathcal{H}_c$, and Clifford gates on $\mathcal{H}_c\otimes\mathcal{H}$
    	\Statex \hrulefill
        \Statex{\textbf{Procedure:}}
        \Statex \hspace{-1.5em}\textit{Pre-processing:} 
        \State Define $\tilde{f}$ as shown in Eq. (\ref{def_tilde_f}) and compute Fourier coefficients $\tilde{c}_k:=(1/2)\int_{-1}^1 \mathrm{d}x e^{-ik\pi x} \tilde{f}(x)$ for $k\in \{-\grave{K}, -\grave{K}+1,\ldots ,\grave{K}\}$ for a $\grave{K}>0$ satisfying Eq. (\ref{grvK})
        \For{$k\in \{1,\ldots ,\grave{K}\}$} 
        \State Generate $(\hat{A}_k, \hat{\theta}_k)$ by Subroutine \ref{subr::gen} of the SM, App.~\ref{app:subrs} with allowed error set as $\sqrt{3}\epsilon/(12\pi (\sum_{k=-\infty}^{\infty} |\tilde{c}_k||k|))$
        \State $(\hat{A}_{-k}, \hat{\theta}_{-k})\gets (\hat{A}_{k}, -\hat{\theta}_{k})$
        \EndFor
        \State $(\hat{A}_0, \hat{\theta}_0)\gets (1, 0)$
        \State Compute $\grave{N}:=N(\grave{\beta}, t, \epsilon/3)$ for $\grave{\beta} :=\sum_{k=-\grave{K}}^{\grave{K}} (|\tilde{c}_k|/\hat{A}_k)$ 
        \vspace{0.3em}
        \Statex \hspace{-1.5em}\textit{Main Process:}
        \State{Initialize}
        $\ket{\text{current}}\gets \ket{+}\otimes \ket{\psi}$
        \For {$m\in \{1,\ldots ,\grave{N}\}$}
        \State Randomly choose $k$ with probability $\grave{p}_k:=|\tilde{c}_k|/(\hat{A}_k  \grave{\beta})$
        \State Randomly choose $j_k:=\{\vec{v}_{1,10k^2} \}\in (\{0,1,2,3\}^n)^{10k^2}$
        \State $\ket{\text{current}}\gets \grave{W}_{k,j_k}^{\dagger}(e^{-i[\cos\phi_kX-\sin\phi_kY]\grave{\beta} t/ \grave{N}^{(F)}}\otimes I)\grave{W}_{k,j_k}\ket{\text{current}}$
        \EndFor
        \State Trace out $\mathcal{H}_c$ of $\ket{\text{current}}$
        \State {\textbf{Return}} $\ket{\text{current}}$
    \end{algorithmic}
\end{algorithm}

\begin{figure}[htb]
    \begin{center}
        \includegraphics[width=\linewidth]{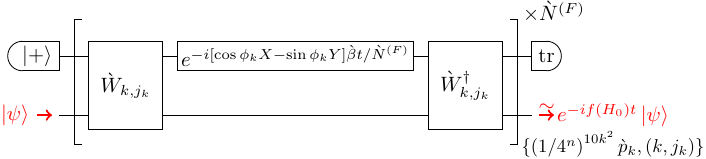}
        \caption{\emph{Compiled version of outer layer (Fig. \ref{fig::uncompiled_circuit}).---}By using modified values $\grave{p}_k,\ \grave{\beta},\  \grave{N}^{(F)}$, and a modified operator $\grave{W}_{k,j_k}$ (see Algorithm \ref{alg::main} for details), the iteration number $N_k^{(C)}=\Theta(\beta^2 t^2 k^2 /\epsilon^2)$ for controlization in Algorithm \ref{alg::umcompi} is reduced to $10k^2=\Theta(k^2)$ independent of $f, n, t$ and $\epsilon$, leading to a more efficient UHET algorithm.\label{fig::cmpld_alg3}}
    \end{center}
\end{figure}

The main part of Algorithm~\ref{alg::main} then makes use of these estimates to compensate the error in simulating the desired transformed dynamics $e^{-if(H_0)t}$; we provide a complete error and time complexity analysis in the SM, App.~\ref{app:al1}. The circuit representation of Algorithm~\ref{alg::main} is depicted in Fig.~\ref{fig::cmpld_alg3}, in which we choose a new cutoff number $\grave{K}$ satisfying
\begin{align}
    \left| \tilde{f}(x)-\sum_{k=-\grave{K}}^{\grave{K}} \tilde{c}_k e^{ik\pi x}\right| <\frac{\epsilon}{6t}
    \quad \forall \, x\in [-1,1]
    \label{grvK},
\end{align}
\noindent and set $j_k\!:=\!(\vec{v}_{1:10k^2})\!\in\!(\{0,1,2,3\}^n)^{10k^2}$, $\!\grave{\beta}\!:=\!\sum_{k=-\grave{K}}^{\grave{K}}(|\tilde{c}_k|/\hat{A}_k)$, $\grave{N}\!:=\!N(\grave{\beta}, t, \epsilon /3)$, $\grave{p}_k\!:=\!|\tilde{c}_k|/(\hat{A}_k\grave{\beta})$, and 
\begin{align}
    \label{graW}
    \grave{W}_{k,j_k}:=W^{(10k^2)}_{k,j_k}(e^{i\hat{\theta}_kZ/2}\otimes I).
\end{align}

\noindent The average number of times that the unknown dynamics $e^{-iH\tau}\ (\tau>0)$ is called in any one sampling of $k$ is reduced via compilation from $\sum_{k} p_k \Theta(\beta^2t^2k^2/\epsilon^2)$ (Algorithm \ref{alg::umcompi}) to $\sum_k\grave{p}_k \Theta(k^2)$ (Algorithm \ref{alg::main}) and the average depth of the overall circuit is reduced accordingly.

The time complexity of the pre-processing of Algorithm~\ref{alg::main} is $O(\grave{K}^3t^3n/\epsilon^3)+T_4$, where $\grave{K}=O[(t/\epsilon)^{1/3}]$ is the cutoff number defined in step 1 and $T_{4}$ refers to the sum of the classical computation time complexity for steps 1 (computation of the Fourier coefficients $\tilde{c}_k$ until Eq.~(\ref{grvK}) is satisfied) and step 7 (computation of $\grave{N}$). The main process has time complexity $O(C_{4,f}t^2n/\epsilon)$ (see the SM, App.~\ref{app:al1}). The scaling of the time complexity of the main process in the $t\to \infty$ and $\epsilon\to 0$ limits are reduced in Algorithm \ref{alg::main} compared to Algorithm \ref{alg::umcompi}. Lastly, the total evolution time for the main process is evaluated in the same way as in Subroutine \ref{subr::fouri} as $O(C_{2,f}t^2/\epsilon)$.

\FloatBarrier

\section{Comparison with a QSVT-based UHET Algorithm}\label{sec::qsvt}

Algorithms~\ref{alg::umcompi} and~\ref{alg::main} above provide two novel methods to implement the task of UHET; we now move to compare these algorithms with another one that achieves the same task via a modified QSVT procedure. In principle, one can combine standard QSVT methods with the ability to simulate Hamiltonian dynamics to achieve said task. Given a classical description of the Hamiltonian, this can be achieved by performing a block-encoding of $H$ into a unitary, using QSVT to approximate $e^{-if(H_0)t}$ (up to a proportionality constant), and then amplifying the block via \emph{robust oblivious amplitude amplification} \cite{gilyen2019quantum,martyn2021grand}. However, such a procedure requires a classical description of the Hamiltonian to be known \emph{a priori}, in contradistinction to the UHET task we have so far considered. Here, instead of a classical description (or block encoding) of $H$, one is only given access to a black-box Hamiltonian dynamics $e^{\pm iHt}$. In the SM, App.~\ref{app::qsvt}, we provide a method that modifies the standard QSVT procedure by controlizing a modified version of the unknown Hamiltonian and then applying appropriate gates to simulate the desired function.

This procedure allows fair comparison between Algorithm \ref{alg::umcompi} (uncompiled), Algorithm \ref{alg::main} (compiled) and the QSVT-based Algorithm~\ref{alg::qsvt} presented in full in the SM, App.~\ref{app::comparison}. Note that in all three algorithms compared here, the scaling of the time complexity dominates that of the total evolution time; as such, the overall runtime behavior is determined by the time complexity. Thus, in this section, we only compare the algorithms in terms of their time complexities. We show that the scaling of the time complexity of the main process (i.e., the part that must be run each time) of these algorithms behaves in the limit of $\epsilon\to 0$ as:
\begin{align}
    &\text{time complexity of Algorithm \ref{alg::main} (compiled)}
    \nonumber\\
    \leq\,&
    \text{time complexity of QSVT-based Algorithm~\ref{alg::qsvt}}
    \nonumber\\
    \leq\,&
    \text{time complexity of Algorithm \ref{alg::umcompi} (uncompiled)}.
    \label{order_three}
\end{align}
In other words, Algorithm \ref{alg::umcompi}, which is slower than the QSVT-based Algorithm~\ref{alg::qsvt} when uncompiled, becomes faster than it via compilation into Algorithm~\ref{alg::main}. In the SM, App.~\ref{app::comparison}, we describe how QSVT techniques can be applied to the task of UHET and present an algorithm that leverages ideas from the Hamiltonian singular value transformation~\cite{lloyd2021hamiltonian} and the QSVT-based Hamiltonian simulation~\cite{low2019hamiltonian,martyn2021grand,gilyen2019quantum}. In the SM, App.~\ref{app::qsvtcomparisondetails}, we calculate the $\epsilon$ dependence of the three studied algorithms, demonstrating the relation~\eqref{order_three}, and explain the technical reasons for the differences in scaling. Put briefly, the resource advantage of Algorithm~\ref{alg::main} stems from the fact that we can remove the $\epsilon$-dependence of the controlization part via compilation and the algorithm efficiently treats high frequency terms due to the fact that low-weight terms in the Fourier series are sampled infrequently.

\section{Conclusions}\label{sec::conclusion}

In this work, we developed universal quantum algorithms for transforming the eigenvalues of any Hamiltonian by any (suitably differentiable) function, while keeping the eigenstates unchanged. Our algorithms are universal in the sense that they do not rely on knowledge of the input Hamiltonian, whose dynamics can be given as a black box. The uncompiled algorithm is constructed by concatenating two subroutines, namely controlization and Fourier series simulation. We furthermore introduced a general framework for compilation, which uses correlated randomness to perform multiple layers of random sampling efficiently. For our algorithm, we showed that the compilation step significantly reduces the time complexity, making it even more efficient than simulation methods based on QSVT. 

Our results have implications across the realm of quantum information processing and beyond. First, the notion of compilation reconciles quantum computing practices with a concept in classical computing, where subroutines are compiled into larger functions in order to be implemented more efficiently. Extending this approach to different tasks could significantly improve our ability to develop complex and efficient quantum algorithms in a modular fashion, similar to that of classical software. Second, we have expanded the class of universal Hamiltonian transformations that can be efficiently performed to all suitably differentiable functions on the space of Hamiltonians, extending previously known results for linear functions of Hamiltonians \cite{odake2023higher}. This generalization provides a crucial step towards simulating complex many-body dynamics and designing processes with desired non-linearities.

In the future, developing a general theoretical framework for higher-order transformation of Hamiltonian dynamics will provide more insight into the possible manipulation of Hamiltonian dynamics for quantum information processing tasks. Indeed, complementing our approach here with other methods for Hamiltonian simulation that adhere to certain assumptions, such as QSVT~\cite{low2019hamiltonian, gilyen2019quantum} or LU-based approaches~\cite{childs2012linear,berry2015hamiltonian,childs2017quantum}, could provide further advantages in many contexts of interest and we look forward to future work in this area. As fault-tolerant quantum computers become more readily available, we envisage that our methods will apply to practical use cases in the simulation and manipulation of quantum systems, such as in quantum chemistry or materials discovery.

\begin{acknowledgments}
We would like to thank Anirban Chowdhury, Toshinori Itoko, Antonio Mezzacapo, Zane Rossi, Kunal Sharma, Satoshi Yoshida, Lei Zhang, and Alexander Zlokapa for helpful discussions. This work was supported by MEXT Quantum Leap Flagship Program (MEXT QLEAP) JPMXS0118069605, JPMXS0120351339, Japan Society for the Promotion of Science (JSPS) KAKENHI Grant No. 21H03394 and 23K21643, and IBM Quantum. P. T. acknowledges support from the Japan Society for the Promotion of Science (JSPS) Postdoctoral Fellowship for Research in Japan. Research at Perimeter Institute is supported in part by the Government of Canada through the Department of Innovation, Science and Economic Development and by the Province of Ontario through the Ministry of Colleges and Universities. 
\end{acknowledgments}

%\bibliography{references.bib}

%apsrev4-2.bst 2019-01-14 (MD) hand-edited version of apsrev4-1.bst
%Control: key (0)
%Control: author (8) initials jnrlst
%Control: editor formatted (1) identically to author
%Control: production of article title (0) allowed
%Control: page (0) single
%Control: year (1) truncated
%Control: production of eprint (0) enabled
%

\clearpage
\onecolumngrid

\begin{center}
    \textbf{\underline{SUPPLEMENTAL MATERIAL}}\vspace{-2em}
\end{center}

\addcontentsline{toc}{section}{Appendices}

\appendix

\section{Preliminary Definitions}\label{app::prelim}

\noindent Here, we provide some basic definitions that will prove useful throughout our work.

\subsection{$\boldsymbol{J}$-Smoothness of Functions}\label{app::jsmooth}

\begin{de}[Piecewise $C^J$]\label{def::piecewisecj}
Let $\mathcal{X}:=\{x_0,\dots, x_n\}$, where $-1=x_0<x_1<\cdots <x_n=1$. 
A function \mbox{$g:[-1,1]\backslash \mathcal{X}' \to \mathbb{R}$}, where $\mathcal{X}' \subseteq \mathcal{X}$, is \emph{\textbf{piecewise} $\mathbf{C^J}$} $(J \in \mathbb{N}_+)$
$\overset{\mathrm{def}}{\Leftrightarrow}$ the derivatives $g^{(m)}$ for $m\in \{0,1,\ldots ,J\}$ are well-defined and continuous everywhere in $[-1,1]\backslash \mathcal{X}'$, and 
additionally, at the \emph{exceptional points} $x_1, \ldots, x_{n-1}$ of $g$, we have that for all $j\in \{1,\ldots ,n-1\}$, $\lim_{x\to x_j^+} g^{(m)}(x)$ and $\lim_{x\to x_j^-} g^{(m)}(x)$ both exist for all $m\in \{0,\ldots ,J\}$ (although the limits from above and below need not coincide), as well as that for $x=\pm 1$, $\lim_{x\to -1^+}g^{(m)}(x)$ and $\lim_{x\to 1^-}g^{(m)}(x)$ exist. In particular, when $J=1$, $g$ is said to be \emph{\textbf{piecewise smooth}}.
\end{de}

\begin{de}[$J$-Smoothness]
    \label{J_sm}
    Here, we extend the notion of piecewise continuity from above to higher orders of smoothness.
    \begin{enumerate}
        \item 
    A function $g:[-1,1]\to \mathbb{R}$ is \emph{$\boldsymbol{J}$\textbf{-smooth}}
    $\overset{\mathrm{def}}{\Leftrightarrow}$
    $g$ is piecewise C$^{J-1}$ and $g^{(J)}$ is piecewise $C^2$.
        \item 
    A function $g:[-1,1]\to \mathbb{R}$ is \emph{\textbf{periodically} $\boldsymbol{J}$\textbf{-smooth}} $:\overset{\mathrm{def}}{\Leftrightarrow}$
    $g$ is $J$-smooth and $g^{(m)}(+1)=g^{(m)}(-1)\ \forall \ m\in \{0,\ldots ,J-1\}$.
    \item
    A function $g:[-1,1]\to \mathbb{R}$ is \emph{\textbf{strictly} $\boldsymbol{J}$\textbf{-smooth}} $\overset{\mathrm{def}}{\Leftrightarrow}$
    $g$ is $J$-smooth and $g^{(J)}$ does not satisfy 
    \begin{align}
    \label{not_satis_notp}
    \lim_{x\to a^+}g^{(J)}(x)=\lim_{x\to a^-}g^{(J)}(x)\quad \forall \textup{ exceptional points } a\in (-1,1).
    \end{align}
    \item
    A function $g:[-1,1]\to \mathbb{R}$ is \emph{\textbf{strictly periodically} $\boldsymbol{J}$\textbf{-smooth}}
    $\overset{\mathrm{def}}{\Leftrightarrow}$
    $g$ is periodically $J$-smooth and $g^{(J)}$ does not satisfy either (or both) of the following conditions:
    \begin{align}
    \label{not_satis}
    &\mathrm{1.}\ 
    \lim_{x\to a^+}g^{(J)}(x)=\lim_{x\to a^-}g^{(J)}(x)\quad \forall \textup{ exceptional points } a\in (-1,1)\nonumber\\
    &\mathrm{2.}\ 
    \lim_{x\to 1^-}g^{(J)}(x)=\lim_{x\to -1^+}g^{(J)}(x).
    \end{align}
    \end{enumerate}
\end{de}

\FloatBarrier

\subsection{Norms for Quantifying Errors}

\noindent When quantifying the error of a simulated quantum operation, we will often make use of the diamond norm, defined as
\begin{align}\label{eq::diamondnorm}
    \|\Phi \|_{\diamond}:=\max_{A \in \mathcal{L}(\mathcal{H}\otimes \mathcal{H});\|A\|_1=1}\|(\Phi\otimes \mathcal{I})(A)\|_1 ,
\end{align}
where $\Phi :\mathcal{L}(\mathcal{H})\to \mathcal{L}(\mathcal{H})$ is a quantum operation, $\mathcal{I}$ is an identity operation on $\mathcal{L}(\mathcal{H})$, and $\|\cdot\|_1$ denotes the 1-norm $\|A\|_1:= {\rm tr}(\sqrt{A^{\dagger}A})$.

\subsection{Scaling Notation}

\noindent Throughout, we use the symbols $O(\cdot),\ \Omega (\cdot),$ and $\Theta (\cdot)$---the definitions of which are provided in Table~\ref{tab::scaling}---to denote the scaling behavior of time complexities of the algorithms presented. Intuitively speaking, $f(x) = O(g(x))$ if $g(x)$ grows at least as fast as $f(x)$ in $\lim x \to \infty$ (i.e., $g$ asymptotically upper bounds $f$); $f(x) = \Omega(g(x))$ if $f(x)$ grows at least as fast as $g(x)$ in $\lim x \to \infty$ (i.e., $g$ asymptotically lower bounds $f$); and $f(x) = \Theta(g(x))$ if $g$ provides both an upper and lower bound of $f$ asymptotically. Furthermore, we consider various limits depending on the parameter of interest. In particular, we consider the limit $\to \infty$ for qubit number $n\in \mathbb{Z}_{>0}$,  simulation time $t\in \mathbb{R}$, and the limit $\to 0$ for the allowed error $\epsilon >0$. For instance $f(t,\epsilon)=O(t^2\epsilon^{-1})$ means that $\lim \sup_{t\to \infty}({f(t,\epsilon)}/{t^2})< \infty$ for all $\epsilon >0$ and $\lim \sup_{\epsilon\to 0}({f(t,\epsilon)}/{\epsilon^{-1}})< \infty$ for all $t\in \mathbb{R}$. 

\begin{table}[t]
\begin{center}
\begin{tabular}{|c|c|}
\hline
Notation&Definition\\
\hline 
$f(x)=O(g(x))$&$\lim \sup_{x\to \infty}({f(x)}/{g(x)})< \infty$\\
\hline 
$f(x)=\Omega(g(x))$&$\lim \inf_{x\to \infty}({f(x)}/{g(x)})>0$\\
\hline 
$f(x)=\Theta(g(x))$&$f(x)=O(g(x))$ and $f(x)=\Omega(g(x))$\\
\hline 
\end{tabular}
\end{center}
\caption{\textit{Scaling symbols.---}For any $g(x)$, we say that a function $f(x)$ is $O(g(x))$, $\Omega(g(x))$, or $\Theta(g(x))$, if the above are satisfied.} 
\label{tab::scaling}
\end{table}

\FloatBarrier

%\newpage

\section{Universal Hamiltonian Eigenvalue Transformation (UHET)}\label{app::uhet}

\subsection{Mean Squared Error Bound}\label{app::uhetvarivari}

\noindent Here we demonstrate the validity of Eq.~\eqref{varivari}, which bounds the mean squared error of an average quantum operation in terms of an original error bound. More precisely, we prove:

\begin{lem}[Mean Squared Error Bound]
    \label{gene_vari}
    Consider an arbitrary unitary operation defined by $\mathcal{U}(\rho ):=U\rho U^{\dagger}$ with a unitary operator $U$ and a density operator $\rho$ on a Hilbert space $\mathcal{H}$. If the error (in terms of the 1-norm) of a set of deterministic quantum operations (completely-positive trace-preserving maps) $\mathcal{F}_j:\mathcal{L}(\mathcal{H})\to \mathcal{L}(\mathcal{H})$ and a probability distribution $\{ p_j \}$ satisfies
    \begin{align}
        \label{weak}
        \underset{
        \substack{\ket{\psi}\in \mathcal{H}\\ ||\ket{\psi}||=1}
        }
        {\mathrm{sup}}
        \| \mathcal{U}(\ket{\psi}\bra{\psi})-\sum_j p_j \mathcal{F}_j(\ket{\psi}\bra{\psi}) \|_1 \leq \Delta,
    \end{align}
    for some $\Delta >0$, then the mean squared error of the average quantum operation $\sum_j p_j \mathcal{F}_j$ is upper bounded by 
    \begin{align}
        \label{vari_in_lem}
        \underset{
        \substack{\ket{\psi}\in \mathcal{H}\\ ||\ket{\psi}||=1}}
        {\mathrm{sup}}
        \sum_j p_j\|\mathcal{U}(\ket{\psi}\bra{\psi})-
        \mathcal{F}_j(\ket{\psi}\bra{\psi}) \| _1 ^2
        \leq 2\Delta,
    \end{align}
    where $\ket{\psi}$ is any pure state on $\mathcal{H}$.
\end{lem}

This lemma is proven in App. B of Ref.~\cite{odake2023higher}. Furthermore, if a modified version of Eq.~(\ref{weak}) is applied to an extended Hilbert space, namely 
\begin{align}
    \label{strong}
        &\underset{\substack{ 
    \mathrm{dim}(\mathcal{H}^{\prime}) \\
    \ket{\psi}\in \mathcal{H}\otimes \mathcal{H}^{\prime}\\
    \|\ket{\psi}\|=1 }}{\mathrm{sup}}
        \| \mathcal{U}\otimes \mathcal{I}_{\mathcal{H}'}(\ket{\psi}\bra{\psi})-\sum_j p_j \mathcal{F}_j\otimes\mathcal{I}_{\mathcal{H}'}(\ket{\psi}\bra{\psi}) \|_1 \leq \Delta
\end{align}
holds, then the same reference showed that  a stronger version of Eq.~(\ref{vari_in_lem}), i.e., 
\begin{align}
    \label{strong_vari_in_lem}
        &\underset{\substack{ 
    \mathrm{dim}(\mathcal{H}^{\prime}) \\
    \ket{\psi}\in \mathcal{H}\otimes \mathcal{H}^{\prime}\\
    \|\ket{\psi}\|=1 }}{\mathrm{sup}}
        \sum_j p_j\|\mathcal{U}\otimes\mathcal{I}_{\mathcal{H}'}(\ket{\psi}\bra{\psi})-
        \mathcal{F}_j\otimes\mathcal{I}_{\mathcal{H}'}(\ket{\psi}\bra{\psi}) \| _1 ^2 \leq 2\Delta \, ,
\end{align}
follows.

\FloatBarrier

\newpage

\section{Uncompiled UHET Algorithm}\label{app::uncompiled}

\noindent Here we provide details relevant to the uncompiled UHET algorithm presented throughout Sec.~\ref{sec::uncompiled} of the main text.

\subsection{Sufficient Number of qDRIFT Iterations}\label{app::qdriftiterations}

\noindent The qDRIFT procedure is a stochastic method for simulating Hamiltonian dynamics~\cite{campbell2019random}. Here, we determine a sufficient number of iterations to ensure a sufficiently small error $\epsilon$.

\begin{lem}
\label{qdri:err}
Suppose that one has access to the dynamics $e^{-iH_j\tau}\ (\tau >0)$ corresponding to a set of Hamiltonians $\{H_j\}_j$ on $\mathcal{L}(\mathcal{H})$ (normalized as $\|H_j\|_{\mathrm{op}}=1$). Then, the dynamics $e^{-iHt}\ (t>0)$ of a Hamiltonian represented as $H=\sum_j h_j H_j$ for a set of positive coefficients $\{h_j\}_j$ can be simulated using qDRIFT with an error of at most $(2\lambda^2 t^2/N) e^{2\lambda t/N}$ where $\lambda := \sum_j h_j$ and $N$ is the number of iterations of random sampling. Here, the error is quantified by
\begin{align}
    \label{qdri_error}
    \frac{1}{2}\| \mathcal{F}_{\mathrm{target}}-{ \mathcal{F}}_{\mathrm{approx}}\|_{\diamond},
\end{align}
where $\mathcal{F}_{\mathrm{target}}(\rho) :=e^{-iHt}\rho e^{iHt}$ and $\mathcal{F}_{\mathrm{approx}}$ is the average quantum operation simulated by the qDRIFT protocol.
\end{lem}

\noindent The proof of this Lemma is provided in Ref.~\cite{campbell2019random}. In particular, if $N$ is chosen as $N(\lambda, t, 2\epsilon)$ [as defined in Eq.~\eqref{eq::iterationnumber}], then the error in terms of Eq.~(\ref{qdri_error}) is upper bounded by $\epsilon$, i.e.,
\begin{align}
    \frac{2\lambda^2 t^2}{N(\lambda, t, 2\epsilon)}e^{2\lambda t/N(\lambda, t, 2\epsilon)}
    &\leq 
    2\lambda^2t^2\cdot \left(
    \frac{5\lambda^2t^2}{\epsilon}
    \right)^{-1}
    e^{2\lambda t\cdot (5\lambda t/2)^{-1}} =\frac{2\epsilon}{5}e^{4/5}<\epsilon .
\end{align}
Moreover, using the property of the diamond norm that for any linear map $\Phi$, $$ \underset{\substack{\ket{\psi}\in\mathcal{H}\otimes \mathcal{H'}, \|\ket{\psi}\|=1, \mathrm{dim}{\mathcal{H}'} }}{\mathrm{sup}}||\Phi \otimes \mathcal{I}_{\mathcal{H}'} (\ket{\psi}\bra{\psi})||_1 \leq ||\Phi \otimes \mathcal{I}_{\mathcal{H}'} ||_1 \leq ||\Phi ||_\diamond$$ for any dimension of Hilbert space $\mathcal{H}'$ \cite{watrous2018theory}, it follows that the error in terms of Eq.~(\ref{measure_ope}), namely
\begin{align}
    \underset{\substack{\ket{\psi}\in\mathcal{H}\otimes \mathcal{H'}\\ \|\ket{\psi}\|=1\\ \mathrm{dim}{\mathcal{H}'} }}{\mathrm{sup}}
    \|\mathcal{F}_{{\rm target}}\otimes \mathcal{I}_{\mathcal{H}'}(\ket{\psi}\bra{\psi})-
    \mathcal{F}_{{\rm approx}}\otimes \mathcal{I}_{\mathcal{H}'}(\ket{\psi}\bra{\psi})
    \|_1
\end{align}
is upper bounded by twice that of Eq.~\eqref{qdri_error}, and can thus be upper bounded by $\epsilon$ by choosing the iteration number as $N(\lambda, t, \epsilon)$.

From the above, we see that for a given allowed error $\epsilon$, the time complexity of $N=N(\lambda, t, \epsilon)$ [as defined in Eq.~\eqref{eq::iterationnumber}] is $O(\lambda^2 t^2 /\epsilon)$. By considering the second-order error terms in the qDRIFT protocol (see Appendix A of Ref.\ \cite{campbell2019random}), we can also show that the lower bound of this time complexity is equal to the upper bound, as long as the $\{H_j\}_j$ are not all equal, in which case, the time complexity of $N=N(\lambda, t, \epsilon)$ [as defined in Eq.~\eqref{eq::iterationnumber}] becomes $\Theta(\lambda^2 t^2 /\epsilon)$.

\FloatBarrier

\subsection{Definition of $g_f$ in Fourier Series Simulation [Subroutine \ref{subr::fouri}, Eq.~(\ref{def_tilde_f})]}\label{app:funcs}

\noindent The first step in the Fourier Series Simulation part of our UHET algorithm is to modify the desired transformation function $f$ to a suitable periodically smooth one $\tilde{f}$, which is in turn defined in terms of the function $g_f$ below [see Eq.~\eqref{def_tilde_f}]: 
\begin{align}\label{eq::def-app-g_f}
g_f(x):=\,&
C_1\cos{(\pi x)}+S_1\sin{(\pi x)}+\frac{1}{2}S_2\sin{(2\pi x)}+
C_2\cos{(2\pi x)} +C_3\cos{(3\pi x)}+\frac{1}{3}S_3\sin{(3\pi x)}\nonumber \\
+&\frac{1}{4}S_4\sin{(4\pi x)}+C_4\cos{(4\pi x)},\end{align}
where the coefficients $C_1,\ C_2,\ C_3,\ C_4,\ S_1,\ S_2,\ S_3,\ S_4$ are defined succinctly in terms of the coefficient matrix
\begin{align}
&\Phi:=
\left(\begin{array}{cccc}
9/16&1/16&-9/16&-1/16\\
2/3&1/24&2/3&1/24\\
-1/16&-1/16&1/16&1/16\\
-1/6&-1/24&-1/6&-1/24\\
\end{array} \right)
\end{align}
as
\begin{align}
&(C_1,C_2,C_3,C_4)^T
:=\Phi\cdot \left( f(-1), \frac{4f^{(2)}(-1)}{\pi^2}, f(1), \frac{4f^{(2)}(1)}{\pi}\right)^T,
\nonumber\\
&(S_1,S_2,S_3,S_4)^T :=\Phi\cdot \left( \frac{2f^{(1)}(-1)}{\pi}, \frac{8f^{(3)}(-1)}{\pi^3}, \frac{2f^{(1)}(1)}{\pi}, \frac{8f^{(3)}(1)}{\pi^3}\right)^T
 .
\end{align}

\FloatBarrier

\subsection{Error Analysis of Fourier Series Simulation (Subroutine \ref{subr::fouri})} \label{app:error_nokori}

\noindent Here, we rigorously analyze the performance of Subroutine \ref{subr::fouri}. In particular, we prove the following Theorem:
\begin{Theorem}
\label{theo:sub2}
    Subroutine \ref{subr::fouri} outputs $e^{-if(H_0)t}\ket{\psi}$ with an error of at most $\epsilon$. Here, the error is defined as: 
    \begin{align}
        \label{err_def2}
        \underset{\substack{ 
    \mathrm{dim}(\mathcal{H}^{\prime}) \\
    \ket{\psi}\in \mathcal{H}\otimes \mathcal{H}^{\prime}\\
    \|\ket{\psi}\|=1 }}{\mathrm{sup}}
        \|\mathcal{G}\otimes \mathcal{I}_{\mathcal{H}'}(\ket{\psi}\bra{\psi})-
        \sum_jp_j(\mathcal{G}_j\otimes \mathcal{I}_{\mathcal{H}'})(\ket{\psi}\bra{\psi})
        \|_1 ,
    \end{align}
    where $\mathcal{G}(\rho ):=e^{-if(H_0)t}\rho e^{if(H_0)t}$, $j$ labels the tuple of all random indices $k$ chosen in $N$ iterations, $p_j$ refers to the probability that $j$ is chosen, which leads to the particular quantum operation $\mathcal{G}_j$ being simulated, and $\mathcal{H}'$ is an auxiliary system of arbitrary dimension.
\end{Theorem}

\noindent We prove this theorem as follows. First, we decompose the error into two contributions: that of approximating the function $\tilde{f}$ via its (truncated) Fourier series and that of the qDRIFT protocol itself. We then upper bound each error contribution to derive an upper bound for the total error of the subroutine. \\

\noindent \textbf{Proof:} First, we define three quantum operations $\mathcal{G}_1, \mathcal{G}_2, \mathcal{G}_3$ $:\mathcal{L}(\mathcal{H}_c\otimes \mathcal{H})\to \mathcal{L}(\mathcal{H}_c\otimes \mathcal{H})$:
\begin{align}
    \mathcal{G}_1(\rho )&:=e^{-i(X\otimes f(H_0))t}\rho e^{i(X\otimes f(H_0))t} \nonumber\\
    \mathcal{G}_2(\rho )&:=e^{-i(X\otimes \tilde{f}_K(H_0))t}\rho e^{i(X\otimes \tilde{f}_K(H_0))t}
    \nonumber\\
    \mathcal{G}_3&:=\sum_j p_j\mathcal{G}'_j.
\end{align}
where $\tilde{f}_K(x):=\sum_{k=-K}^K \tilde{c}_k e^{ik\pi (x+1)/2}$ is the truncated Fourier representation of $f$ 
(which is a real function due to $\tilde{c}_{-k}=\tilde{c}_k^*$) and $\mathcal{G}'_j: \mathcal{L}(\mathcal{H}_c\otimes \mathcal{H})\to \mathcal{L}(\mathcal{H}_c\otimes \mathcal{H})$ is the quantum operation simulated between steps 5 and 9 of Subroutine \ref{subr::fouri}. By applying $\mathcal{G}_1$ to the initial state $\ket{+}\otimes\ket{\psi}$ and finally tracing over the control Hilbert space $\mathcal{H}_c$, one yields the desired transformation, i.e., the ideal dynamics. The expression $\mathcal{G}_2$ corresponds to the simulated dynamics of a truncated Fourier series in the absence of any Trotterization error (i.e., a perfectly accurate simulation of the finite Fourier series). Lastly, $\mathcal{G}_3$ denotes the actual dynamics simulated in Subroutine \ref{subr::fouri}, which includes errors due to both finite Fourier series cutoff and Trotterization.

Consider the norm $E$ defined as 
\begin{align}
    \label{our_err3}
    E(\mathcal{F}):=    \underset{\substack{\ket{\psi}\in\mathcal{H}_c\otimes \mathcal{H}\otimes \mathcal{H'}\\ \|\ket{\psi}\|=1\\ \mathrm{dim}{\mathcal{H}'} }}{\mathrm{sup}}
    \|\mathcal{F}\otimes \mathcal{I}_{\mathcal{H}'}(\ket{\psi}\bra{\psi})\|_1 ,
\end{align}
for any quantum operation $\mathcal{F}:\mathcal{L}(\mathcal{H}_c\otimes \mathcal{H})\to \mathcal{L}(\mathcal{H}_c\otimes \mathcal{H})$. The difference $E(\mathcal{G}_3-\mathcal{G}_1)$, therefore quantifies the total error between the simulated dynamics and the ideal case.

Since $\mathcal{G}_3$ is a qDRIFT protocol approximating the ideal dynamics by simulating a Hamiltonian $X\otimes \tilde{f}_K(H_0)$ with finite cutoff number $K$ as shown in Eq.~(\ref{qdrift_out}),  it follows from Lemma \ref{qdri:err} that $E(\mathcal{G}_3-\mathcal{G}_2) \leq \epsilon /2$  (since the precision is chosen as $\epsilon /2$). Furthermore, we can make use of the identity $\|\ket{\beta}\bra{\beta}-\ket{\gamma}\bra{\gamma}\|_1=2\sqrt{1-|\braket{\beta|\gamma}|^2}$ (see, e.g., Eq.~(1.186) of \cite{watrous2018theory}) where $\ket{\beta},\ \ket{\gamma}$ are unit vectors in the same Hilbert space, as well as the triangle inequality to yield
\begin{align}
    E(\mathcal{G}_3-\mathcal{G}_1)&\leq E(\mathcal{G}_3-\mathcal{G}_2)+E(\mathcal{G}_2-\mathcal{G}_1)
    \nonumber\\
    &\leq
    \frac{\epsilon}{2}+\underset{\substack{\ket{\psi}\in \mathcal{H}_c\otimes \mathcal{H}\otimes \mathcal{H}'\\ \|\ket{\psi}\|=1\\ {\rm dim} (\mathcal{H}')}}{{\rm sup}}
    2\left[
    1-
    |\bra{\psi}
    e^{-iX\otimes (f(H_0)-\tilde{f}_K (H_0))t}\otimes I
    \ket{\psi}
    |^2
    \right]^{1/2}.
    \label{eq::b_inter}
\end{align}
We now lower bound the r.h.s.\ by decomposing $\ket{\psi}=:\sum_{s,m}a_{s,m}\ket{s}\ket{E_m}\ket{\psi_{s,m}}\ (\sum_{s,m}|a_{s,m}|^2=1)$, where $E_m$ and $\ket{E_m}$ are the eigenvalues and eigenvectors of $H_0$ respectively, $\ket{s}\in \{\ket{+}, \ket{-}\}$ are eigenvectors of the operator $X$ in $\mathcal{H}_c$, and $\ket{\psi_{s,m}}$ is a unit vector in $\mathcal{H}'$. With this, it follows that $|\bra{\psi}e^{-iX\otimes (f(H_0)-\tilde{f}_K (H_0))t}\otimes I\ket{\psi}|$ is lower bounded by
\begin{align}
    \label{eq::psipsi}
    |\bra{\psi}e^{-iX\otimes (f(H_0)-\tilde{f}_K (H_0))t}\otimes I\ket{\psi}|
    &=
    \left|\sum_{s,m}|a_{s,m}|^2e^{-i\bra{s}X\ket{s}\cdot (f(E_m)-\tilde{f}_K(E_m))t}\right|
    \nonumber\\
    &\geq
    \left|\sum_{s,m}|a_{s,m}|^2{\rm Re}(e^{-i\bra{s}X\ket{s}\cdot (f(E_m)-\tilde{f}_K(E_m))t})\right|
    \nonumber\\
    &\geq
    \left|\sum_{s,m}|a_{s,m}|^2
    \cos(\tfrac{1}{2}R(f-\tilde{f}_K)t)
    \right|
    \nonumber\\
    &=\cos(\tfrac{1}{2}R(f-\tilde{f}_K)t),
\end{align}
where $R(g):=2\underset{x\in [-1,1]}{{\rm max}}|g(x)|$ for a function $g:[-1,1]\to \mathbb{R}$ [also noting that $| t \max(|f-\tilde{f}_K|) | \leq \epsilon/4 \leq \pi$ by Eq.\ \eqref{stop2}]. Substituting Eq.~(\ref{eq::psipsi}) into Eq.~(\ref{eq::b_inter}) and invoking Eq.~(\ref{stop2}), we finally have
\begin{align}
    E(\mathcal{G}_3-\mathcal{G}_1)
    \leq 
    \frac{\epsilon}{2}+
    2\sin [R(f-\tilde{f}_K)t/2]\leq
    \frac{\epsilon}{2}+R(f-\tilde{f}_K)t \leq \frac{\epsilon}{2}+\frac{\epsilon}{2}=\epsilon.
\end{align}
By substituting this back into Eq.~(\ref{our_err3}), we have that
\begin{align}
    \label{siage2}
    \epsilon
    \geq E(\mathcal{G}_3-\mathcal{G}_1)
    &\geq
    \underset{\substack{\ket{\psi}\in\mathcal{H}_c\otimes \mathcal{H}\otimes \mathcal{H'}\\ \|\ket{\psi}\|=1\\ \mathrm{dim}{\mathcal{H}'} }}{{\rm sup}}
    \|
    {\rm tr}_{\mathcal{H}_c}
    [(\mathcal{G}_3-\mathcal{G}_1)\otimes \mathcal{I}_{\mathcal{H'}}]
    (\ket{\psi}\bra{\psi})
    \|_1
    \nonumber\\
    &\geq
    \underset{\substack{\ket{\psi}\in \mathcal{H}\otimes \mathcal{H'}\\ \|\ket{\psi}\|=1\\ \mathrm{dim}{\mathcal{H}'} }}{{\rm sup}}
    \|
    {\rm tr}_{\mathcal{H}_c}
    [(\mathcal{G}_3-\mathcal{G}_1)\otimes \mathcal{I}_{\mathcal{H}'}]
    (\ket{+}\bra{+}\otimes \ket{\psi}\bra{\psi})
    \|_1,
\end{align}
which is equal to the expression in Eq. \eqref{err_def2},
thus asserting our claim. \qed

\FloatBarrier

\subsection{Fourier Series Convergence}\label{app::fourierconvergence}
\begin{lem}
    \label{4_smo}
    For an arbitrary $f:[-1,1]\to \mathbb{R}$ that belongs to class $C^3$ with $f^{(4)}$ being piecewise $C^2$, the function $\tilde{f}$ defined in Eq. (\ref{def_tilde_f}) is periodically 4-smooth.
\end{lem}

\noindent {\bf Proof:} We will demonstrate that (a) $\tilde{f}$ belongs to class $C^3$, (b) the periodicity condition $\tilde{f}^{(j)}(1)=\tilde{f}^{(j)}(-1)$ holds $ \forall \ j\in \{0,1,2,3\}$, and (c)  $\tilde{f}^{(4)}$ is piecewise $C^2$. 

Let $S\subset [-1,1]$ denote the finite set of points outside of which $f^{(4)}$ is defined and is continuously differentiable. It is straightforward to see that $\tilde{f}^{(4)}$ is piecewise $C^2$ with exceptional points $x\in \tilde{S}\cup \{0\}$ where $\tilde{S}:=\{(x+1)/2\mid x\in S\}$. Thus, it suffices to show that $\tilde{f}$ belongs to class $C^3$ and $\tilde{f}^{(j)}(1)=\tilde{f}^{(j)}(-1)\ \forall \ j\in \{0,1,2,3\}$. This follows directly from the following equations, which in turn follow from the definition of $g_f$ [see Eq.~\eqref{eq::def-app-g_f}]
\begin{align}
    &\frac{\mathrm{d}^j}{\mathrm{d}x^j}g_f(x)|_{x=0}=\frac{\mathrm{d}^j}{\mathrm{d}x^j}f(2x-1)|_{x=0}
    \nonumber\\
    &\frac{\mathrm{d}^j}{\mathrm{d}x^j}g_f(x)|_{x=-1}=\frac{\mathrm{d}^j}{\mathrm{d}x^j}f(2x-1)|_{x=1},
\end{align}
where $j\in \{0,1,2,3\}$.\qed \\

We now move on to analyze the convergence rate of the Fourier series being simulated, which leads to a relationship between the allowed error $\epsilon$ and the truncation number $K$. Whenever a function  $g:[-1,1]\to \mathbb{R}$ is piecewise smooth, its Fourier series converges (except for at a finite number of discontinuities; see, e.g., Ref.~\cite{bocher1906introduction}). In general, for a periodically $J$-smooth function, the rate of convergence can be determined via the following Lemma.

\begin{lem}
    \label{Jsmo_fou_conv}
    For any periodically $J$-smooth function $g:[-1,1]\to \mathbb{R}$, its Fourier coefficients $c_k:= (1/2)\int_{-1}^1 \mathrm{d}x \, g(x)e^{-i\pi k x}$ converge at the rate
    \begin{align}\label{eq::fourierconvergenceapp}
    \lim_{|k|\to \infty} |c_k||k|^{J}< \infty \, ,
\end{align}
which further implies that 
    \begin{equation}
        \sum_{k=K+1}^\infty |c_k| = O [K^{-(J-1)}] \, ,
    \end{equation}
    for $J\ge 2$ and any positive integer $K \ge 1$. Moreover, if $g$ is \emph{strictly} periodically $J$-smooth, then
    \begin{align}
    &\sum_{k=K}^{\infty}|c_k|^2=\Omega [K^{-(2J+1)}]
    \quad \text{and} \quad \sum_{k=-\infty}^{-K}|c_k|^2=\Omega [K^{-(2J+1)}]
    \label{convomega}
    \end{align}
    holds for any positive integer $K>0$.
\end{lem}

\noindent {\bf Proof:}
If $g$ is periodically $J$-smooth, then each coefficient $c_k$ can be rewritten using the following Fourier asymptotic coefficient expansion method~\cite{boyd2001chebyshev}:
\begin{align}
\label{FACE}
c_k&=\frac{1}{2}\int_{-1}^1 \mathrm{d}x \, g(x)e^{-i\pi k x}
\nonumber\\
&= \frac{1}{2} \frac{i}{\pi k} \left[ (-1)^k
\left\{g(1) - g(-1) \right\}
-\int_{-1}^1 \mathrm{d}x \, g^{(1)}(x)e^{-i\pi kx}
\right]
\nonumber\\
&=\cdots =\frac{1}{2}\sum_{j=0}^{J-1}
(-1)^{k+j}\left(\frac{i}{\pi k}\right)^{j+1}
\left\{g^{(j)}(1)-g^{(j)}(-1)\right\}
+\frac{(-1)^J}{2}
\left(\frac{i}{\pi k}\right)^{J}
\int_{-1}^1 \mathrm{d}x \, g^{(J)}(x)e^{-i\pi kx}.
\end{align}
The final line follows by successively applying integration by parts on the relevant integral term; this technique is valid since the interval of integration $[-1,1]$ can be decomposed into smaller intervals $[x_0,x_1],\hdots,[x_{n-1},x_n]$ (with $x_0=-1$ and $x_n=1$) between the exceptional points, upon which all derivatives of $g$ (up to the $J$-th derivative) are $C^2$ by assumption. The rightmost part of Eq. \eqref{FACE} is $O(1/k^J)$ because all of the terms in the summation vanish due to periodicity and the remaining integral in the second term is bounded. This proves the first part of our claim, i.e., the rate of convergence according to Eq.~\eqref{eq::fourierconvergenceapp}.

The second part of the claim is proven using the following inequality which is valid for $K \ge 1$:
\begin{equation}
    \sum_{k=K+1}^\infty \frac{1}{\left|k^J \right| } 
    \leq \int_K^\infty \frac{1}{x^J} {\rm d} x 
    = \frac{1}{(J-1) K^{J-1}} \,.
\end{equation}
This inequality, together with Eq.\ \eqref{eq::fourierconvergenceapp}, implies that
\begin{equation}
    \sum_{k=K+1}^\infty |c_k| \leq \sum_{k=K+1}^\infty \frac{R}{|k|^J} =  O [K^{-(J-1)}] \, ,
\end{equation}
where $0 < R < \infty$.

Furthermore, when $g$ is \emph{strictly} periodically $J$-smooth, the integral term on the r.h.s.\ of Eq.~\eqref{FACE} can be written as:  
\begin{align}
    \label{J1deta}
    \int_{-1}^1 \mathrm{d}x \, g^{(J)}(x)e^{-i\pi kx}
    &=
    \int_{-1}^{x_1} \mathrm{d}x \, g^{(J)}(x)e^{-i\pi kx}+\cdots +
    \int_{x_{n-1}}^{1} \mathrm{d}x \,g^{(J)}(x)e^{-i\pi kx}
    \nonumber\\
    &=\frac{i}{\pi k}\left[
    e^{i\pi k}\{ g^{(J)}(1^-)-g^{(J)}(-1^+) \} \right. +
    e^{-i\pi kx_1}\{ g^{(J)}(x_1^-)-g^{(J)}(x_1^+) \} +
    \nonumber\\
    & \left. \cdots +
    e^{-i\pi kx_{n-1}}\{ g^{(J)}(x_{n-1}^-)-g^{(J)}(x_{n-1}^+) \}
    \right] -\frac{i}{\pi k}
    \int_{-1}^{1} \mathrm{d}x g^{(J+1)}(x)e^{-i\pi kx}.
\end{align}
By rewriting the final integral similarly, we can show that this expression is of $O(1/k^2)$. Specifically, let us define $g_m$ for $m\in \{0,\ldots n-1\}$ as $g_0:=g^{(J)}(1^-)-g^{(J)}(-1^+)$ and $g_m:=g^{(J)}(x_m^-)-g^{(J)}(x_m^+)$ for $m>1$; with this, Eq.~\eqref{J1deta} can be expressed succinctly as $\tfrac{i}{\pi k}\sum_{m=0}^{n-1}g_m e^{-i\pi kx_m}+O(1/k^2)$. Substituting Eq.~\eqref{J1deta} into Eq.~\eqref{FACE} then yields
\begin{align}
    \label{ck_simp}
    c_k=
    \frac{(-1)^J}{2}\left(
    \frac{i}{\pi k}
    \right)^{J+1}
    \left(
    \sum_{m=0}^{n-1}g_me^{-i\pi kx_m}\right)
    +O(k^{-(J+2)}).
\end{align}
In order to assert the claim, we now seek to lower bound $\left|c_k\right|$. For any $k_0\in \mathbb{Z}$ and $M>0$, we have that
\begin{align}
    \label{eq::abssq}
    \sum_{k=k_0}^{k_0+M-1}\left|
    \sum_{m=0}^{n-1}g_me^{-i\pi kx_m}
    \right|^2
    &=\sum_{k=k_0}^{k_0+M-1}\left[
    \sum_{m=0}^{n-1}|g_m|^2+
    \sum_{m_1\neq m_2} g_{m_1}^*g_{m_2}e^{-i\pi k(x_{m_2}-x_{m_1})}
    \right] 
    \nonumber\\
    &\geq 
    \sum_{k=k_0}^{k_0+M-1}
    \sum_{m=0}^{n-1}|g_m|^2
    -
    \left|
    \sum_{m_1\neq m_2} g_{m_1}^*g_{m_2}
    \sum_{k=k_0}^{k_0+M-1}
    e^{-i\pi k(x_{m_2}-x_{m_1})}
    \right|
    \nonumber\\
    &\geq 
    \sum_{k=k_0}^{k_0+M-1}
    \sum_{m=0}^{n-1}|g_m|^2
    -\sum_{m_1\neq m_2} |g_{m_1}||g_{m_2}|
    \left|\sum_{k=k_0}^{k_0+M-1}
    e^{-i\pi k(x_{m_2}-x_{m_1})}\right|
    \nonumber\\
    &\geq
    M\sum_{m=0}^{n-1}|g_m|^2-
    \frac{1}{\Delta}
    \left(
    \sum_{m=0}^{n-1}|g_m|
    \right)^2 ,
\end{align}
where $\Delta:=\sin \left[\frac{\pi}{2}\left\{\underset{0\leq m_1< m_2\leq n-1}{\min}\left(x_{m_2}-x_{m_1}, 2-(x_{m_2}-x_{m_1})\right)\right\} \right]>0$. The last inequality above follows by evaluating the geometric series and then bounding it appropriately, i.e.,  $|\sum_{k=k_0}^{k_0+M-1}e^{-i\pi kq}|=|(e^{-i\pi (k_0+M)q}-e^{-i\pi k_0q})/(e^{-i\pi q}-1)|\leq 2/|e^{-i\pi q}-1|= 1/\sin (\pi q/2)$ for $q>0$.
Since $1/\sin (\pi q/2)=1/ \sin (\pi (2-q)/2)$ and $1/\sin (\pi q/2)$ is a decreasing function in $0 < q \leq 1$, it follows that $\left|\sum_{k=k_0}^{k_0+M-1} e^{-i\pi k(x_{m_2}-x_{m_1})}\right|$ can be upper bounded by $1/\Delta$ for $m_1\neq m_2$.

In particular, when $M$ is chosen as $M':=\mathrm{ceil}\{[2(\sum_{m=0}^{n-1}|g_m|)^2]/[ \Delta (\sum_{m=0}^{n-1} |g_m|^2)]\}$, the last line of Eq.~(\ref{eq::abssq}) yields
\begin{align}
    M'\sum_{m=0}^{n-1}|g_m|^2-
    \frac{1}{\Delta}
    \left(
    \sum_{m=0}^{n-1}|g_m|
    \right)^2&= 
    M'\left[
    \sum_{m=0}^{n-1}|g_m|^2
    -
    \frac{1}{M'\Delta} \left(
    \sum_{m=0}^{n-1}|g_m|
    \right)^2
    \right]
    \nonumber\\
    &\geq M'\left[
    \sum_{m=0}^{n-1}|g_m|^2
    -
    \frac{1}{\Delta}
    \left(
    \sum_{m=0}^{n-1}|g_m|
    \right)^2\cdot 
    \frac{\Delta (\sum_{m=0}^{n-1} |g_m|^2)}{2(\sum_{m=0}^{n-1}|g_m|)^2}
    \right]
    \nonumber\\
    &=
    M'\cdot \frac{1}{2}
    \sum_{m=0}^{n-1}|g_m|^2.
\end{align}
Therefore, the average value of $\left|\sum_{m=0}^{n-1}g_me^{-i\pi kx_m}\right|^2$ for $k\in \{k_0,\ldots ,k_0+M'-1\}$ is lower bounded by $(\sum_{m=0}^{n-1}|g_m|^2)/2$. Since there exists at least one $k\in \{k_0,\ldots ,k_0+M'-1\}$ such that $|\sum_{m=0}^{n-1}g_me^{-i\pi kx_m}|^2\geq (\sum_{m=0}^{n-1}|g_m|^2)/2$ for all $k_0>0$ and the $O(k^{-(J+2)})$ term converges to $0$ quicker than $1/k^{J+1}$, there exists constants $C>0$ and $k_0'>0$ such that for all $k_0>k_0'$, there exists at least one $k\in \{k_0,\ldots ,k_0+M'-1\}$ such that $|c_k|>C/k^{J+1}\geq C/(k_0+M'-1)^{J+1}$.
Since $\sum_{k=K}^{\infty}|c_k|^2=\sum_{l=1}^{\infty}\sum_{j=0}^{M'-1}|c_{K+(l-1)M'+j}|^2\geq \sum_{l=1}^{\infty}|C/(K+lM'-1)^{J+1}|^2= \Theta (K^{-(2J+1)})$ for sufficiently large $K$, it follows that $\sum_{k=K}^{\infty}|c_k|^2=\Omega (K^{-(2J+1)})$ and $\sum_{k=-\infty}^{-K}|c_k|^2=\Omega (K^{-(2J+1)})$ for $K>0$, asserting the second part of our claim.
\qed \\

%%%%%%%%%%%%%%%%%%%%%%%%%%%%%%%%%%%%%%%%%%%%%%%

\newpage \section{Compiled UHET Algorithm}\label{app::compiled}

\subsection{Proof of Eq.~(\ref{Atheta}) for the Compiled Algorithm (Algorithm \ref{alg::main})}
\label{app::pratheta}

\noindent Here, we prove the validity of Eq.~(\ref{Atheta}) of the main text. To be precise with notation, we denote the order of matrix multiplication as $\Pi_{j=1}^{n}M_j:= M_1\cdots M_n$. Formally, we show:

\begin{lem}
    \label{zatu_approx} Let $k\in \mathbb{Z}$, $N\in \mathbb{Z}_{\geq 0}$, and $j:=(\vec{v}_1,\ldots ,\vec{v}_N)\in (\{0,1,2,3\}^n)^N$. For any unitary $W^{(N)}_{k,j}$ defined as per Eq.~\eqref{our_two}, namely 
    \begin{align}
        \label{our_two2}
        &W^{(N)}_{k,j}:=\left[
        \prod_{m=1}^{N}{\tt ctrl}(\sigma_{\vec{v}_m})(I\otimes e^{-i \tfrac{k\pi}{2N}H}){\tt ctrl}(\sigma_{\vec{v}_m})
        \right] 
        (e^{-i\tfrac{k\pi}{4}Z}\otimes I),
    \end{align}
    there exist parameters
    $A_{k,N}>0$ and $\theta_{k,N}\in [0,2\pi)$ such that
    \begin{align}
        &\left( \frac{1}{4}
        \right)^{nN}
        \sum_{\vec{v}_1,\ldots ,\vec{v}_{N}}({W}^{(N)}_{k,j})^{\dagger}( [\cos \phi_k X -\sin \phi_k Y]\otimes I) {W}^{(N)}_{k,j}
        \nonumber\\
        =& A_{k,N} (e^{i\tfrac{\theta_{k,N}}{2}Z}\otimes I)
        \left(
        \begin{array}{cc}
            0&e^{i\phi_k}e^{i\tfrac{k\pi}{2} (H_0+I)}\\
            e^{-i\phi_k}e^{-i\tfrac{k\pi}{2} (H_0+I)}&0\\
        \end{array}
        \right)(e^{-i\tfrac{\theta_{k,N}}{2}Z}\otimes I),
        \label{Atheta2}
    \end{align}
    where $H_0$ is the traceless part of $H$, i.e., $H_0:=H-(\mathrm{tr}[H]/2^n)I$. 
    Furthermore, $A_{k,N}$ and $\theta_{k,N}$ satisfy
    \begin{align}
        1-\frac{\pi^2k^2}{8N}\leq A_{k,N} \leq 1
        \quad \textup{and} \quad |\theta_{k,N}|\leq \frac{\pi^3 |k|^3}{32N^2}\quad (N\geq 0.625\pi |k|) \, .
        \label{eq::Atheta_bound}
    \end{align}
\end{lem}

\noindent {\bf Proof:} For the special case $N=0$ (in which case we employ the convention that for any set of matrices $\{ X_m\}_m$, we have $\Pi_{m=1}^{N=0} X_m = I$), it is straightforward to verify that $(A_{k, N}, \theta_{k, N})=(1,0)$ satisfies Eq.~(\ref{Atheta2}); thus, the remainder of the proof concerns $N>0$.

For convenience, we define
    \begin{align}\label{eq::upsilon}
        \Upsilon_{k, N, j}&:=\prod_{m=1}^{N}{\tt ctrl}(\sigma_{\vec{v}_m})(I\otimes e^{-i\tfrac{k\pi}{2N}H}){\tt ctrl}(\sigma_{\vec{v}_m})= 
        \left(
        \begin{array}{cc}
            e^{-i\tfrac{k\pi}{2} H}&0\\
            0&\prod_{m=1}^N\sigma_{\vec{v}_m}e^{-i\tfrac{k\pi}{2N}H}\sigma_{\vec{v}_m}\\
        \end{array}
        \right).
    \end{align}
With this, Eq.~\eqref{Atheta2} can be rewritten as
\begin{align}
    &\left( \frac{1}{4}
    \right)^{nN}
    \sum_{\vec{v}_1,\ldots ,\vec{v}_{N}}
    \Upsilon_{k,N,j}^{\dagger}([\cos \phi_k X -\sin \phi_k Y]\otimes I) \Upsilon_{k,N,j}
    \nonumber\\
    &=\left( \frac{1}{4}
    \right)^{nN}\sum_{\vec{v}_1,\ldots ,\vec{v}_{N}}
    \left(
    \begin{array}{cc}
        0&e^{i\phi_k}e^{i\tfrac{k\pi}{2}H}(\prod_{l=1}^N\sigma_{\vec{v}_l}e^{-i\tfrac{k\pi}{2N} H}\sigma_{\vec{v}_l})\\
        e^{-i\phi_k}(\prod_{l=1}^N\sigma_{\vec{v}_l}e^{-i\tfrac{k\pi}{2N}H}\sigma_{\vec{v}_l})^{\dagger}e^{-i\tfrac{k\pi}{2} H}&0\\
    \end{array}
    \right)
    \nonumber\\
    &=
    \left(
    \begin{array}{cc}
        0&e^{i\phi_k}e^{i\tfrac{k\pi}{2}H}[\prod_{l=1}^N\{(\frac{1}{4})^n \sum_{\vec{v}_l} \sigma_{\vec{v}_l}e^{-i\tfrac{k\pi}{2N}H }\sigma_{\vec{v}_l}\}]\\
        e^{-i\phi_k}[\prod_{l=1}^N\{(\frac{1}{4})^n \sum_{\vec{v}_l} \sigma_{\vec{v}_l}e^{-i\tfrac{k\pi}{2N} H}\sigma_{\vec{v}_l}\}]^{\dagger}e^{-i\tfrac{k\pi}{2}H}&0\\
    \end{array}
    \right)
    \nonumber\\
    &=
    \left( \frac{1}{2}
    \right)^{nN}\left(
    \begin{array}{cc}
        0&e^{i\phi_k}\{\mathrm{tr}[e^{-i\tfrac{k\pi}{2N}H}]\}^Ne^{i\tfrac{k\pi}{2} H}\\
        e^{-i\phi_k}\{\mathrm{tr}[e^{-i\tfrac{k\pi}{2N} H}]^*\}^Ne^{-i\tfrac{k\pi}{2}H)}&0\\
    \end{array}
    \right)
    ,
    \label{sikihenkei}
\end{align}
where we have made use of the matrix identity $\sum_{\vec{v}_1,\ldots ,\vec{v}_N}(\prod_{l=1}^NM_{\vec{v}_l})=\prod_{l=1}^N(\sum_{\vec{v}_l}M_{\vec{v}_l})$. By now invoking $\{\mathrm{tr}(e^{-i\frac{\pi k}{2N}H})\}^Ne^{i\frac{\pi k}{2} H}=\{\mathrm{tr}(e^{-i\frac{\pi k\alpha}{2N}}e^{-i\frac{\pi k}{2N}H_0})\}^Ne^{i\frac{\pi k\alpha}{2}}e^{i\frac{\pi k}{2} H_0}=\{\mathrm{tr}(e^{-i\frac{\pi k}{2N}H_0})\}^Ne^{i\frac{\pi k}{2} H_0}$, where $\alpha:={\rm tr}(H)/2^n$, the final line of Eq.~(\ref{sikihenkei}) can be rewritten as
\begin{align}
    &\left( \frac{1}{2}
    \right)^{nN}\left(
    \begin{array}{cc}
        0&e^{i\phi_k}\{\mathrm{tr}(e^{-i\frac{\pi k}{2N}H})\}^Ne^{i\frac{\pi k}{2} H}\\
        e^{-i\phi_k}\{\mathrm{tr}(e^{-i\frac{\pi k}{2N}H})^*\}^Ne^{-i\frac{\pi k}{2} H}&0\\
    \end{array}
    \right)
    \nonumber\\
    =&
    \left( \frac{1}{2}
    \right)^{nN}\left(
    \begin{array}{cc}
        0&e^{i\phi_k}\{\mathrm{tr}(e^{-i\frac{\pi k}{2N}H_0})\}^Ne^{i\frac{\pi k}{2} H_0}\\
         e^{-i\phi_k}\{\mathrm{tr}(e^{-i\frac{\pi k}{2N}H_0})^*\}^Ne^{-i\frac{\pi k}{2} H_0}&0\\
    \end{array}
    \right)
    .
\end{align}
Thus, by setting 
\begin{align}
A_{k,N}e^{i\theta_{k,N}}:= \left[
\frac{\mathrm{tr}(e^{-i\frac{\pi k}{2N}H_0})}{2^n}
\right]^N 
\label{eq::def_Atheta},
\end{align}
where $A_{k,N}\geq 0$ and $\theta_{k,N}\in \mathbb{R}$, we obtain

\begin{align}
     &\left( \frac{1}{4}
        \right)^{nN}
        \sum_{\vec{v}_1,\ldots ,\vec{v}_{N}}\Upsilon_{k,N,j}^{\dagger}([\cos \phi_k X -\sin \phi_k Y]\otimes I) \Upsilon_{k,N,j} =
        A_{k,N}
        \left(
        \begin{array}{cc}
            0&e^{i(\phi_k+\theta_{k,N})}e^{i\tfrac{k\pi}{2} H_0}\\
            e^{-i(\phi_k+\theta_{k,N})}e^{-i\tfrac{k\pi}{2} H_0}&0\\
        \end{array}
        \right)
        \label{Atheta3},
\end{align}
which is equivalent to Eq.~(\ref{Atheta2}).

For the second part of the claim, note that when $H_0$ is diagonalized as $H_0=\sum_{j=0}^{2^n-1}E_j\ket{j}\!\bra{j}$, then the expression $\{\mathrm{tr}(e^{-i\frac{\pi k}{2N}H_0})/2^n\}^N$ reduces to
\begin{align}
\{\mathrm{tr}(e^{-i\frac{\pi k}{2N}H_0})/2^n\}^N
=\left(\frac{\sum_{j=0}^{2^n-1} e^{-i\frac{\pi k}{2N}E_j}}{2^n}\right)^N 
=\left(\frac{\sum_{j=0}^{2^n-1} \cos(\frac{\pi k}{2N}E_j)-i\sum_{j=0}^{2^n-1}\sin(\frac{\pi k}{2N}E_j)}{2^n}\right)^N ,
\label{alth}
\end{align}
from which $A_{k,N}\leq 1$ follows. Further invoking $|E_j|\leq 1$ (which follows from our assumption that $\|H_0\|_{\rm op} \leq 1$), the assumption $N\geq 0.625\pi |k|$, and the inequality $\cos (x)\geq 1-x^2/2\ (x\in \mathbb{R})$, it follows that $\cos(\tfrac{\pi k}{2N} E_j)\geq \cos{(\tfrac{\pi k}{2N})} \geq 1-\tfrac{1}{2}\left(\tfrac{\pi k}{2N}\right)^2$ and subsequently
\begin{align}
    \label{A_k_low}
    A_{k,N}
    &\geq \left|\frac{\sum_{j=0}^{2^n-1} \cos(\frac{\pi k}{2N}E_j)}{2^n} \right|^N
    \geq \left[1-\frac{1}{2}\left(\frac{\pi k}{2N}\right)^2\right]^N \geq 1-\frac{\pi ^2 k^2}{8N} .
\end{align}
Finally, due to the fact that $\sum_{j=0}^{2^n-1}E_j=0$ (which follows from $\mathrm{tr}(H_0)=0$) as well as the inequality $|\sin x- x|\leq (|x|^3/6)\ (x\in \mathbb{R})$, we have that
\begin{align}
    \left|\sum_{j=0}^{2^n-1}\sin\left(\frac{\pi k}{2N}E_j\right)\right|
    =
    \left|
    \sum_{j=0}^{2^n-1}\left[ \sin\left(\frac{\pi k}{2N}E_j\right) - \frac{\pi k}{2N}E_j\right]
    \right|\leq
    \frac{1}{6}
    \sum_{j=0}^{2^n-1}\left| \frac{\pi k}{2N}E_j\right|^3
    \leq \frac{2^n}{6}\left(\frac{\pi |k|}{2N}\right)^3 \,.
\end{align}\\
Then, since $\tfrac{1}{2^n}\sum_{j=0}^{2^n-1} \cos(\tfrac{\pi k}{2N}E_j)\geq 
\tfrac{1}{2^n}\sum_{j=0}^{2^n-1} \cos(\tfrac{\pi k}{2N})=\cos(\tfrac{\pi k}{2N})\geq 1-\tfrac{1}{2}\left(\tfrac{\pi k}{2N}\right)^2$, it follows that $\theta_{k,N}$ is upper bounded by
 \begin{align}
    |\theta_{k,N}|\leq
    N\tan^{-1} \left[ \left\{
    \frac{1}{6}\left(
    \frac{\pi |k|}{2N}
    \right)^3\right\}
    \left\{
    1-\frac{1}{2}
    \left(
    \frac{\pi k}{2N}
    \right)^2
    \right\}^{-1}
    \right] \leq \frac{\pi^3 |k|^3}{32N^2} \quad 
    (N\geq 0.625\pi |k|).
 \end{align}
The final inequality follows from $\tan^{-1}[\tfrac{1}{6}x^3/(1-\tfrac{1}{2}x^2)]\leq \tfrac{1}{4}x^3\ (0\leq x\leq 0.8)$. 
\qed

\FloatBarrier

\subsection{Parameter Estimation for Compiled Algorithm (Algorithm \ref{alg::main})}
\label{app:subrs}

\noindent The compiled UHET Algorithm \ref{alg::main} makes use of ``correction'' parameters $(\hat{A}_k, \hat{\theta}_k)$ to compensate the error of the main process and reduce the time complexity from that of Algorithm \ref{alg::umcompi}. Here, we will develop two subroutines, namely Subroutines \ref{subr::gen} and \ref{subr::forgen}, that allow one to estimate such parameters without any knowledge of the seed Hamiltonian. Subroutine \ref{subr::forgen} is used in step 2 of Subroutine \ref{subr::gen} to generate a state $\ket{\phi (N', l, k\pi/N', 25k, \Phi)}$ that is used for robust phase estimation, from which the parameters of interest can be estimated. 

\begin{algorithm}[H]
    \renewcommand{\thealgorithm}{\arabic{algorithm}}
    \floatname{algorithm}{Subroutine}
    \caption{Generating parameters $(\hat{A}_l, \hat{\theta}_l)$}
    \label{subr::gen}
    \begin{algorithmic}[1]
        \Statex{\textbf{Input:}}
        \begin{itemize}
            \item A finite number of queries to a black-box Hamiltonian dynamics $e^{\pm iH\tau}$ of a seed Hamiltonian $H$ with $\tau>0$ on an $n$-qubit system $\mathcal{H}$
            \item Parameter $l\in \mathbb{Z}_{>0}$
            \item Allowed error $\epsilon >0$
            \item Time $t>0$
        \end{itemize}
        \Statex{\textbf{Output:}}
        Estimates 
        $\hat{A}_l>\tfrac{1}{2}$ and $\hat{\theta}_l\in [0, 2\pi)$ of $A_{l,10l^2}$ and ${\theta}_{l,10l^2}$, respectively, with root mean square error of $|1-(A_{l,10l^2}e^{i\theta_{l,10l^2}})/(\hat{A}_l e^{i\hat{\theta}_l})|$ that is upper bounded by $\epsilon/t$.
        \Statex{\textbf{Time complexity:}} $O(l^2t^3n/\epsilon^3)$
    \Statex \hrulefill
        \Statex{\textbf{Procedure:}}
        \For{$\Phi\in \{0, \tfrac{\pi}{2}\}$}
        \State Perform robust phase estimation \cite{kimmel2015robust} with allowed root mean square of error set as $\epsilon/(2\sqrt{2}t)$. Here, success of $\ket{0}-$measurements and $\ket{+}-$measurements are defined as obtaining outcomes for $\ket{0}\otimes I$ and $\ket{+}\otimes I$ when performing $Z-$ and $X-$basis measurements respectively on the first qubit of the state $\ket{\phi (N', l, k\pi/N', 25k, \Phi)}$ (for the required various values of $k$), where $N':=N(1, k\pi, 1/ (4\sqrt{2}))$, generated by Subroutine \ref{subr::forgen}. This provides an estimate $\hat{v}_{\Phi}$ of the quantity $2\pi A_{l,10l^2}\cos(\theta_{l,10l^2}+\Phi)$, which depends on $H$ and $l$.
        \EndFor
        \If{$(1/2\pi)\sqrt{\hat{v}_0^2+\hat{v}_{\pi/2}^2}\leq \tfrac{1}{2}$}
        \State Return to step 1
        \EndIf
        \State Compute $\hat{A}_l,\ \hat{\theta}_l$ by
        \begin{align}
            \hat{A}_l \cos(\hat{\theta}_l) &=\frac{1}{2\pi}\hat{v}_{0}
            \nonumber\\
            \hat{A}_l \sin(\hat{\theta}_l) &=\frac{1}{2\pi}\hat{v}_{\pi/2}
            \nonumber
        \end{align}
        \State {\textbf{Return}} $(\hat{A}_l, \hat{\theta}_l)$
    \end{algorithmic}
\end{algorithm}

\noindent Step 2 of this routine makes use of the following subroutine to generate an appropriate state.

\begin{algorithm}[H]
    \renewcommand{\thealgorithm}{\arabic{algorithm}}
    \floatname{algorithm}{Subroutine}
    \caption{Generating a state used in Subroutine \ref{subr::gen}}
    \label{subr::forgen}
    \begin{algorithmic}[1]
        \Statex{\textbf{Input:}}
        \begin{itemize}
            \item A finite number of queries to a black-box Hamiltonian dynamics $e^{\pm iH\tau}$ of a seed Hamiltonian $H$ with $\tau>0$  on an $n$-qubit system $\mathcal{H}$
            \item Parameters $N, l, M\in \mathbb{Z}_{>0}$
            \item $\gamma \in [0,1]$
            \item $\Phi\in [0,2\pi)$
        \end{itemize}
        \Statex{\textbf{Output:}}
        A random state $\ket{\phi(N,l,\gamma, M, \Phi)}$
        \Statex{\textbf{Time complexity:}} $\Theta(Nl^2nM)$
    \Statex \hrulefill
        \Statex{\textbf{Procedure:}}
        \State{Initialize:}
        \Statex \hskip1.0em$\ket{\text{current}}\gets\ket{0}\otimes \ket{0}\in \mathcal{H}_c\otimes \mathcal{H}$
        \For{$m\in \{1,\ldots ,N\}$}
        \State From $j':=\{\vec{u}_1,\ldots ,\vec{u}_{10l^2}\}$, $j'':=\{\vec{w}_1,\ldots ,\vec{w}_{l^2M}\}$  where $\vec{u}_1,\ldots ,\vec{u}_{10l^2}, \vec{w}_1,\ldots ,\vec{w}_{l^2M}\in \{0,1,2,3\}^n$, and $s\in \{+1, -1\}$, uniformly randomly choose $j=(j', j'', s)$
        \State $\ket{\text{current}}\gets \widetilde{W}_{l, j, \Phi}(e^{-i\gamma Y}\otimes I)\widetilde{W}_{l, j, \Phi}^{\dagger}\ket{\text{current}}$ for 
        \begin{align}
            \widetilde{W}_{l, j,\Phi}&:=
            (e^{-i\tfrac{s \Phi}{2}Z}\otimes I)\left[\prod_{m''=1}^{l^2M}{\tt ctrl}(\sigma_{\vec{w}_{m''}})(I\otimes e^{i \tfrac{s \pi}{2lM}H}){\tt ctrl}(\sigma_{\vec{w}_{m''}})
            \right]\left[\prod_{m'=1}^{10l^2}{\tt ctrl}(\sigma_{\vec{u}_{m'}})(I\otimes e^{-i \tfrac{s \pi}{20l}H}){\tt ctrl}(\sigma_{\vec{u}_{m'}})
            \right]
            \nonumber
        \end{align}
        \EndFor
        \State {\textbf{Return}} $\ket{\phi(N,l,\gamma, M, \Phi)}:=\ket{\text{current}}$
    \end{algorithmic}
\end{algorithm}
\noindent The time complexity of Subroutine \ref{subr::forgen} scales as (number of iterations $N$)$\times$(time complexity of $\widetilde{W}_{l,j,\Phi}$)=$N\times \Theta (l^2nM)=\Theta(Nl^2nM)$.

\begin{figure*}[t]
        \includegraphics[width=\linewidth]{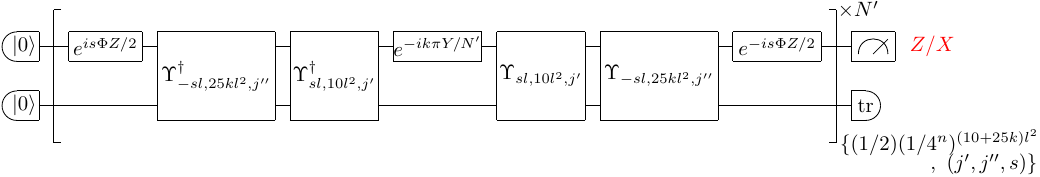}
        \caption{\emph{Compensation parameter estimation.---}The circuit used to obtain estimates $(\hat{A}_l, \hat{\theta}_l)$ that are subsequently used in Algorithm~\ref{alg::main}. $\Upsilon_{l,N,j}$ is defined in Eq.\ \eqref{eq::upsilon}\label{fig::estimation} and its relation to $\widetilde{W}_{l,j,\Phi}$ is provided in Eq.\ \eqref{eq:Wtilde-upsilon}.}
\end{figure*}

A circuit depiction of the combination of Subroutines \ref{subr::gen} and \ref{subr::forgen} is provided in Fig. \ref{fig::estimation}. In particular, the preparation of the state $\ket{\phi (N', l, k\pi/N', 25k, \Phi)}$ in Subroutine \ref{subr::forgen} corresponds to the part of Fig. \ref{fig::estimation} before the measurement, with $M, \gamma$ set to $25k, k\pi/N'$ respectively. For many values of $k$, Subroutine \ref{subr::gen} implements this circuit and performs appropriate measurements on the state in order to estimate $(\hat{A}_l, \hat{\theta}_l)$ via robust phase estimation~\cite{kimmel2015robust}. We now formalize the validity of this algorithm and demonstrate its time complexity.

\begin{Theorem}
    \label{alg2_vari}
    Subroutine \ref{subr::gen} outputs estimates $(\hat{A}_l, \hat{\theta}_l)$ of the parameters $(A_{l,10l^2}, \theta_{l,10l^2})$ defined in Lemma \ref{zatu_approx} for $l>0$ with a root mean squared error of $|1-(A_{l,10l^2}e^{i\theta_{l,10l^2}})/(\hat{A}_le^{i\hat{\theta}_l})|$ upper bounded by $\epsilon /t$ with a time complexity $O(l^2t^3n/\epsilon^3)$. 
\end{Theorem}

\noindent In order to prove Theorem~\ref{alg2_vari}, we combine the results of Lemma \ref{zatu_approx} with Lemma \ref{rob_ph} below, which concerns the robust phase estimation procedure and is proven in Ref.~\cite{kimmel2015robust}. We first show that steps 2 to 5 of Subroutine \ref{subr::forgen} generates the state $\ket{\phi(N,l,\gamma, M, \Phi)}$ by simulating a particular Hamiltonian that is proportional to $\hat{v}_{\Phi}$  [which is, in turn, a function of $(A_{l,10^2}, \theta_{l,10l^2})$ and is defined below] via qDRIFT. We then show that the success probabilities of $\ket{0}$-measurements and $\ket{+}$-measurements for each $k$ (respectively, denoted $(0,k)-$measurements and $(+,k)-$measurements in Lemma \ref{rob_ph}) differ from $\tfrac{1}{2}(1+\cos(k\hat{v}_{\Phi}))$ and $\tfrac{1}{2}(1+\sin(k\hat{v}_{\Phi}))$, respectively, at most by $\tfrac{1}{\sqrt{8}}$, demonstrating that $\hat{v}_{\Phi}$ can be well-estimated by robust phase estimation. Finally, we prove that $A_{l,10l^2}$ and $\theta_{l,10l^2}$ can be obtained with a root mean square of error smaller than or equal to $\epsilon /t$.\\

\begin{lem}[Robust Phase Estimation~\cite{kimmel2015robust}]
    \label{rob_ph}
    Let $k\in \mathbb{Z}_{>0}$. Suppose that one can perform two families of measurements, $(0,k)$-measurements and $(+,k)$-measurements, whose success probabilities for obtaining outcomes $0$ and $+$ respectively are given in terms of $\theta \in (-\pi, \pi]$ as
    \begin{alignat*}{2}
    \textup{$(0,k)$-measurement:}& \quad
    p_{0,k}(\theta)&&:=\frac{1+\cos{(k\theta})}{2}+\delta_0(k)\nonumber\\
    \textup{$(+,k)$-measurement:}& \quad
    p_{+,k}(\theta)&&:=\frac{1+\sin{(k\theta})}{2}+\delta_+(k),
    \nonumber
    \end{alignat*}
    where $\delta_0(k)$ and $\delta_+(k)$ satisfy
    \begin{align}
    \underset{k}{\mathrm{sup}}\{
    |\delta_0(k)|, |\delta_+(k)|
    \} =:\delta_{\mathrm{sup}}<\frac{1}{\sqrt{8}} .
    \nonumber
    \end{align}
    
\noindent Then for any allowed standard deviation $s>0$, an estimate $\hat{ \theta}$ of $\theta$ can be obtained with a root mean squared error smaller than or equal to $s$ with a time complexity of $O(1/s)$.
This computation is a function of the numbers of successful $(0,2^{j-1})$-measurements and $(+,2^{j-1})$-measurements $(j\in \{1,\ldots ,K\})$, with both measurement choices being implemented $M_j$ times. Here, $M_j$ and $K$ are defined as
    \begin{align}
    K&:=\mathrm{ceil}\left[\log_2{\left(\frac{3\pi}{s}\right)}\right]\nonumber\\
    M_j&:=F(\delta_{\mathrm{sup}})(3(K-j)+1)\nonumber\\
F(\delta_{\mathrm{sup}})&:=\mathrm{ceil}\left[
    \frac{\log \left( \frac{1}{2}(1-\sqrt{8}\delta_{\mathrm{sup}})^\frac{1}{3(K-j)+1}\right)}{\log \left( 1-\frac{1}{2}(1-\sqrt{8}\delta_{\mathrm{sup}})^2\right)}
    \right] .\label{eq::KMF}
    \end{align}
\end{lem}

\noindent This lemma is proven in Ref.~\cite{kimmel2015robust} (see Theorem 1 therein). We are now in a position to prove Theorem~\ref{alg2_vari}. \\

\noindent{\bf Proof of Theorem \ref{alg2_vari}:} To begin, note that steps 2 to 5 of Subroutine \ref{subr::forgen} simulates the dynamics $e^{-iH'}$ using qDRIFT for the Hamiltonian $H'$ defined as
\begin{align}   
    \label{sikihenkeimae}
    H'&=\frac{N\gamma}{2} \left(
    \frac{1}{4}
    \right) ^{n(10+M)l^2}
    \sum_{\substack{\vec{u}_{1},\ldots ,\vec{u}_{10l^2} \\ \vec{w}_{1},\ldots ,\vec{w}_{l^2M} \\ s\in \{1,-1\}}}
    \widetilde{W}_{l, j, \Phi}(Y\otimes I)\widetilde{W}_{l, j, \Phi}^{\dagger}=:
    t\sum_j h_j H_j
    ,
\end{align}
where $j:=\{(\vec{u}_{1},\ldots ,\vec{u}_{10l^2}),(\vec{w}_{1},\ldots ,\vec{w}_{l^2M}), s\}$, $t:=N\gamma$, $h_j:=\tfrac{1}{2}(\tfrac{1}{4})^{n(10+M)l^2}$, and 
$H_j:=\widetilde{W}_{l, j, \Phi}(Y\otimes I)\widetilde{W}_{l, j, \Phi}^{\dagger}$. With respect to Lemma \ref{qdri:err}, we can evaluate $\lambda=\sum_j h_j=1$ and thus the error of simulating this dynamics via qDRIFT in terms of Eq. (\ref{qdri_error}) is upper bounded by $(2t^2/N)e^{2t/N}$.

We can further simplify Eq.~(\ref{sikihenkeimae}) as follows. First, note that $\widetilde{W}_{l,j,\Phi}$ can be rewritten as
\begin{align}\label{eq:Wtilde-upsilon}
    \widetilde{W}_{l, j,\Phi}=(e^{-i\tfrac{s \Phi}{2}Z}\otimes I)
    \Upsilon_{-sl, l^2M, j''}
    \Upsilon_{sl, 10l^2, j'}
\end{align}
by using $\Upsilon_{k,N,j}$ defined in Eq.~\eqref{eq::upsilon}. Moreover, according to Eq.~\eqref{Atheta3}, the equality
\begin{align}
     \left( \frac{1}{4}
        \right)^{nN}
        \sum_{\vec{v}_1,\ldots ,\vec{v}_{N}}\Upsilon_{k,N,j}^{\dagger}
        \left(
        \begin{array}{cc}
            0&e^{-i\phi}I\\
            e^{i\phi}I&0\\
        \end{array}
        \right)
        \Upsilon_{k,N,j}=A_{k,N}
        \left(
        \begin{array}{cc}
            0&e^{-i(\phi-\theta_{k,N})}e^{i\tfrac{k\pi}{2} H_0}\\
            e^{i(\phi-\theta_{k,N})}e^{-i\tfrac{k\pi}{2}H_0}&0\\
        \end{array}
        \right)
\end{align}
holds for $\phi\in [0,2\pi)$. Thus, Eq.~\eqref{sikihenkeimae} can be rewritten as 
\begin{align}
    &\frac{N\gamma}{2} \left(
    \frac{1}{4}
    \right) ^{n(10+M)l^2}
    \sum_{\substack{\vec{u}_{1},\ldots ,\vec{u}_{10l^2} \\ \vec{w}_{1},\ldots ,\vec{w}_{l^2M} \\ s\in \{1,-1\}}}
    \widetilde{W}_{l, j, \Phi}(Y\otimes I)\widetilde{W}_{l, j, \Phi}^{\dagger} \notag \\
    &=
     \frac{t}{2} \left(
    \frac{1}{4}
    \right) ^{n(10+M)l^2}
    \sum_{\substack{\vec{u}_{1},\ldots ,\vec{u}_{10l^2} \\ \vec{w}_{1},\ldots ,\vec{w}_{l^2M} \\ s\in \{1,-1\}}} (e^{-i \tfrac{s \Phi}{2}Z}\otimes I)
    \Upsilon_{-sl, l^2M, j''}
    \Upsilon_{sl, 10l^2, j'}
    \left(
    \begin{array}{cc}
        0&-iI\\
        iI&0\\
    \end{array}
    \right)
    \Upsilon_{sl, 10l^2, j'}^{\dagger}
    \Upsilon_{-sl, l^2M, j''}^{\dagger}
    (e^{i\tfrac{s \Phi}{2}Z}\otimes I)
    \nonumber\\
    &=
    \frac{t}{2} \left(
    \frac{1}{4}
    \right) ^{n(10+M)l^2}
    \sum_{\substack{\vec{u}_{1},\ldots ,\vec{u}_{10l^2} \\ \vec{w}_{1},\ldots ,\vec{w}_{l^2M} \\ s\in \{1,-1\}}}
     (e^{-i\tfrac{s \Phi}{2}Z}\otimes I)
    \Upsilon_{sl, l^2M, j''}^{\dagger}
    \Upsilon_{-sl, 10l^2, j'}^{\dagger}
    \left(
    \begin{array}{cc}
        0&-iI\\
        iI&0\\
    \end{array}
    \right)
    \Upsilon_{-sl, 10l^2, j'}
    \Upsilon_{sl, l^2M, j''}
    (e^{i\tfrac{s \Phi}{2}Z}\otimes I)
    \nonumber\\
    &=
    \frac{t}{2} A_{-sl, 10l^2}
    \left(
    \frac{1}{4}
    \right) ^{nMl^2}
    \sum_{\substack{\vec{w}_{1},\ldots ,\vec{w}_{l^2M} \\ s\in \{1,-1\}}} (e^{-i\tfrac{s \Phi}{2}Z}\otimes I)
    \Upsilon_{sl, l^2M, j''}^{\dagger}
    \left(
    \begin{array}{cc}
        0&-ie^{i\theta_{-sl, 10l^2}}e^{-i\tfrac{sl\pi}{2}H_0}\\
        ie^{-i\theta_{-sl, 10l^2}}e^{i\tfrac{sl\pi}{2}H_0}&0\\
    \end{array}
    \right)
    \Upsilon_{sl, l^2M, j''}
    (e^{i\tfrac{s \Phi}{2}Z}\otimes I)
    \nonumber\\
    &=
    \frac{t}{2} A_{-sl, 10l^2}A_{sl, l^2M}
    \sum_{s\in \{1,-1\}}(e^{-i\tfrac{s \Phi}{2}Z}\otimes I)
    \left(
    \begin{array}{cc}
        0&-ie^{i(\theta_{-sl, 10l^2}+\theta_{sl,l^2M})}I\\
        ie^{-i(\theta_{-sl, 10l^2}+\theta_{sl,l^2M})}I&0\\
    \end{array}
    \right) 
    (e^{i\tfrac{s \Phi}{2}Z}\otimes I)
    \label{eq::upsupssand},
\end{align}
where in the third line we made use of the fact that $\Upsilon_{-k,N,j} = (\Upsilon_{k,N,j'})^\dagger$, where $j':= (\vec{v}_N,\hdots, \vec{v}_1)$. By substituting Eq.~\eqref{eq::def_Atheta} into Eq.~\eqref{eq::upsupssand}, we have
\begin{align}
    t\sum_j h_jH_j=
    \frac{t}{2}\sum_{s\in \{1,-1\}}
    (e^{-i\tfrac{s \Phi}{2}Z}\otimes I)
    \left(
    \begin{array}{cc}
        0&-ia'_{l,M,s}a'_{l,10,-s} I\\
        ia'_{l,M,-s}a'_{l,10,s}I&0\\
    \end{array}
    \right)(e^{i\tfrac{s \Phi}{2}Z}\otimes I)=
    t a_{l,M,\Phi}
    Y\otimes I ,
\end{align}
where $a'_{l, m, s}:= [\tfrac{1}{2^n}\mathrm{tr}\{e^{-is(\pi /(2ml))H_0}\}]^{ml^2}$ and $a_{l,M,\Phi}:= \tfrac{1}{2}(e^{i\Phi}a'_{l,M,-1}a'_{l,10,1}+e^{-i\Phi}a'_{l,M,1}a'_{l,10,-1})$ (note that $a_{l,M,\Phi}\in \mathbb{R}$).

In terms of the parameters $A_{k,N}$ and $\theta_{k,N}$, it is straightforward to show that $a_{l,M,\Phi}$ can be expressed as 
\begin{align}
    \label{motome_err}
    a_{l,M,\Phi}=A_{l,Ml^2}A_{l,10l^2}\cos(\theta_{l,10l^2}-\theta_{l,Ml^2}+\Phi) .
\end{align}
In Subroutine \ref{subr::gen}, the input parameters of the state are specifically chosen as $(N,l,\gamma,M,\Phi)=(N', l, k\pi/N', 25k,\Phi)$ where $N'=N(1,k\pi,1/(4\sqrt{2}))$, with respect to which, $ta_{l,M,\Phi}$ is expressed as
\begin{align}
    ta_{l,25k,\Phi}=k\pi A_{l,25kl^2}A_{l,10l^2}\cos(\theta_{l,10l^2}-\theta_{l,25kl^2}+\Phi) .
\end{align}
Using the fact that $A_{l,10l^2}\leq 1$, we can upper bound the difference between the above equation and $k\pi A_{l,10l^2}\cos(\theta_{l,10l^2}+\Phi)$ by
\begin{align}
\label{eq::acosphi}
    |ta_{l,25k,\Phi}-&k\pi A_{l,10l^2}\cos(\theta_{l,10l^2}+\Phi)|
    =
    k\pi
    |
    A_{l,25kl^2}A_{l,10l^2}\cos(\theta_{l,10l^2}-\theta_{l,25kl^2}+\Phi)
    -A_{l,10l^2}\cos(\theta_{l,10l^2}+\Phi)
    |
    \nonumber\\
    &\leq
    k\pi |
    A_{l,25kl^2}\cos (\theta_{l,10l^2}-\theta_{l,25kl^2}+\Phi)
    -\cos (\theta_{l,10l^2}+\Phi)
    |
    \nonumber\\
    &\leq
    k\pi\{|
    A_{l,25kl^2}[\cos (\theta_{l,10l^2}-\theta_{l,25kl^2}+\Phi)
    -
    \cos (\theta_{l,10l^2}+\Phi)
    ]
    | +
    |
    (1-A_{l,25kl^2})
    \cos (\theta_{l,10l^2}+\Phi)
    |\}.
\end{align}
Invoking the inequality $|\cos(\theta +\theta')-\cos(\theta)|=|-\int_{\theta}^{\theta+\theta'}{\rm d}x\sin (x)|\leq \theta'$ and Eq.~(\ref{eq::Atheta_bound}), we then have that
\begin{align}\label{eq::athetafinalbound}
    k\pi\{|
    A_{l,25kl^2}&[\cos (\theta_{l,10l^2}-\theta_{l,25kl^2}+\Phi)
    -
    \cos (\theta_{l,10l^2}+\Phi)
    ]
    |+
    |
    (1-A_{l,25kl^2})
    \cos (\theta_{l,10l^2}+\Phi)
    |\}\nonumber\\
    &\leq
    k\pi(
    |\theta_{l,25kl^2}|+
    |1-A_{l,25kl^2}|
    )\leq
    k\pi \left(
    \left|
    \frac{\pi^3}{20000 k^2 l}
    \right|+
    \left|
    \frac{\pi^2}{200k}
    \right| \right)< 0.16< \frac{1}{4\sqrt{2}} .
\end{align}

Now, let us define $\mathcal{F}_{1},\ \mathcal{F}_{2},\ \mathcal{F}_{3}:\mathcal{L}(\mathcal{H}_c\otimes \mathcal{H})\to \mathcal{L}(\mathcal{H}_c\otimes \mathcal{H})$ as 
\begin{align}
    \mathcal{F}_1(\rho)&:=
    e^{-ik\pi A_{l,10l^2}\cos(\theta_{l,10l^2}+\Phi)Y\otimes I}\rho e^{ik\pi A_{l,10l^2}\cos(\theta_{l,10l^2}+\Phi)Y\otimes I}
    \nonumber\\
    \mathcal{F}_{2}(\rho)&:=e^{-ik\pi a_{l,25k,\Phi}Y\otimes I}\rho e^{ik\pi a_{l,25k,\Phi}Y\otimes I}
    \nonumber\\
    \mathcal{F}_3&:=\text{quantum operation simulated by qDRIFT in steps 2 to 5 of Subroutine \ref{subr::forgen}  } \\
    &\quad ~~ \text{(with parameters as defined in Subroutine \ref{subr::gen})}
\end{align}
With this, the success probability of $\ket{0}$-measurements and $\ket{+}$-measurements, which is given by $\bra{0}\mathrm{tr}_{\mathcal{H}}[\mathcal{F}_3(\ket{0}\!\bra{0}\otimes \ket{0}\!\bra{0})]\ket{0}$ and $\bra{+}\mathrm{tr}_{\mathcal{H}}[\mathcal{F}_3(\ket{0}\!\bra{0}\otimes \ket{0}\!\bra{0})]\ket{+}$ respectively, satisfies
\begin{align}
    |S(\mathcal{F}_3,\ket{\psi})-S(\mathcal{F}_1,\ket{\psi})|\leq
    |S(\mathcal{F}_3,\ket{\psi})-S(\mathcal{F}_2,\ket{\psi})|+
    |S(\mathcal{F}_2,\ket{\psi})-S(\mathcal{F}_1,\ket{\psi})|< 
    \frac{1}{4\sqrt{2}}+\frac{1}{4\sqrt{2}}=\frac{1}{\sqrt{8}},
\end{align}
where, for any quantum operation $\mathcal{F}:\mathcal{L}(\mathcal{H}_c\otimes \mathcal{H})\to \mathcal{L}(\mathcal{H}_c\otimes \mathcal{H})$ and state $\ket{\psi}\in \mathcal{H}_c$, we define $S(\mathcal{F},\ket{\psi}):=\bra{\psi}\mathrm{tr}_{\mathcal{H}}[\mathcal{F}(\ket{0}\!\bra{0}\otimes \ket{0}\!\bra{0})]\ket{\psi}$. Here, the upper bound of $|S(\mathcal{F}_3,\ket{\psi})-S(\mathcal{F}_2,\ket{\psi})|$ is obtained using Lemma \ref{qdri:err}, and that of $ |S(\mathcal{F}_2,\ket{\psi})-S(\mathcal{F}_1,\ket{\psi})|$ is obtained by noting that ${\rm tr}_{\mathcal{H}}[\mathcal{F}_1(\ket{0}\!\bra{0}\otimes \ket{0}\!\bra{0})]$ and ${\mathrm{tr}_{\mathcal{H}}}[\mathcal{F}_2(\ket{0}\!\bra{0}\otimes \ket{0}\!\bra{0})]$ are pure states, and so we can use the fact that the operator norm of $\ket{\beta}\!\bra{\beta}-\ket{\gamma}\!\bra{\gamma}$ for unit vectors $\ket{\beta}$ and $\ket{\gamma}$ is $\sqrt{1-|\braket{\beta|\gamma}|^2}$ (see Eq.~(1.185) of Ref.~\cite{watrous2018theory}).
In particular, $\braket{\beta | \gamma}$ is evaluated as $\bra{0}e^{-ik\pi( A_{l,10l^2}\cos(\theta_{l,10l^2}+\Phi)- a_{l,25k,\Phi})Y}\ket{0}$ and $t=k\pi$. Invoking Eqs.~(\ref{eq::acosphi}) and (\ref{eq::athetafinalbound}), as well as a similar discussion to Eq.~(\ref{eq::psipsi}), we have  $|S(\mathcal{F}_2,\ket{\psi})-S(\mathcal{F}_1,\ket{\psi})|\leq 1/4\sqrt{2}$.

Setting the appropriate measurements as $\ket{0}$ and $\ket{+}$, we can express the success probabilities as
\begin{align}
    S(\mathcal{F}_1,\ket{0})&=\frac{1+\cos(2k\pi A_{l,10l^2}\cos(\theta_{l,10l^2}+\Phi))}{2}
    \nonumber\\
    S(\mathcal{F}_1,\ket{+})&=\frac{1+\sin(2k\pi A_{l,10l^2}\cos(\theta_{l,10l^2}+\Phi))}{2}.
\end{align}
\noindent Thus, an estimate $\hat{v}_{\Phi}$ of $2\pi A_{l,10l^2}\cos(\theta_{l,10l^2}+\Phi)$ can be successfully obtained by robust phase estimation using Lemma \ref{rob_ph}. In particular, by setting $K$ in Eq.~\eqref{eq::KMF} as $K={\rm ceil}[\log_2 (3\sqrt{2}t/ \epsilon)]$, one can estimate $A_{l,10l^2}\cos(\theta_{l,10l^2}+\Phi)$ with root mean square of error upper bounded by $\epsilon /(2\sqrt{2}t)$ with a time complexity $O(l^2t^3n/\epsilon^3)$. This follows from the fact that the time complexity of generating $\ket{\phi(N', l, k\pi/N', 25k, \Phi)}$ is $\Theta (l^2 n k^3)$ and invoking 
\begin{align}
    \sum_{j=1}^K(K-j)r^j=\frac{r^{K+1}-Kr^2+(K-1)r}{(r-1)^2} \quad \quad \quad (r>1).
\end{align}
Thus, the total time complexity of the robust phase estimation procedure is given by $\sum_{j=1}^K M_j O(l^2n(2^{j-1})^3)=O(l^2 n t^3/\epsilon^3)$. 

Finally, by setting $\Phi \in \{ 0 ,\tfrac{\pi}{2}\}$, one can obtain estimates $\hat{v}_0, \hat{v}_{\pi/2}$ of $2\pi A_{l,10l^2}\cos(\theta_{l,10l^2})$ and $2\pi A_{l,10l^2}\sin(\theta_{l,10l^2})$, respectively with the aforementioned error and time complexity.
When $A_{l,10l^2}\cos(\theta_{l,10l^2})$ and $A_{l,10l^2}\sin(\theta_{l,10l^2})$ are estimated with an error of $\delta_1$ and $\delta_2$, respectively, then the quantity $|1-(A_{l,10l^2}e^{i\theta_{l,10l^2}})/(\hat{A}_le^{i\hat{\theta}_l})|$ is upper bounded by
\begin{align}
     \left| 1-\frac{A_{l,10l^2}e^{i\theta_{l,10l^2}}}{\hat{A}_le^{i\hat{\theta}_l}}\right| &=
     \frac{1}{|\hat{A}_le^{i\hat{\theta}_l}|}|A_{l,10l^2}e^{i\theta_{l,10l^2}}-\hat{A}_le^{i\hat{\theta}_l}|
     \nonumber\\
     &=
     \frac{1}{|\hat{A}_le^{i\hat{\theta}_l}|}
     \left|\left(A_{l,10l^2}\cos(\theta_{l,10l^2})-\frac{\hat{v}_0}{2\pi}\right)+i\left(A_{l,10l^2}\sin(\theta_{l,10l^2})-\frac{\hat{v}_{\pi/2}}{2\pi}\right)\right|
     \nonumber\\
     &\leq 
     2\sqrt{\delta_1^2+\delta_2^2} \, ,
\end{align}
where the inequality comes from the fact that Subroutine \ref{subr::gen} returns values for the estimators only if $\hat{A}_{l}=\sqrt{\hat{v}_0^2+\hat{v}_{\pi/2}^2}>1/2$. 
Since the root mean square errors of $\delta_1$ and $\delta_2$ are upper bounded by $\epsilon/(2\sqrt{2}t)$, it follows that 
the root mean square error of $|1-(A_{l,10l^2}e^{i\theta_{l,10l^2}})/(\hat{A}_le^{i\hat{\theta}_l})|$ is upper bounded by $\epsilon /t$, as claimed in the output of Subroutine \ref{subr::gen}.
Finally, note that including the rejection condition of steps 4 and 5 of Subroutine \ref{subr::gen} only increases the average time complexity of generating $\hat{v}_0, \hat{v}_{\pi/2}$ by a constant factor. This is because the root mean squared error of $|1-(A_{l,10l^2}e^{i\theta_{l,10l^2}})/(\hat{A}_le^{i\hat{\theta}_l})|$ is upper bounded by $\epsilon /t$ and according to Eq.~(\ref{eq::Atheta_bound}), we have that $A_{l, 10l^2}\geq 1-\pi^2/80 >1/2$, and thus the probability that $\sqrt{\hat{v}_0^2+\hat{v}_{\pi/2}^2} \leq 1/2$ is smaller than $1/2$ for sufficiently small $\epsilon /t$. 
\qed\\

Above, we have demonstrated the ability to accurately obtain estimates $(\hat{A}_l, \hat{\theta}_l)$ for the case of $l >0$. We now show how these estimates can be used to provide estimates for cases $l \leq 0$. 

\begin{lem}
\label{le::lnegand0}
With respect to estimates $(\hat{A}_l, \hat{\theta}_l)$ with $l>0$ obtained by Subroutine \ref{subr::gen}, define $(\hat{A}_l, \hat{\theta}_l)$ for $l\leq 0$ as
\begin{align}
    (\hat{A}_l, \hat{\theta}_l):=
    \begin{cases}
        (\hat{A}_{-l}, -\hat{\theta}_{-l})&\quad l<0\\
        (1,0)&\quad l=0.
    \end{cases}
    \label{eq::lnegand0}
\end{align}
Thus defined, these provide estimates of $(A_{l,10l^2}, \theta_{l,10l^2})$ in Lemma \ref{zatu_approx} with a root mean squared error of $|1-(A_{l,10l^2}e^{i\theta_{l,10l^2}})/(\hat{A}_le^{i\hat{\theta}_l})|$ upper bounded by $\epsilon /t$.
\end{lem}

\noindent {\bf Proof:} For $l=0$, it is shown in the proof of Lemma \ref{zatu_approx} that $(A_{l,10l^2}, \theta_{l,10l^2})=(1,0)$, thus Eq.~(\ref{eq::lnegand0}) provides an exact estimate of $(A_{l,10l^2}, \theta_{l,10l^2})$. For $l< 0$, note that by Eq.~(\ref{eq::def_Atheta}), we have
\begin{align}
    A_{l,10l^2}e^{i\theta_{l,10l^2}}:= \left[
\frac{\mathrm{tr}(e^{-i\frac{\pi}{20l}H_0})}{2^n}
\right]^{10l^2},
\label{eq::def_Atheta_again}
\end{align}
and so $(A_{l,10l^2},\theta_{l,10l^2})=(A_{-l,10(-l)^2}, -\theta_{-l, 10(-l)^2})$ holds. We can then write
\begin{align}
    \left| 1-
    \frac{A_{l,10l^2}e^{i\theta_{l,10l^2}}}{\hat{A}_le^{i\hat{\theta}_l}}
    \right|
    &=
    \left| 1-
    \frac{A_{-l,10(-l)^2}e^{-i\theta_{-l,10(-l)^2}}}{\hat{A}_{-l}e^{-i\hat{\theta}_{-l}}}
    \right|=
    \left|\left( 1-
    \frac{A_{-l,10(-l)^2}e^{i\theta_{-l,10(-l)^2}}}{\hat{A}_{-l}e^{i\hat{\theta}_{-l}}}
    \right) ^*\right|=
    \left| 1-
    \frac{A_{-l,10(-l)^2}e^{i\theta_{-l,10(-l)^2}}}{\hat{A}_{-l}e^{i\hat{\theta}_{-l}}}
    \right| ,
    \nonumber
\end{align}
whose root mean squared error is upper bounded by $\epsilon /t$, thus asserting our claim.
\qed

\begin{figure*}[b]
    \begin{center}
        \includegraphics[scale=0.7,angle=90]{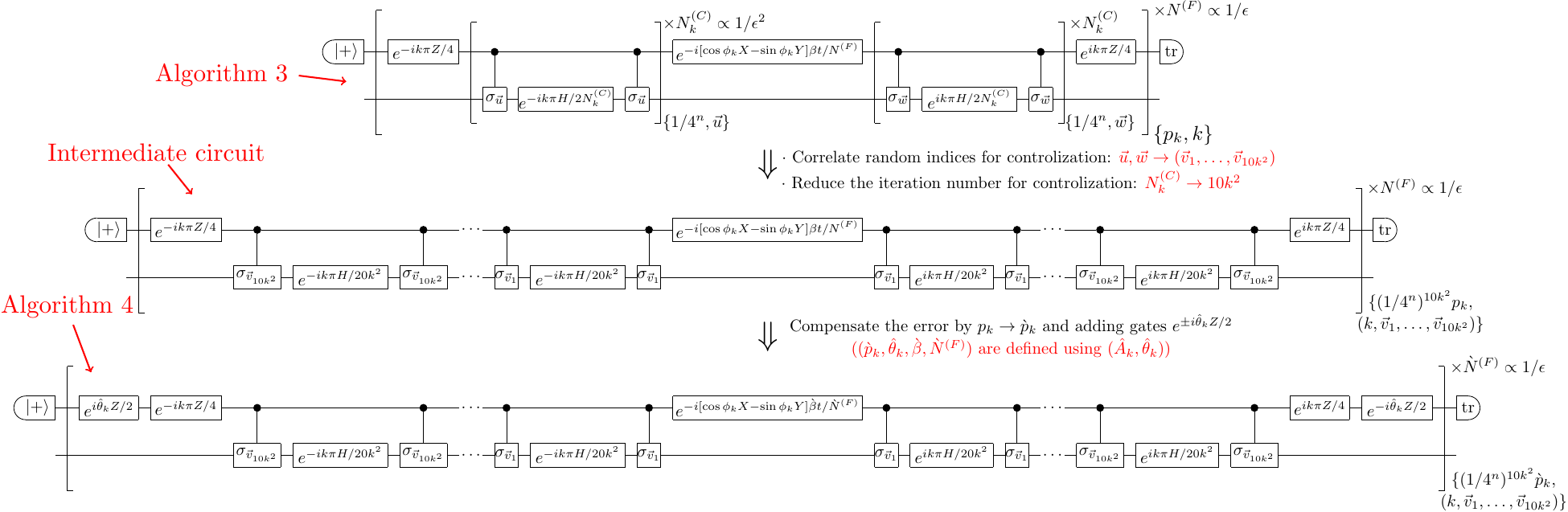}
        \caption{\emph{Summary of compilation of Algorithm \ref{alg::umcompi}.---}First, the random variables $\vec{u},\ \vec{w}\in \{0,1,2,3\}^n$ of the controlization become correlated. Since the error of this intermediate circuit for finite iteration number $10k^2$ of the controlization happens as in Eq.~(\ref{our_two}), this error is compensated in the final step by modifying $N^{(F)}\to \grave{N}^{(F)}$, $\beta \to \grave{\beta}$ and $p_k\to \grave{p}_k$, and introducing an additional gate $e^{\pm i\hat{\theta}_kZ/2}$ according to the parameters $(\hat{A}_{k}, \hat{\theta}_k)$ obtained by Subroutine~\ref{subr::gen} in App.~\ref{app:subrs}.\label{fig::compilation_summary}}
    \end{center}
\end{figure*}

In summary, we have shown that one can estimate the parameters $(\hat{A}_l, \hat{\theta}_l)$, which are necessary to construct appropriate corrections to the ``intermediate circuit'' in order to build the compiled Algorithm \ref{alg::main}, as depicted in Fig.~\ref{fig::compilation_summary}.

\FloatBarrier

\subsection{Error and Time Complexity Analysis of Compiled Algorithm (Algorithm \ref{alg::main})} \label{app:al1}

Here, we will prove a theorem on the error, the mean square of the error, and the time complexity of Algorithm \ref{alg::main}.

\begin{Theorem}
\label{alg1:all_proof}
\begin{enumerate}~
    \item Algorithm \ref{alg::main} outputs $e^{-if(H_0)t}\ket{\psi}$ with an error (in terms of the 1-norm) upper bounded by $\epsilon$, i.e.,
    \begin{align}
        \label{err_def}
        \underset{\substack{ 
    \mathrm{dim}(\mathcal{H}^{\prime}) \\
    \ket{\psi}\in \mathcal{H}\otimes \mathcal{H}^{\prime}\\
    \|\ket{\psi}\|=1 }}{\mathrm{sup}}
        \|\mathcal{F}\otimes \mathcal{I}_{\mathcal{H}'}(\ket{\psi}\bra{\psi})-
        \sum_jp_j(\mathcal{F}_j\otimes \mathcal{I}_{\mathcal{H}'})(\ket{\psi}\bra{\psi})
        \|_1 \leq \epsilon,
    \end{align}
    where 
    $j\in [\bigcup_{k=-\grave{K}}^{\grave{K}} (\{k\}\times (\{0,1,2,3\}^n)^{10k^2})]^{{\grave{N}^{(F)}}}\times (\mathbb{R}\times [0,2\pi))^{\grave{K}}$ is chosen from the set of all random indices, namely $(k, j_k)$ where $k\in \{-\grave{K},\ldots ,\grave{K}\}$ and $j_k=(\vec{v}_{1:10k^2})\in (\{0,1,2,3\}^n)^{10k^2}$ for each of the $\grave{N}^{(F)}$ iterations and parameters $\Phi:=\{(\hat{A}_1, \hat{\theta}_1),\ldots, (\hat{A}_{\grave{K}}, \hat{\theta}_{\grave{K}})\}$ where $\hat{A}_l\in \mathbb{R}$ and $\hat{\theta}_l\in [0,2\pi)$ (which are random due to their dependence on measurement outcomes) with $l\in \{1,\ldots ,\grave{K}\}$, $p_j$ is the probability that $j$ is chosen, $\mathcal{F}(\rho):=e^{-if(H_0)t}\rho e^{if(H_0)t}$, ${ \mathcal{F}}_j$ is the unitary performed when $j$ is chosen, and $\mathcal{H}'$ is an auxiliary system of arbitrary dimension.
    \item Algorithm \ref{alg::main} outputs $e^{-if(H_0)t}\ket{\psi}$ with mean squared error upper-bounded by $2\epsilon$, i.e.,
\begin{align}
    \label{vari_def}
    \underset{\substack{ 
    \mathrm{dim}(\mathcal{H}^{\prime}) \\
    \ket{\psi}\in \mathcal{H}\otimes \mathcal{H}^{\prime}\\
    \|\ket{\psi}\|=1 }}{\mathrm{sup}}
    \sum_{j} p_{j} \|
    ({\mathcal{F} \otimes \mathcal{I}}_{\mathcal{H}^{\prime}})(\ket{\psi}\bra{\psi}) -
    ({\mathcal{F}}_j\otimes { \mathcal{I}}_{\mathcal{H}^{\prime}})(\ket{\psi}\bra{\psi})
    \|_1^2 \leq 2\epsilon.
\end{align}
\item The time complexity of Algorithm \ref{alg::main} comprises of a pre-processing step and a main process. The pre-processing step has a time complexity of $O(\grave{K}^3t^3n/\epsilon^3)+T_{4}$, where $T_{4}$ is the sum of computation time complexities (on a classical computer) for step 1 (calculation of Fourier coefficients $\{\tilde{c}_k\}$ until Eq.~(\ref{grvK}) is satisfied) and step 7 (computation of $\grave{N}^{(F)}$), and $\grave{K}=O[(t/\epsilon)^{1/3}]$. The main process has a time complexity of $O(C_{4,f} t^2n/\epsilon)$, where $C_{4,f}$ is a (function-dependent) constant. 
\end{enumerate}
\end{Theorem}
\noindent Part 1 of Theorem \ref{alg1:all_proof} implies in particular that for an arbitrary input state $\ket{\psi}\in \mathcal{H}$, we have
\begin{align}
    \label{sta_err}
    \|\mathcal{F}(\ket{\psi}\bra{\psi})-\sum_j p_j\mathcal{F}_j(\ket{\psi}\bra{\psi})\|_1
    \leq \epsilon .
\end{align}
In addition, part 2 implies that the mean square of the difference between the ideal state and the single-shot output state is upper bounded by
\begin{align}
    \label{sta_ms}
    \sum_j p_j
    \|\mathcal{F}(\ket{\psi}\bra{\psi})-\mathcal{F}_j(\ket{\psi}\bra{\psi})\|_1^2
    \leq 2\epsilon .
\end{align}
Therefore, the following proof implies the error bounds stated in Algorithm \ref{alg::main}. \\

\noindent \textbf{Proof:} We begin with the first statement. 

\textbf{1.} Steps 2 to 13 of Algorithm \ref{alg::main} perform the following quantum operation 
\begin{align}
    \mathcal{F}_6:=\sum_{\Phi}
    p_{\Phi}\mathcal{F}_{\Phi}
\end{align}
applied to the input state $\ket{+}\otimes\ket{\psi}$, where $\Phi:=\{(\hat{A}_1,\hat{\theta}_1),\ldots ,(\hat{A}_{\grave{K}},\hat{\theta}_{\grave{K}})\}$, $p_{\Phi}$ is the probability $\Phi$ is obtained, and $\mathcal{F}_{\Phi}$ is the quantum operation performed by qDRIFT in steps 9 to 13 of Algorithm \ref{alg::main} whenever $\Phi$ is chosen. The operation $\mathcal{F}_{\Phi}$ approximates the dynamics $e^{-iH_{\Phi}t}$ for the Hamiltonian
\begin{align}
    \label{H_overall}
    H_{\Phi}&:=
    \sum_{k=-\grave{K}}^{\grave{K}}\frac{|\tilde{c}_k|}{\hat{A}_k}\left(\frac{1}{4}\right)^{10nk^2}
    \sum_{j_k\in (\{0,1,2,3\}^n)^{10k^2}}\grave{W}_{k,j_k}^{\dagger}([\cos(\phi_k) X-\sin(\phi_k)Y]\otimes I)\grave{W}_{k, j_k}
    \nonumber\\
    &
    =
    \sum_{k=-\grave{K}}^{\grave{K}}\frac{|\tilde{c}_k|}{\hat{A}_k}
    (e^{-i\tfrac{\hat{\theta}_k}{2}Z}\otimes I)\left[
    \left(\frac{1}{4}\right)^{10nk^2}
    \sum_{j_k\in (\{0,1,2,3\}^n)^{10k^2}}
    (W^{(10k^2)}_{k,j_k})^{\dagger}( [\cos(\phi_k) X-\sin(\phi_k)Y]\otimes I)W^{(10k^2)}_{k, j_k}
    \right]
    (e^{i\tfrac{\hat{\theta}_k}{2}Z}\otimes I)
    \nonumber\\
    &=
    \sum_{k=-\grave{K}}^{\grave{K}}\frac{|\tilde{c}_k|}{\hat{A}_k}A_{k,10k^2}
    (e^{i\tfrac{\theta_{k, 10k^2}-\hat{\theta}_k}{2}Z}\otimes I)\left(
    \begin{array}{cc}
        0&{e^{i\phi_k}}e^{i\tfrac{k\pi}{2} (H_0+I)}\\
        {e^{-i\phi_k}}e^{-i\tfrac{k\pi}{2}(H_0+I)}&0\\
    \end{array}\right)
    (e^{-i\tfrac{\theta_{k, 10k^2}-\hat{\theta}_k}{2}Z}\otimes I).
\end{align}
By defining $\Delta_{k}:=\frac{A_{k,10k^2}e^{i\theta_{k,10k^2}}}{\hat{A}_ke^{i\hat{\theta}_k}}-1$, the last line of Eq.~(\ref{H_overall}) can be rewritten as
\begin{align}
\label{eq::Delta_appears}
    \sum_{k=-\grave{K}}^{\grave{K}}|\tilde{c}_k| 
    \left(
    \begin{array}{cc}
        0&(1+{ \Delta_k}){e^{i\phi_k}}{e^{i\tfrac{k\pi}{2} (H_0+I)}}\\
        (1+{\Delta_k^*}){e^{-i\phi_k}}{e^{-i\tfrac{k\pi}{2} (H_0+I)}}&0\\
    \end{array}
    \right).
\end{align}
This expression can be further simplified by defining a function $f_{\Phi}:[-1,1]\to \mathbb{R}$ as 
\begin{align}\label{eq::f_phi}
f_{\Phi}(x):=\sum_{k=-\grave{K}}^{\grave{K}} \tilde{c}_k\Delta_ke^{ik\pi x}.
\end{align}
Note that the output is real since $\Delta_{-k}=\Delta_k^*$. With this, we have that
\begin{align}
    \label{eq::H_Phi_simpl}
    H_{\Phi}&=
    X\otimes \left[f_{\grave{K}}\left(\frac{H_0+I}{2}\right)+f_{\Phi}\left(\frac{H_0+I}{2}\right)\right]
\end{align}
where $f_{\grave{K}}(x):=\sum_{k=-\grave{K}}^{\grave{K}}\tilde{c}_k e^{ik\pi x}$. 

Now, for any quantum operation $\mathcal{F}:\mathcal{L}(\mathcal{H}_c\otimes \mathcal{H})\to \mathcal{L}(\mathcal{H}_c\otimes \mathcal{H})$, we define the norm
\begin{align}
    \label{our_err}
    E(\mathcal{F}):=    \underset{\substack{\ket{\psi}\in\mathcal{H}_c\otimes \mathcal{H}\otimes \mathcal{H'}\\ \|\ket{\psi}\|=1\\ \mathrm{dim}{\mathcal{H}'} }}{\mathrm{sup}}
    \|\mathcal{F}\otimes \mathcal{I}_{\mathcal{H}'}(\ket{\psi}\bra{\psi})\|_1 ,
\end{align}
where $\mathcal{I}_{\mathcal{H}'}$ is the identity operation in $\mathcal{L}(\mathcal{H}')$. Moreover, we define
\begin{align}
    \mathcal{F}_4(\rho)&:=e^{-i(X\otimes f(H_0))t}\rho e^{i(X\otimes f(H_0))t}
    \nonumber\\    \mathcal{F}_5(\rho)&:=\sum_{\Phi}p_{\Phi}e^{-iH_{\Phi}t}\rho e^{iH_{\Phi}t}=:\sum_{\Phi}p_{\Phi}\mathcal{G}_{\Phi}.
\end{align}
With these definitions at hand, we can upper bound the simulation error $E(\mathcal{F}_6-\mathcal{F}_4)$ using Lemma \ref{qdri:err} and similar arguments to those presented in Eqs.~(\ref{eq::b_inter}) and~(\ref{eq::psipsi}) as follows. 
\begin{align}
    \label{naga_eq}
    &E(\mathcal{F}_6-\mathcal{F}_4)\leq 
    E(\mathcal{F}_6-\mathcal{F}_5)+E(\mathcal{F}_5-\mathcal{F}_4)
    \nonumber\\
    &=    \underset{\substack{\ket{\psi}\in\mathcal{H}_c\otimes \mathcal{H}\otimes \mathcal{H}'\\ \|\ket{\psi}\|=1\\\mathrm{dim}\mathcal{H}'}}{\mathrm{sup}}
    \left\|\sum_{\Phi}p_{\Phi} [\mathcal{F}_{\Phi}\otimes\mathcal{I}_{\mathcal{H}'}(\ket{\psi}\bra{\psi})-
    \mathcal{G}_{\Phi}\otimes \mathcal{I}_{\mathcal{H}'}(\ket{\psi}\bra{\psi})]\right\|_1 
    \nonumber\\    &+\underset{\substack{\ket{\psi}\in\mathcal{H}_c\otimes \mathcal{H}\otimes \mathcal{H}'\\ \|\ket{\psi}\|=1\\ \mathrm{dim}\mathcal{H}'}}{\mathrm{sup}}
    \left\|\sum_{\Phi}p_{\Phi} [(e^{-iH_{\Phi}t}\otimes I)\ket{\psi}\bra{\psi}(e^{iH_{\Phi}t}\otimes I)
    - (e^{-i(X\otimes f(H_0))t}\otimes I)\ket{\psi}\bra{\psi}(e^{i(X\otimes f(H_0))t}\otimes I)]\right\|_1
    \nonumber\\
    &\leq \sum_{\Phi} p_{\Phi} \|\mathcal{F}_{\Phi}-\mathcal{G}_{\Phi}\|_{\diamond}+\sum_{\Phi}p_{\Phi}\underset{\substack{\ket{\psi}\in\mathcal{H}_c\otimes \mathcal{H}\otimes \mathcal{H}'\\\|\ket{\psi}\|=1 \\\mathrm{dim}\mathcal{H}'}}{\mathrm{sup}}2[1-|\bra{\psi}
    (e^{-i\{X\otimes (\tilde{f}((H_0+I)/2)-f_{\grave{K}}((H_0+I)/2)-f_{\Phi}((H_0+I)/2))\}t}\otimes I)\ket{\psi}|^2]^{1/2}
    \nonumber\\
    &\leq \frac{\epsilon}{3} +\sum_{\Phi} p_{\Phi}2\sin(R(\tilde{f}-f_{\grave{K}}-f_{\Phi})t/2) \leq \frac{\epsilon}{3} +\sum_{\Phi} p_{\Phi}[R(\tilde{f}-f_{\grave{K}})+R(f_{\Phi})]t
    \leq \frac{\epsilon}{3} +\frac{\epsilon}{3} +\sum_{\Phi} p_{\Phi}R(f_{\Phi})t
    ,
\end{align}
where for any function $g:[-1,1]\to \mathbb{R}$, we have that $R(g):=2\underset{x\in [-1,1]}{\max}|{g(x)}|$. 

The final term $\sum_{\Phi} p_{\Phi}R(f_{\Phi})t$ in the above equation can be upper bounded as follows. First note that, for $f_\Phi$ defined in Eq.~\eqref{eq::f_phi}, we have
\begin{align}
    \label{R_upp}
    R(f_{\Phi})\leq 2\sum_{k=-\grave{K}}^{\grave{K}} |\tilde{c}_k||\Delta_k|.
\end{align}
Now, the allowed error of Subroutine \ref{subr::gen} in step 3 of Algorithm \ref{alg::main} is set as $\epsilon'=\sqrt{3}\epsilon/[12\pi (\sum_{k=-\infty}^{\infty}|\tilde{c}_k||k|)]$ and the root mean squared error of $|\Delta_k|$ is upper bounded by the allowed error $\epsilon'$ divided by $t$ (Theorem \ref{alg2_vari}).  
Consider the Chebyshev inequality for the random variable $| \Delta_k |$, which we assume w.l.o.g. to have zero mean $\mu$ and whose variance $\sigma_k^2$ is non-zero and finite, which is
\begin{align}
    \textup{Pr}\left[ |\Delta_k | \geq x \right] \leq \frac{\sigma_k^2}{x^2} \, , 
\end{align}
where $x \in [0, \infty)$. Employing the Chebyshev inequality, by setting $x = |k|(\sqrt{3}\epsilon C)/[12\pi (\sum_{k=-\infty}^{\infty}|\tilde{c}_k||k|)t]$ and noting that $\sigma_k^2 \leq \epsilon' /t$, the probability that $|\Delta_k|\leq |k|(\sqrt{3}\epsilon C)/(12\pi (\sum_{k=-\infty}^{\infty}|\tilde{c}_k||k|)t)$ for all $k$ and a fixed positive value $C>0$ is lower bounded by
\begin{align}
    \label{eq::basel_used}
    \prod_{\substack{k\in \{-\grave{K},\ldots,-1\\1,\ldots ,\grave{K}\}}} \left(1-\frac{1}{k^2C^2}\right)\geq 1-2\left(\sum_{k=1}^{\infty}\frac{1}{k^2C^2}\right)= 1-\frac{\pi^2}{3C^2} \,,
\end{align}
where the final equality follows from the identity $\sum_{k=1}^{\infty}(1/k^2)=\pi^2/6$. 

Independently, assuming that $|\Delta_k|\leq |k|(\sqrt{3}\epsilon C)/(12\pi (\sum_{k=-\infty}^{\infty}|\tilde{c}_k||k|)t)$ holds for all $k$, it follows that the r.h.s.\ of Eq. (\ref{R_upp}) is upper bounded by $(\sqrt{3}C\epsilon)/(6\pi t)$. 

Combining these two results, we have that
\begin{align}\label{eq::smprobabilityboundr}
    \mathrm{Pr}\left[R(f_{\Phi})\geq \frac{\sqrt{3}C\epsilon}{6\pi t}\right] \leq \frac{\pi^2}{3C^2}.
\end{align}
Making use of the ``tail expectation formula'',
\begin{align}
    \sum_{\Phi}p_{\Phi} R(f_{\Phi})=
    \int_{0}^{\infty} \mathrm{d}x \,\mathrm{Pr}[R(f_{\Phi})\geq x],
\end{align}
we can upper bound $\sum_{\Phi}p_{\Phi}R(f_{\Phi})$ (by setting $x = \tfrac{\sqrt{3}C\epsilon}{6\pi t}$ above) to yield
\begin{align}
    \label{eq::chebbound}
    \sum_{\Phi}p_{\Phi} R(f_{\Phi})\leq 
    \int_{0}^{\infty} \mathrm{d}x \min\left[1, \left(\frac{\epsilon}{6tx}\right)^2\right]= \int_{0}^{\tfrac{\epsilon}{6t}} \textup{d}x + \int_{\tfrac{\epsilon}{6t}}^\infty\left(\frac{\epsilon}{6tx}\right)^2 \textup{d}x =\frac{\epsilon}{6t}+\frac{\epsilon}{6t}=\frac{\epsilon}{3t} ,
\end{align}
where the minimization is included since probabilities cannot exceed one.

Finally substituting Eq.~(\ref{eq::chebbound}) into Eq.~(\ref{naga_eq}), we obtain
\begin{align}
    E(\mathcal{F}_6-\mathcal{F}_4)\leq \epsilon .
\end{align}
Therefore, 
\begin{align}
    \label{siage}
    \epsilon
    \geq E(\mathcal{F}_6-\mathcal{F}_4)
    &\geq
    \underset{\substack{\ket{\psi}\in\mathcal{H}_c\otimes \mathcal{H}\otimes \mathcal{H'}\\ \|\ket{\psi}\|=1\\ \mathrm{dim}{\mathcal{H}'} }}{{\rm sup}}
    \|
    {\rm tr}_{\mathcal{H}_c}[
    (\mathcal{F}_6-\mathcal{F}_4)\otimes \mathcal{I}_{\mathcal{H}'}]
    (\ket{\psi}\bra{\psi})
    \|_1
    \nonumber\\
    &\geq
    \underset{\substack{\ket{\psi}\in \mathcal{H}\otimes \mathcal{H'}\\ \|\ket{\psi}\|=1\\ \mathrm{dim}{\mathcal{H}'} }}{{\rm sup}}
    \|
    {\rm tr}_{\mathcal{H}_c}[
    (\mathcal{F}_6-\mathcal{F}_4)\otimes \mathcal{I}_{\mathcal{H}'}]
    (\ket{+}\bra{+}\otimes \ket{\psi}\bra{\psi})
    \|_1,
\end{align}
which is equal to the expression in \ref{err_def},
as required. \\

\textbf{2.} This statement follows directly by combining the above result with Lemma \ref{gene_vari}. \\

\textbf{3.} In order to prove this statement regarding the time complexity, we use Lemma \ref{4_smo}, which shows that $\tilde{f}$ defined for any input function $f$ of Algorithm \ref{alg::main} is $4$-smooth.
\\

Based on Lemmas \ref{4_smo} and \ref{Jsmo_fou_conv}, we have that $\sum_{k=K+1}^{\infty}|\tilde{c}_k|=O(1/K^3)$ and $\sum_{k=-\infty}^{-(K+1)}|\tilde{c}_k|=O(1/K^3)$ for $K\ge 1$. Thus, the truncation number $\grave{K}$ defined in Eq.~(\ref{grvK}) is shown to be $O[(t/\epsilon)^{1/3}]$, as claimed. Furthermore, the time complexity can be split into one for the pre-processing stage and one for the main process as follows.\\

{\bf Pre-processing: }For each $k\in \{-\grave{K},\ldots ,\grave{K}\}$, the time complexity of generating $(\hat{A}_k, \hat{\theta}_k)$ is shown in Theorem \ref{alg2_vari} to be $O(k^2t^3n/\epsilon^3)$. Thus, by summing this over all $k$, the total time complexity of steps 2 to 6 in Algorithm \ref{alg::main} is $O(\grave{K}^3t^3n/\epsilon^3)$. Thus the total time complexity of the pre-processing step is $O(\grave{K}^3t^3n/\epsilon^3)+T_{4}$. 
\\

{\bf Main process:} The average time complexity of the main process in Algorithm \ref{alg::main} is evaluated as (number of iterations $\grave{N}^{(F)}$)$\times$(average time complexity of steps 10 to 12) $=N(\grave{\beta}, t, \epsilon/3)\cdot \Theta(\sum_{k=-\grave{K}}^{\grave{K}}\grave{p}_k nk^2) =\Theta( \grave{\beta}^2 t^2 / \epsilon ) \cdot \Theta(\sum_{k=-\grave{K}}^{\grave{K}}|\tilde{c}_k|k^2 n/ \grave{\beta})
= O[ 
(\sum_{k=-\infty}^{\infty}|\tilde{c}_k|)
(\sum_{k=-\infty}^{\infty}|\tilde{c}_k|k^2) t^2 n/ \epsilon
]
=: O(C_{4, f} t^2n/\epsilon)$ for some $f$-dependent constant $C_{4, f}$. (Note that $|\hat{A}_k|\geq 1/2$
due to steps 4 and 5 of Subroutine \ref{subr::gen}). 
\qed

\FloatBarrier

%%%%%%%%%%%%%%%%%%%%%%%%%%%%%%%%%%%%%%%%%%%%%%%%%

\section{QSVT-based UHET Algorithm}\label{app::qsvt}

\noindent We will now present and analyze an alternative procedure to achieve UHET based upon a QSVT procedure. We will finally compare this method to Algorithms \ref{alg::umcompi} and \ref{alg::main}.

\subsection{QSVT-based UHET Algorithm (Algorithm~\ref{alg::qsvt})}\label{app::comparison}

\noindent The QSVT-based algorithm is given in Algorithm~\ref{alg::qsvt} and detailed below. We begin by defining functions $f_{0}$ and $f_{1}$ as
\begin{align}
    \label{f_01}
    f_0(x):=
    \begin{cases}
        f_{0,-}(x)&(0\leq x\leq \frac{1}{2})\\
        \cos\left[f\left( \frac{12\cos^{-1}(x)}{\pi}-3 \right)t\right]&(\frac{1}{2}\leq x \leq \frac{\sqrt{3}}{2})\\
        f_{0,+}(x)&(\frac{\sqrt{3}}{2}\leq x \leq 1)\\
        f_0(-x)&(-1\leq x\leq 0)
    \end{cases}\nonumber\\
    f_1(x):=
    \begin{cases}
        f_{1,-}(x)&(0\leq x\leq \frac{1}{2})\\
        \sin\left[f\left( \frac{12\cos^{-1}(x)}{\pi}-3 \right)t\right]&(\frac{1}{2}\leq x \leq \frac{\sqrt{3}}{2})\\
        f_{1,+}(x)&(\frac{\sqrt{3}}{2}\leq x \leq 1)\\
        f_1(-x)&(-1\leq x\leq 0)
    \end{cases} ,
\end{align}
where $f_{0,-},f_{0,+},f_{1,-},f_{1,+}$ are any functions that are infinitely differentiable on their domains and lead to $f_0$ and $f_1$ such that: 
\begin{enumerate}
    \item $2\sin(\tfrac{\pi}{10})f_0(x)$ and $2\sin(\tfrac{\pi}{10})f_1(x)$ is bounded in $[-1,1]$ for all $x\in [-1,1]$. This ensures that they can be constructed using QSVT (in particular, using the technique presented in Theorem 10 of Ref.~\cite{martyn2021grand}).
    \item $f^{(n)}_{s}(\tfrac{1}{2}^+)=f^{(n)}_{s}(\tfrac{1}{2}^-)$,  $f^{(n)}_{s}(\tfrac{\sqrt{3}}{2}^+)=f^{(n)}_{s}(\tfrac{\sqrt{3}}{2}^-)$, and $f^{(n)}_{s}(0^+)=f^{(n)}_{s}(1^-)=0$ for $s\in \{0,1\}$ and $n\in\{0,1,\ldots ,J-1\}$ for any integer $J\geq 4$. 
\end{enumerate}

\noindent The Hamiltonian $H^Q$ in step 4 is defined as
\begin{align}\label{eq::app-qsvt-Hq}
    H^Q:=\frac{\pi (H_0+3I)}{12} \, ,
\end{align}
the definition of which is designed to shift the spectrum of $H_0$ into the range $[\frac{1}{2},\frac{\sqrt{3}}{2}]$.

The functions $f_{0}$ and $f_{1}$ are defined in such a way that 
\begin{align}
    f_0(\cos H^Q)-if_1(\cos H^Q)=e^{-if(H_0)t}.
\end{align}
Furthermore, as we discuss in more detail below, $f_{0}$ and $f_{1}$ are implementable via QSVT. Since the time complexity of QSVT scales polynomially on the cutoff number $K^Q_s$ \cite{martyn2021grand},  the functions $f_{0}(x)$ and $f_{1}(x)$ are chosen to be class $C^3$ in $x\in [-1,1]$ so that the sum in Eq.~\eqref{kdef} converges rapidly to $2\sin(\tfrac{\pi}{10})f_s(\cos(\pi x))$ and $K^Q_s$ scales slowly with $\epsilon$. 

\setcounter{algorithm}{2}
\begin{algorithm}[H]
    \caption{QSVT-based algorithm}
    \label{alg::qsvt}
    \begin{algorithmic}[1]
        \Statex{\textbf{Input:}}
        \begin{itemize}
            \item A finite number of queries to a black-box Hamiltonian dynamics $e^{\pm iH\tau}$ of a seed Hamiltonian $H$ normalized as $\|H_0\|_{\rm op}=1$ where $H_0$ is the traceless part of $H$, i.e., $H_0:=H-(1/2^n)\mathrm{tr}(H)I$, with $\tau>0$
            \item A class C$^3$ function $f:[-1, 1]\to \mathbb{R}$ such that $f^{(4)}$ is piecewise $C^2$ (see SM, App.~\ref{app::jsmooth})
            \item Input state $\ket{\psi}\in \mathcal{H}$
            \item Allowed error $\epsilon >0$
            \item Time $t>0$
        \end{itemize}
        \Statex{\textbf{Output:}}
         A state approximating $e^{-if(H_0)t}\ket{\psi}\ (t>0)$ with an error in terms Eq.~(\ref{measure_sta}) upper bounded by $\epsilon$ 
    \Statex \hrulefill
    	\Statex{\textbf{Time complexity:}}
            \Statex \hskip1.0em Pre-processing (only once): $O\left[(K_{f,t,\epsilon}^Q)^3 \right]$
            \Statex \hskip1.0em Main Process: $\Theta \left[(K^Q_{f,t,\epsilon})^2/\epsilon \right]$ for $K^Q_{f,t,\epsilon}$ depending on $f$, $t$, and $\epsilon$
            \Statex{\textbf{Total evolution time (main process):}} $\Theta (K^Q_{f,t,\epsilon})$
    	\Statex{\textbf{Used Resources:}}
    	\Statex \hskip1.0em System: $\mathcal{H}$ and two auxiliary qubit $\mathcal{H}_b,\  \mathcal{H}_c$
    	\Statex \hskip1.0em Gates: $e^{\pm iH\tau}\ (\tau>0)$ and controlled-Pauli gates on $\mathcal{L}(\mathcal{H}_b\otimes\mathcal{H}_c\otimes\mathcal{H})$
    	\Statex \hrulefill
        \Statex{\textbf{Procedure:}} 
        \Statex \hspace{-1.5em} {\it Pre-processing:}
        \State Define functions $f_{0}, f_{1}:[-1,1]\mapsto \mathbb{R}$ as per Eq.~(\ref{f_01})
        \State Compute $c^{(s)}_k:=\int_{-1}^1\mathrm{d}x \, 2\sin(\tfrac{\pi}{10})f_{s}[\cos{(\pi x)}]\cos(k\pi x)$ for $k>0$, $c^{(s)}_0:=\int_{-1}^1\mathrm{d}x \, \sin(\tfrac{\pi}{10})f_{s}[\cos{(\pi x)}]$, and $K_s^Q$ ($s\in\{0,1\}$)  satisfying
        \begin{align}
        \label{kdef}
        \left|2\sin(\tfrac{\pi}{10})f_{s}[\cos{(\pi x)}]-\sum_{k=0}^{K_{s}^Q}c^{(s)}_k\cos({\pi k x})\right|< \Theta(\epsilon)
        \end{align}
        for all $x\in [0,1]$. Note that one can recast the above bound in terms of Chebyshev polynomials $T_k$ [defined by $T_k (\cos \theta) := \cos(k\theta)$] as $\left|2\sin(\pi/10)f_s(x')-\sum_{k=0}^{K_{s}^Q}c^{(s)}_kT_k(x')\right|<\Theta(\epsilon)$ for $x=:(1/\pi)\cos^{-1}(x')\ (x'\in [-1,1])$, as we will employ in the coming steps.
        \State Set $K^Q_{f,t,\epsilon}\gets K_{0}^Q+K_{1}^Q$ 
        \State For $s\in \{0,1\}$, find gate sequence of QSVT \cite{martyn2021grand} for a block-encoding unitary $U_s$  of $\sum_{k=0}^{K_{s}^Q}c^{(s)}_kT_{k}[\cos (H^Q)]$ on $\ket{s0}\!\bra{s0}\in \mathcal{L}(\mathcal{H}_b\otimes\mathcal{H}_c)$ (with two extra qubits) using $B_s$ defined Eq.~(\ref{B_s_def}) (see Fig.~\ref{fig::B_s})
        \vspace{0.3em}
        \Statex \hspace{-1.5em} {\it Main Process:}
        \State Initialize $U_{\text{current}}\gets I$
        \For{$s\in \{0,1\}$}
        \State Construct unitary $U_s'$ by replacing $B_s$ in the gate sequence for $U_s$ by the random unitary $B_s'$ which approximate $B_{s}$ up to an error of $\Theta(\epsilon /K^Q_{f,t,\epsilon})$ (see Fig.~\ref{fig::fig::B_s})
        \State $U_{\text{current}}\gets (S^{-s}\otimes I\otimes I)U_{s}'U_{\text{current}}$\Comment{$S:= \ket{0}\bra{0} + i \ket{1}\bra{1}$ is the phase gate}
        \EndFor
        \State Set $U_{\text{current}}\gets (\text{HAD} \otimes I\otimes I)U_{\text{current}}(\text{HAD} \otimes I\otimes I)$
        \State Perform robust oblivious amplitude amplification (see Theorem 28 of \cite{gilyen2019quantum}) with $n=5$ iterations  on the block of $U_{\text{current}}$ specified by $\ket{00}\bra{00}\in \mathcal{L}(\mathcal{H}_b\otimes \mathcal{H}_c)$ 
        (n.b. the unitary $U_{\rm current}^\dagger$ needed for this procedure can be made by inverting all the gates in the gate sequence for $U_{\rm current}$, noting that both black-box dynamics $e^{\pm i H \tau}$ are available)
        \State {\textbf{Return} $U_{\text{current}}(\ket{0}^{\otimes 2}\otimes \ket{\psi})$ } 
    \end{algorithmic}
\end{algorithm}

\begin{figure}[t]
        \includegraphics[width=0.8\linewidth]{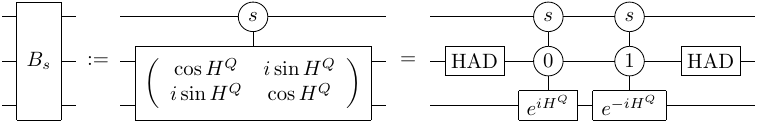}
        \caption{\emph{Circuit representation of $B_s$.---}The circled number in the controlled qubits means that the unitary on the target qubit is applied if the control qubit is that state and otherwise identity is applied.\label{fig::B_s}}
\end{figure}

\noindent With respect to the Hamiltonian defined in Eq.~\eqref{eq::app-qsvt-Hq}, the unitary $B_{s}$ in step 4 is defined as
\begin{align}
B_s:=&\ket{s}\bra{s}\otimes 
        \left(
        \begin{array}{cc}
            \cos(H^Q)&i\sin(H^Q)\\
            i\sin(H^Q)&\cos(H^Q)\\
        \end{array}
        \right)
        +\ket{\bar{s}}\bra{\bar{s}}\otimes I\otimes I.
\label{B_s_def}
\end{align}
See Fig.~\ref{fig::B_s} for its circuit representation.

By defining the following function for $s_b,s_c\in \{0,1\}$ and unitary $U$
\begin{align}
&{\tt ctrl}_2[U, (s_b,s_c)]:= \left[ I-(\ket{s_b s_c}\!\bra{s_b s_c}) \right]
\otimes I + \ket{s_b s_c}\!\bra{s_b s_c}\otimes U,
\end{align}
the unitary $B_s$ can be expressed as 
\begin{align}
B_s=&(I\otimes \mathrm{HAD}\otimes I)
{\tt ctrl}_2[e^{-iH^Q}, (s,1)]
{\tt ctrl}_2[e^{iH^Q}, (s,0)]
(I\otimes \mathrm{HAD}\otimes I).
\end{align}

\begin{figure}[t]
        \includegraphics[width=0.9\linewidth]{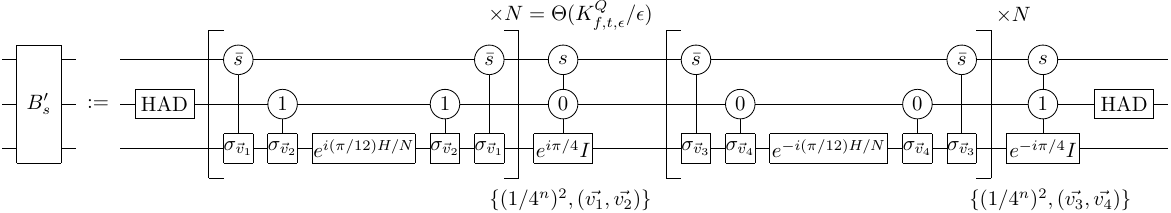}
        \caption{\emph{Definition of $B'_s$.---}${\tt ctrl}_2(e^{\pm iH^Qt}, (s_b,s_c))$ in Fig. \ref{fig::B_s} is replaced by its approximation by controlization.\label{fig::fig::B_s}}
\end{figure}

The operator ${\tt ctrl}_2[e^{\pm iH^Qt}, (s_b,s_c)]$ can be constructed from $e^{\pm iHt}$ via double controlization \cite{dong2019controlled}, which makes use of qDRIFT. In this way, the unitary $B_s$ can be approximated by the circuit shown in Fig. \ref{fig::fig::B_s}.

In summary, the Algorithm \ref{alg::qsvt} simulates the dynamics $e^{-if(H_0)t}$ by the following sequence of steps.
\begin{align}
\label{qsvt_steps}
    &I\otimes I\otimes e^{\pm iHt}\xrightarrow{\text{double controlization}} {\tt ctrl}_2[e^{\pm iH^Qt}, (s_b, s_c)]
    \xrightarrow{\text{Eq.~(\ref{B_s_def})}}
    \nonumber\\
    &B_s=\ket{s}\bra{s}\otimes\left(
        \begin{array}{cc}
            \cos H^Q&i\sin H^Q\\
            i\sin H^Q&\cos H^Q\\
        \end{array}
        \right)+\ket{\bar{s}}\bra{\bar{s}}\otimes I\otimes I
        \xrightarrow{\text{QSVT }\& \text{ post-application of }(S^{{-s}}\otimes I\otimes I})\nonumber\\
    &
    \left(
    \begin{array}{cc|cc}
        2\sin (\tfrac{\pi}{10})f_0(\cos H^Q)&\cdot&\raisebox{-1.2em}{\mbox{\LARGE 0}}&\\
        \cdot&\cdot&&\\
        \hline
        \raisebox{-1.2em}{\mbox{\LARGE 0}}&&{-2i}\sin (\tfrac{\pi}{10})f_1(\cos H^Q)&\cdot\\
        &&\cdot&\cdot\\
    \end{array}
    \right) \xrightarrow{\text{pre- and post-application of }{\rm HAD}\otimes I\otimes I}
    \nonumber\\
    &\left(
    \begin{array}{cccc}
        \sin (\tfrac{\pi}{10})e^{ -if(H_0)t}&\cdot&\cdot&\cdot\\
        \cdot&\cdot&\cdot&\cdot\\
        \cdot&\cdot&\cdot&\cdot\\
        \cdot&\cdot&\cdot&\cdot\\
    \end{array}
    \right)\xrightarrow{\substack{\text{robust oblivious}\\\text{amplitude amplification}}}\left(
    \begin{array}{cccc}
        e^{-if(H_0)t}&0&0&0\\
        0&\cdot&\cdot&\cdot\\
        0&\cdot&\cdot&\cdot\\
        0&\cdot&\cdot&\cdot\\
    \end{array}
    \right).
\end{align}
The leftmost column and the top row of the final matrix in Eq.~(\ref{qsvt_steps}) are both filled with 0 except for the top-left block because both $e^{-if(H_0)t}$ and the entire matrix are unitary.

The procedure of Algorithm \ref{alg::qsvt} ensures that the approximation error of simulating $e^{-if(H_0)t}$ is $O(\epsilon)$. There are two sources of errors in the overall procedure: the error due to the approximation in Eq.~(\ref{kdef}) and that due to the approximation of $B_s$ by the circuit given in Fig. \ref{fig::fig::B_s}. The former error is $O(\epsilon)$ by definition and the latter is upper bounded by $\text{(approximation error of $B_s'$)}\times \text{(number of queries to $B'_s$)}=\Theta (\epsilon /K^Q_{f,t,\epsilon})\times K^Q_{f,t,\epsilon}=\Theta (\epsilon)$. Thus, the sum of these errors is $O(\epsilon)$. 

The time complexity of the main process is asymptotically proportional to $\text{(time complexity of implementing $B_s'$)}\times \text{(number of queries to $B_s'$)}=\Theta(K_{f,t,\epsilon}^Q/\epsilon)\times K_{f,t,\epsilon}^Q=\Theta((K_{f,t,\epsilon}^Q)^2/\epsilon)$ (indeed, the total time complexity of gates other than $B_s'$ grows slower than $(K_{f,t,\epsilon}^Q)^2/\epsilon$ and can thus be ignored). Moreover, the time complexity for the pre-processing step of Algorithm \ref{alg::qsvt} is $O((K^Q_{f,t,\epsilon})^3)$, which can be evaluated by noting that obtaining the gate sequence for QSVT implementing a function of polynomial of degree $d$ requires a time complexity of $O(d^3)$ \cite{chao2020finding}. The total evolution time can be evaluated as $\text{(time evolution of approximating $B_s'$)}\times \text{(number of queries to $B_s'$)}=\Theta(1)\times K_{f,t,\epsilon}^Q=\Theta(K_{f,t,\epsilon}^Q)$.

\FloatBarrier

\subsection{Algorithm Comparison}\label{app::qsvtcomparisondetails}

\noindent We now compare the time complexities of the three algorithms: Algorithm \ref{alg::umcompi} (uncompiled), Algorithm \ref{alg::main} (compiled), and Algorithm \ref{alg::qsvt} (QSVT-based). Recall that for each algorithm here, the time complexity dominates the total evolution time regarding the scaling behavior, and thus we only focus on the former in this section. The time complexities of the pre-processing and the main processes are summarized in Table \ref{runtimes}.\\

\begin{table}[t]
\begin{center}
\begin{tabular}{|c|c|c|}
\hline
&Pre-processing Step&Main Process\\
\hline 
Algorithm \ref{alg::umcompi}&$T_{3}$&$ O(C_{3,f}t^4n/\epsilon^3)$\\
\hline 
Algorithm \ref{alg::main}& $O(\grave{K}^3t^3n/\epsilon^3)+T_{4}$&$O(C_{4,f}t^2n/\epsilon)$\\
\hline
Algorithm \ref{alg::qsvt}&$O[(K_{f,t,\epsilon}^Q)^3]$&$\Theta [(K_{f,t,\epsilon}^Q)^2/\epsilon]$ \\
\hline
\end{tabular}
\end{center}
\caption{\emph{Comparison of time complexities of Algorithm \ref{alg::umcompi} (uncompiled), Algorithm \ref{alg::main} (compiled), and Algorithm \ref{alg::qsvt} (QSVT-based).---}The times $T_{3}$ and $T_{4}$ refer to the total classical computation time complexity for computing the Fourier coefficients and the values $N^{(F)}$ and $\grave{N}^{(F)}$ in Algorithms \ref{alg::umcompi} and \ref{alg::main} respectively. $C_{3, f}$ and $C_{4, f}$ refer to function-dependent constants in Algorithms \ref{alg::umcompi} and \ref{alg::main} respectively. The quantity $\grave{K}$ is upper bounded as $O[(t/\epsilon)^{1/3}]$. The quantity $K_{f,t,\epsilon}^Q$, which depends on the function $f$, the time $t$, and the allowed error $\epsilon$, is upper bounded as  $O(1/\epsilon^{1/3})$ in general, and is lower bounded as $K^Q_{f,t,\epsilon}=\Omega (1/\epsilon^{2/(2J+1)})$  in the special case of  $f$ being strictly periodically $J$-smooth.
\label{tab::comparion}} 
\label{runtimes}
\end{table}

\noindent {\bf Pre-processing: }Rigorous comparison of pre-processing time complexities is difficult in general due to the difficulty in analyzing the quantities $T_3$ and $T_4$. Nevertheless, the time complexity for Algorithm \ref{alg::main} increases faster in terms of $\epsilon$ and $n$ than that of Algorithm \ref{alg::qsvt} because $K^{Q}_{f,t,\epsilon}$ does not depend on $n$ and its $\epsilon$ dependence is given by $K^{Q}_{f,t,\epsilon}=O(1/\epsilon^{1/3})$ (this can be seen by considering that $f_s(x)$ defined in Eq.~\eqref{f_01} is periodically 4-smooth and applying the result of Lemma \ref{Jsmo_fou_conv}, and subsequently noting that $K^Q_{f,t,\epsilon}=K^Q_0+K^Q_1$).\\ 

\noindent {\bf Main Process: }The main process scaling coefficients for Algorithms~\ref{alg::umcompi} and \ref{alg::main}, i.e., 
\begin{align}
    C_{3, f}:=\left(\sum_{k=-\infty}^{\infty}|\tilde{c}_k|\right)^3\left(\sum_{k=-\infty}^{\infty}|\tilde{c}_k|k^2\right) \quad \text{and}\quad
    C_{4, f}:=\left(
    \sum_{k=-\infty}^{\infty}|\tilde{c}_k|
    \right)\left(
    \sum_{k=-\infty}^{\infty}
    |\tilde{c}_k|k^2
    \right),
\end{align}
only depend upon the function $f$ and not on $n,\ t,$ and $\epsilon$. In terms of $t$ and $1/\epsilon$, the time complexity of Algorithm \ref{alg::main} scales slower than that of Algorithm \ref{alg::umcompi}. Furthermore, both algorithms are linear in terms of $n$. Therefore, it follows that compilation reduces the time complexity of the main process. For completeness, the explicit value of $C_{4, f}$ is calculated in the proof of Theorem \ref{alg1:all_proof}. 

In contrast, the scaling coefficient for the main process of Algorithm~\ref{alg::qsvt}, $K^Q_{f,t,\epsilon}$, depends on $f,\ t,$ and $\epsilon$, and its explicit expression is difficult to obtain in general. Nonetheless, below we show that for a particular class of functions, $K^Q_{f,t,\epsilon}$ depends on the allowed error $\epsilon$ as $\Omega (1/\epsilon^{2/9})$. Moreover, we show that an even larger class of functions satisfies $K^Q_{f,t,\epsilon}=\Omega (1/\epsilon^{q})$ for some $0<q\leq 2/9$. Also, since $K^Q_s$ are the cutoff numbers  used to approximate functions $2\sin(\tfrac{\pi}{10})f_s[\cos (\pi x)]$ via their Fourier series and $f_s(x)$ oscillates with frequency proportional to $t$ in the range $x\in [1/2, \sqrt{3}/2]$ as can be seen from Eq.~(\ref{f_01}), it is expected that $K^Q_{f,t,\epsilon}=K^Q_0+K^Q_1$ increases as $t$ grows. The scaling of $K^Q_{f,t,\epsilon}$ can be obtained as an instance of the following Lemma.

\begin{lem}
    \label{nec_cond}
    For a periodically $4$-smooth function $g:[-1,1]\to \mathbb{R}$, the following inequality is a necessary condition for $g_K(x):=\sum_{k=-K}^{K}c_k e^{-i\pi k x}$ ($c_k$ are Fourier coefficients of $g$) to satisfy $|g(x)-g_K(x)|\leq \epsilon$ for all $x\in [-1, 1]$:
    \begin{align}
        \label{nec_eq}
        \sum_{k=-\infty}^{-K-1}|c_k|^2+
        \sum_{k=K+1}^{\infty} |c_k|^2
        \leq \epsilon ^2.
    \end{align}
\end{lem}

\noindent {\bf Proof: }$\int_{-1}^1\mathrm{d}x\, |g(x)-g_K(x)|^2\leq 2\epsilon^2$ is a necessary condition for $|g(x)-g_K(x)|\leq \epsilon$ for all $x\in [-1, 1]$. Due to Parseval's identity, $\int_{-1}^1\mathrm{d}x |g(x)-g_K(x)|^2=2[\sum_{k=-\infty}^{-K-1}|c_k|^2+
\sum_{k=K+1}^{\infty} |c_k|^2]$, thus Eq. (\ref{nec_eq}) is a necessary condition of $|g(x)-g_K(x)|\leq \epsilon$ for all $x\in [-1, 1]$.\qed

Assuming that $f$ is strictly periodically $J$-smooth, $f_0$ and $f_1$ in Eq.~(\ref{f_01}) can also be defined to be strictly periodically $J$-smooth, thus $K^Q_{f,t,\epsilon}$ satisfies $K^Q_{f,t,\epsilon}=\Omega (1/\epsilon^{2/(2J+1)})$ by Lemma \ref{Jsmo_fou_conv}. In some cases, e.g., when $f^{(4)}$ has some jump (i.e., non-removable) discontinuities, $f$ becomes strictly periodically 4-smooth, and thus for some class of functions $f$, $K^Q_{f,t,\epsilon}=\Omega (1/\epsilon^{2/9})$ holds.

We can now compare the scaling of time complexities of all three algorithms in terms of $\epsilon$. For any class of functions $f$ such that $K^Q_{f,t,\epsilon}$ scales as $\Omega (1/\epsilon ^q)\ (0<q\leq 2/9)$, the time complexity scaling of the main processes of Algorithms \ref{alg::umcompi}, \ref{alg::main}, and \ref{alg::qsvt} in terms of $\epsilon$ behaves as $O(1/\epsilon^3),\ O (1/\epsilon)$, and $\Omega(1/\epsilon^{1+2q})$, respectively. By noting that the $\epsilon$ dependence of the time complexity of the main process of Algorithm \ref{alg::qsvt} is $O(1/\epsilon^{5/3})$ due to the relation $K^Q_{f,t,\epsilon}=O(1/\epsilon^{1/3})$, the hierarchy presented in Eq.~(\ref{order_three}) follows. \\

\noindent Finally, we describe the intuition behind two technical factors that make Algorithm \ref{alg::main} more efficient than Algorithm \ref{alg::qsvt} in terms of its $\epsilon$ dependence. \\

\noindent {\bf Efficient inputting of high-frequency terms: } QSVT requires $d$ queries in total to the block-encoding unitary and its inverse when implementing a polynomial function of degree $d$. Consequently, the time complexity of the main process of Algorithm \ref{alg::qsvt} increases proportionally to the total cutoff number $K^Q_{f,t,\epsilon}$, which gets larger as the precision increases. On the other hand, the time complexity of the main process of Algorithm \ref{alg::main} has no explicit dependence on the cutoff number $\grave{K}$. This feature is enabled because the information of Fourier coefficients $\tilde{c}_k$ is input by random sampling according to the magnitude of $\tilde{c}_k$. As can be seen from step 12 of Algorithm \ref{alg::main}, applying unitaries corresponding to high frequency (i.e., large $k$) requires $\Theta(k^2)$ time complexity and is an obstacle in reducing the total time complexity in general. Fortunately, such high-frequency terms $\tilde{c}_k$ for larger $k$ typically have smaller magnitude $|\tilde{c}_k|$ than their low-frequency counterparts, and therefore costly unitaries (i.e., for large $k$) are rarely chosen. Moreover, the average time complexity of one iteration (steps 10 to 12 of Algorithm \ref{alg::main}) given by $\Theta [(\sum_{k=-\grave{K}}^{\grave{K}}|\tilde{c}_k|k^2)/(\sum_{k=-\grave{K}}^{\grave{K}}|\tilde{c}_k|)]$ is tailored to converge in the limit $\grave{K}\to \infty$ by modifying the input function $f$ to a periodically smooth function $\tilde{f}$ with rapidly converging Fourier coefficients, and thus has no dependence on $\grave{K}$.\\

\noindent {\bf Iteration number in controlization procedure is independent of $\epsilon$: }In Algorithm \ref{alg::qsvt}, the allowed error of {\tt ctrl}($e^{-iHt})$ for making $B'_s$ is proportional to $1/K^Q_{f,t,\epsilon}$ because $B'_s$ and its inverse is called $K^Q_{f,t,\epsilon}$ times in total and the error accumulates with each query. Since the iteration number of controlization increases with increasing accuracy, the time complexity of main process of Algorithm \ref{alg::qsvt} gains an additional dependence on $K^Q_{f,t,\epsilon}$, which in turn depends upon $\epsilon$. This same logic holds for Algorithm \ref{alg::umcompi}. On the other hand, the iteration number $10k^2$ of controlization in Algorithm \ref{alg::main} is independent of $ \grave{N}^{(F)}$ and consequently on $\epsilon$ due to compilation, and thus the controlization does not introduce any additional $\epsilon$ dependence.

\end{document}